\documentclass[authoryear]{elsarticle}
\usepackage{soul}
\usepackage{pstricks,graphicx,color}
\usepackage{enumerate}
\usepackage{amsmath,amssymb,latexsym,url,amsthm}
\usepackage{subfigure}
\usepackage{longtable}
\usepackage{array}
\usepackage{enumitem}

\newtheorem{myalg}{Algorithm}

\newcommand{\be}{\begin{equation}}
\newcommand{\ee}{\end{equation}}
\newcommand{\ba}{\begin{eqnarray}}
\newcommand{\ea}{\end{eqnarray}}
\newcommand{\ban}{\begin{eqnarray*}}
\newcommand{\ean}{\end{eqnarray*}}
\newcommand{\bitem}{\begin{itemize}}
\newcommand{\eitem}{\end{itemize}}
\newcommand{\benum}{\begin{enumerate}}
\newcommand{\eenum}{\end{enumerate}}

\newcommand{\diff}{\text{d}}

\definecolor{subtler}{rgb}{1,0,0.1}    
\def\tt{\texttt}
\def\scats{SCATS}
\def\scatsl{SCATS-L}
\def\scatsf{SCATS-F}
\def\sotl{SOTL}
\def\sotla{SOTL}

\def\mp{\mathcal{P}} 
\def\mc{{C}} 
\def\mv{{V}} 
\def\ms{{S}} 
\def\ml{l} 

\def\smallest{0.15}
\def\smaller{0.32}
\def\oneup{0.33}
\def\twoup{0.45}
\def\threeup{0.3}
\def\fourup{0.475}

\def\intersectionturnlanes		{\includegraphics[scale=0.14]{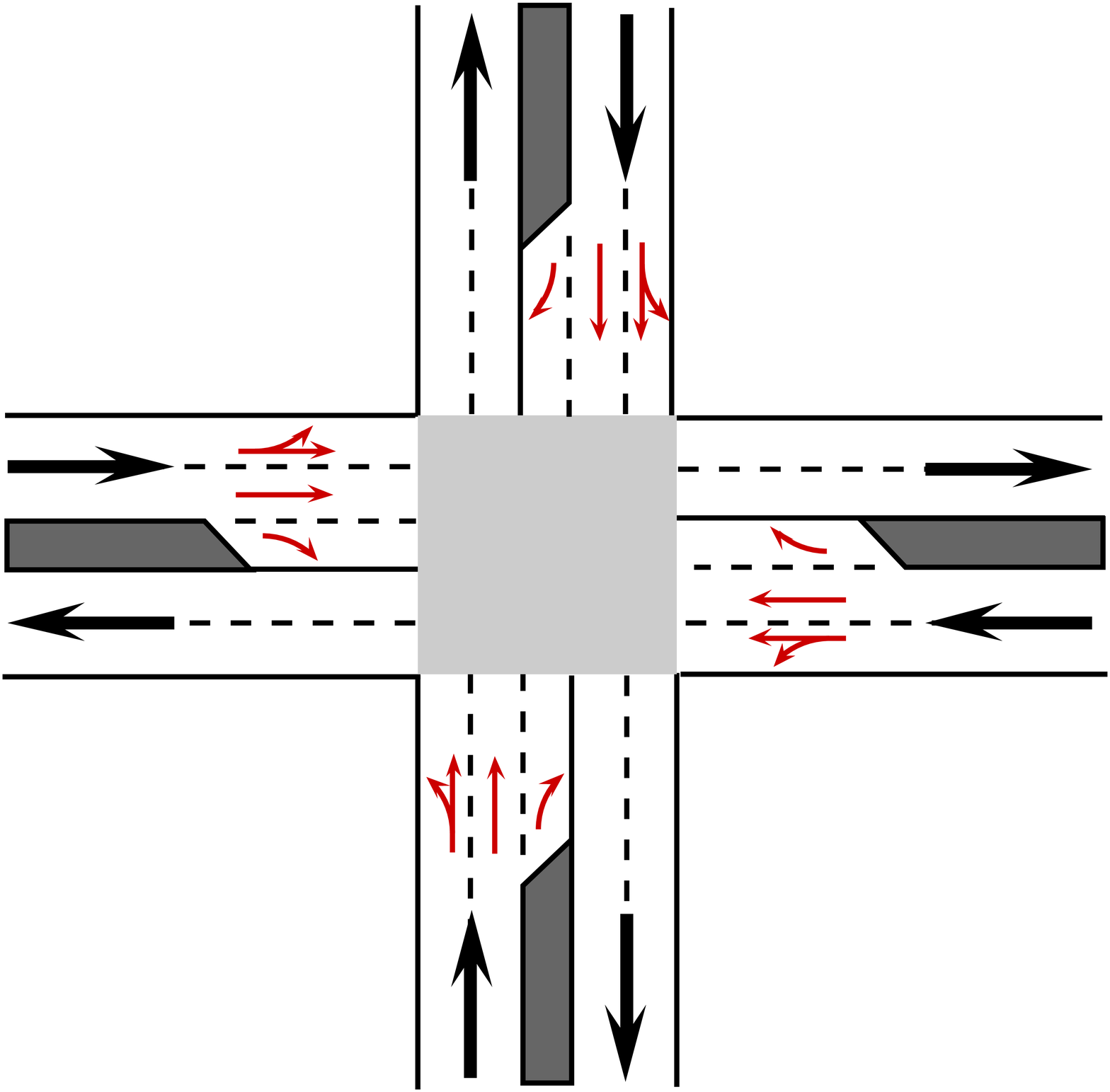}}
\def\phasea				{\includegraphics[scale=\smallest]{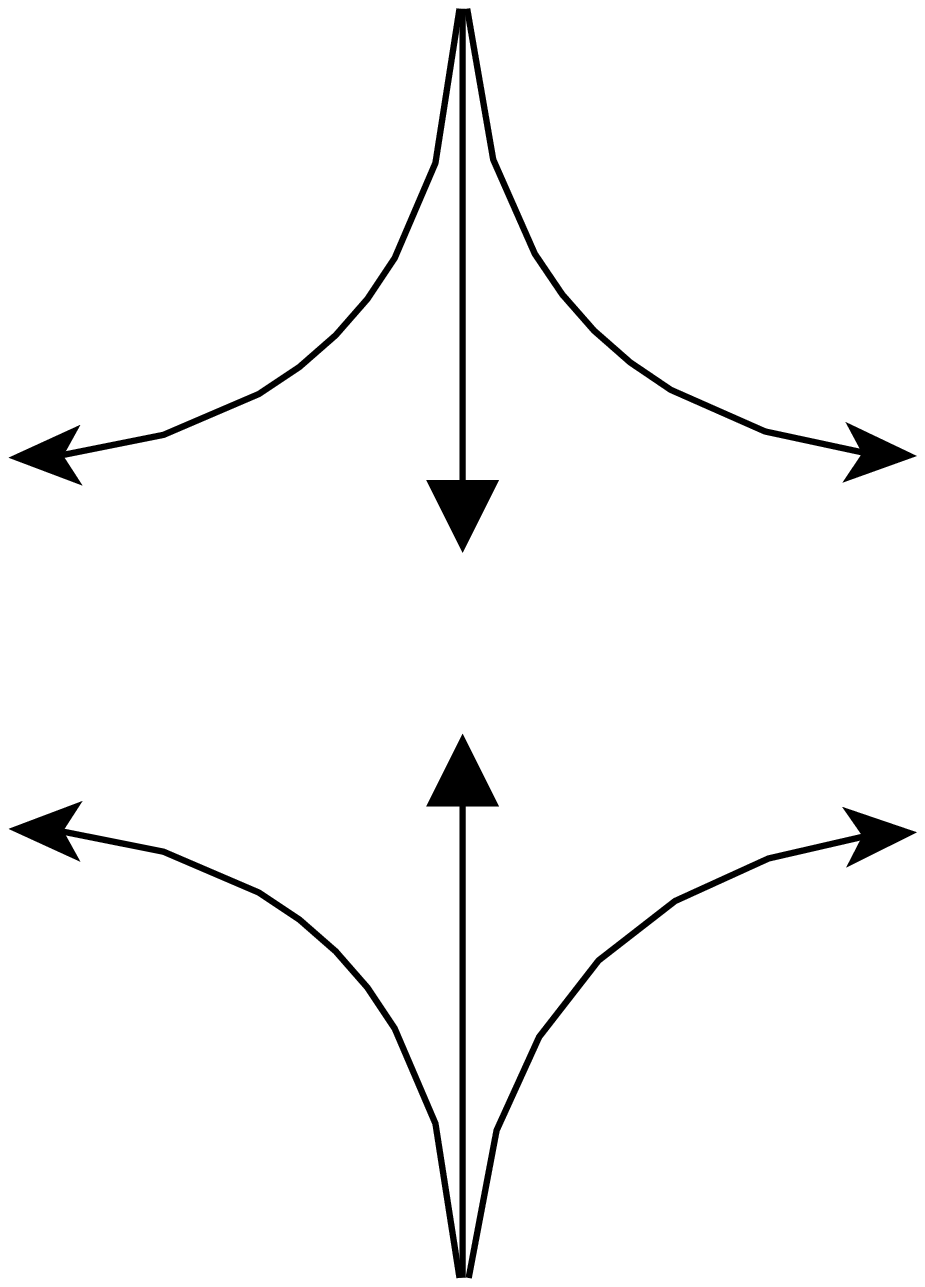}}
\def\phaseb				{\includegraphics[scale=\smallest, angle=90]{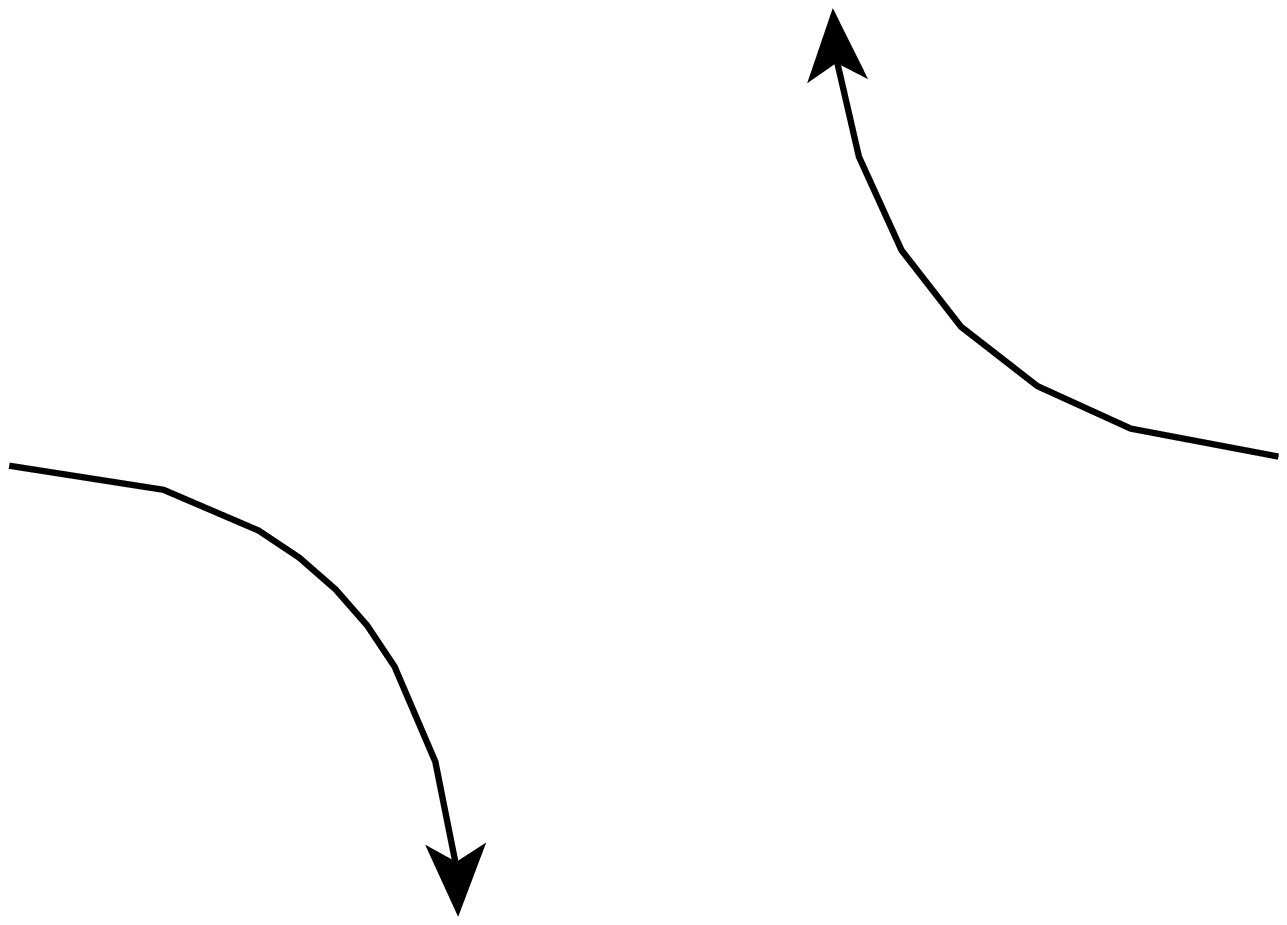}}
\def\phasec				{\includegraphics[scale=\smallest, angle=90]{fig2a.eps}}
\def\phased				{\includegraphics[scale=\smallest]{fig2b.eps}}
\def\latticeeighta 			{\includegraphics[scale=\twoup]{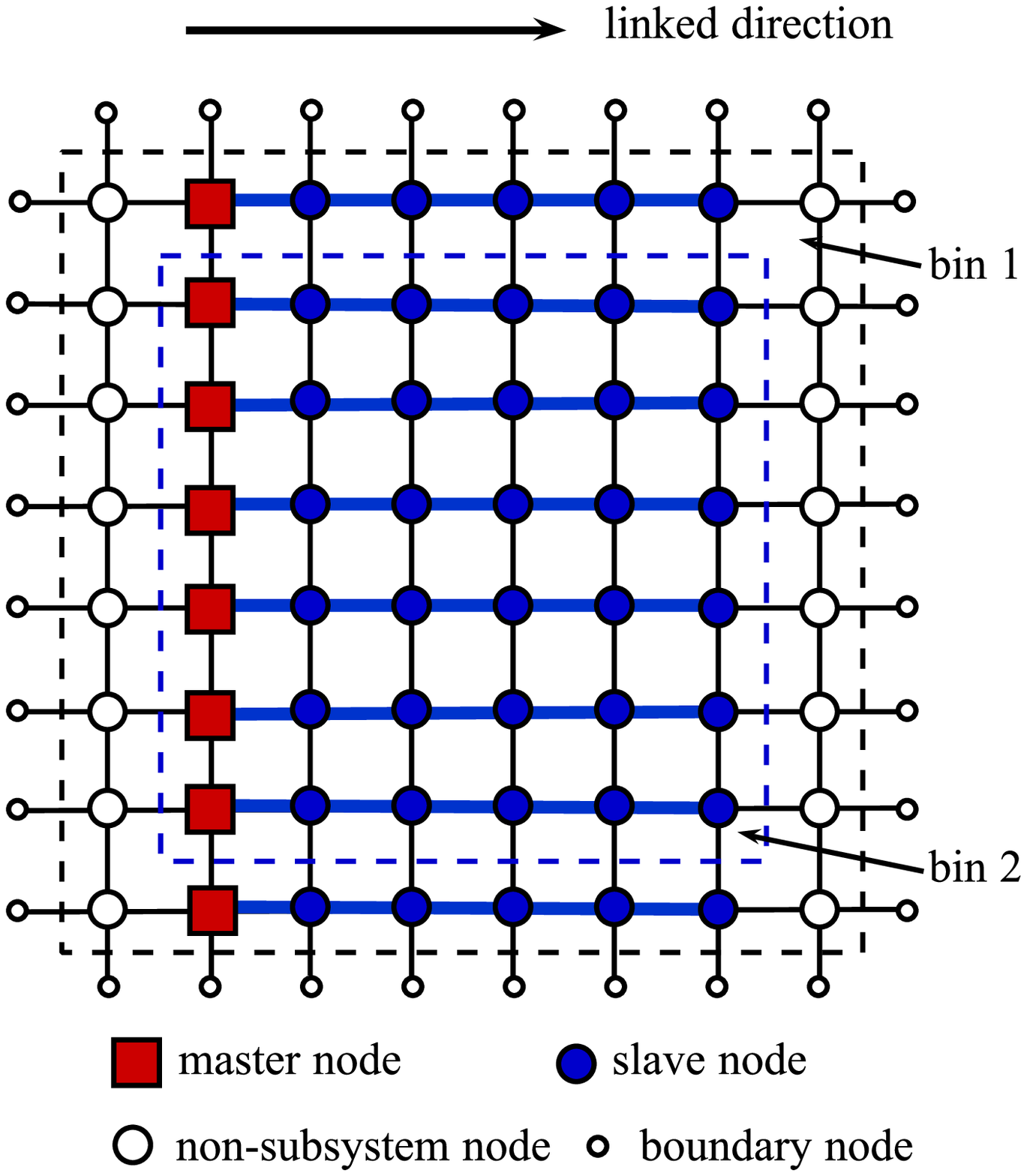}}

\def\fdcuscatsfhoursturnten	{\includegraphics[scale=\oneup]{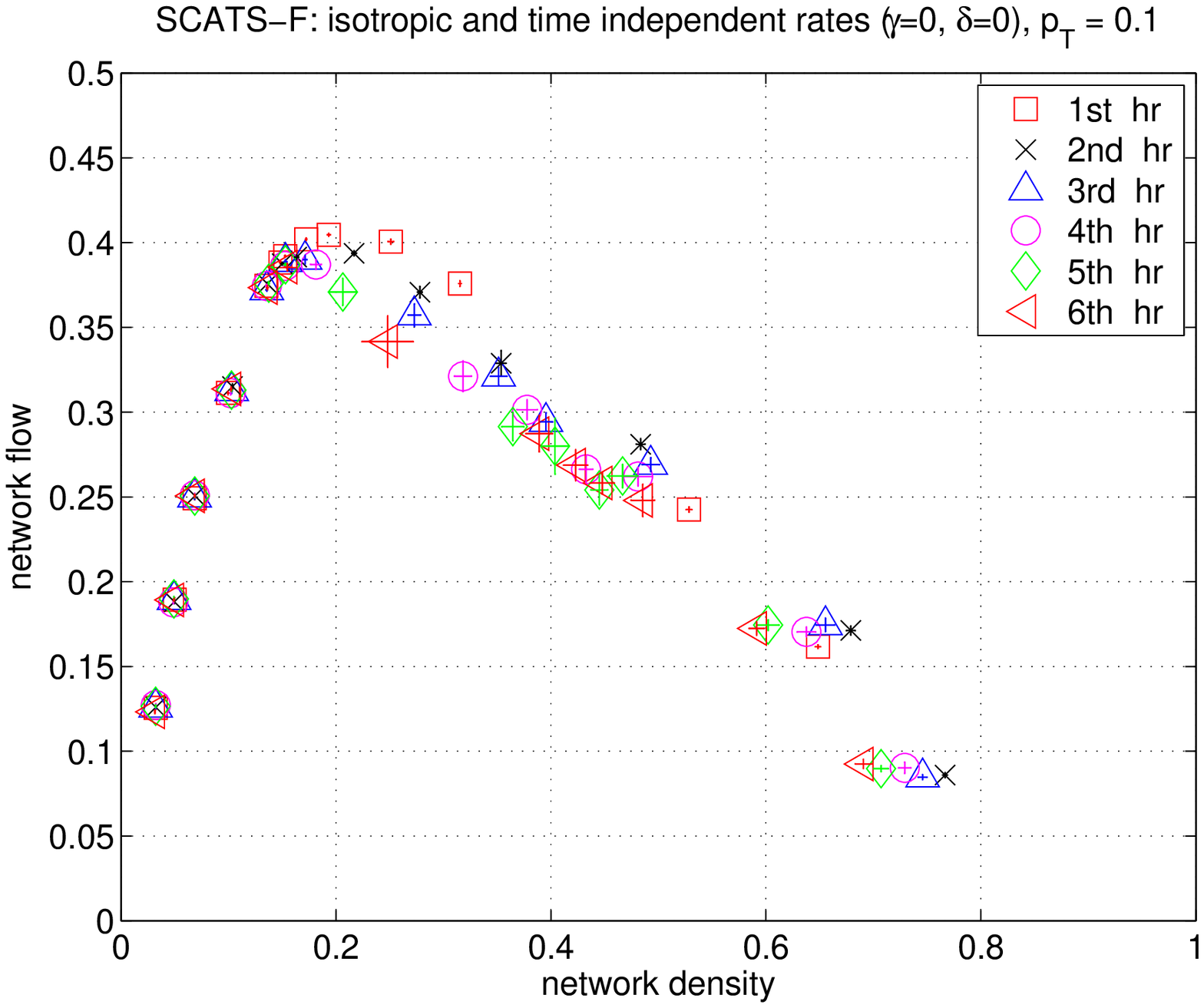}}
\def\fdcuscatslhoursturnten	{\includegraphics[scale=\oneup]{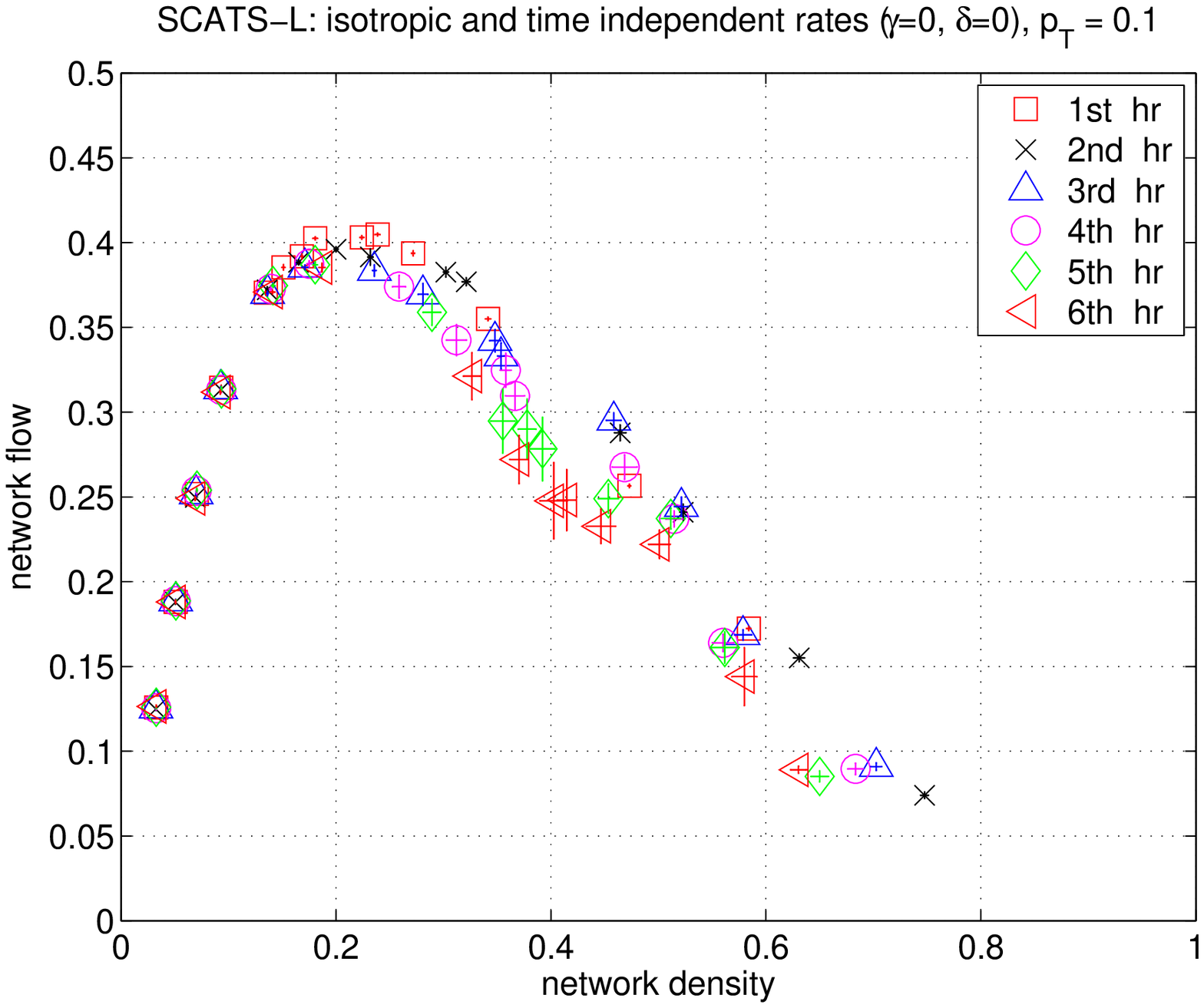}}
\def\fdcusotlhoursturnten		{\includegraphics[scale=\oneup]{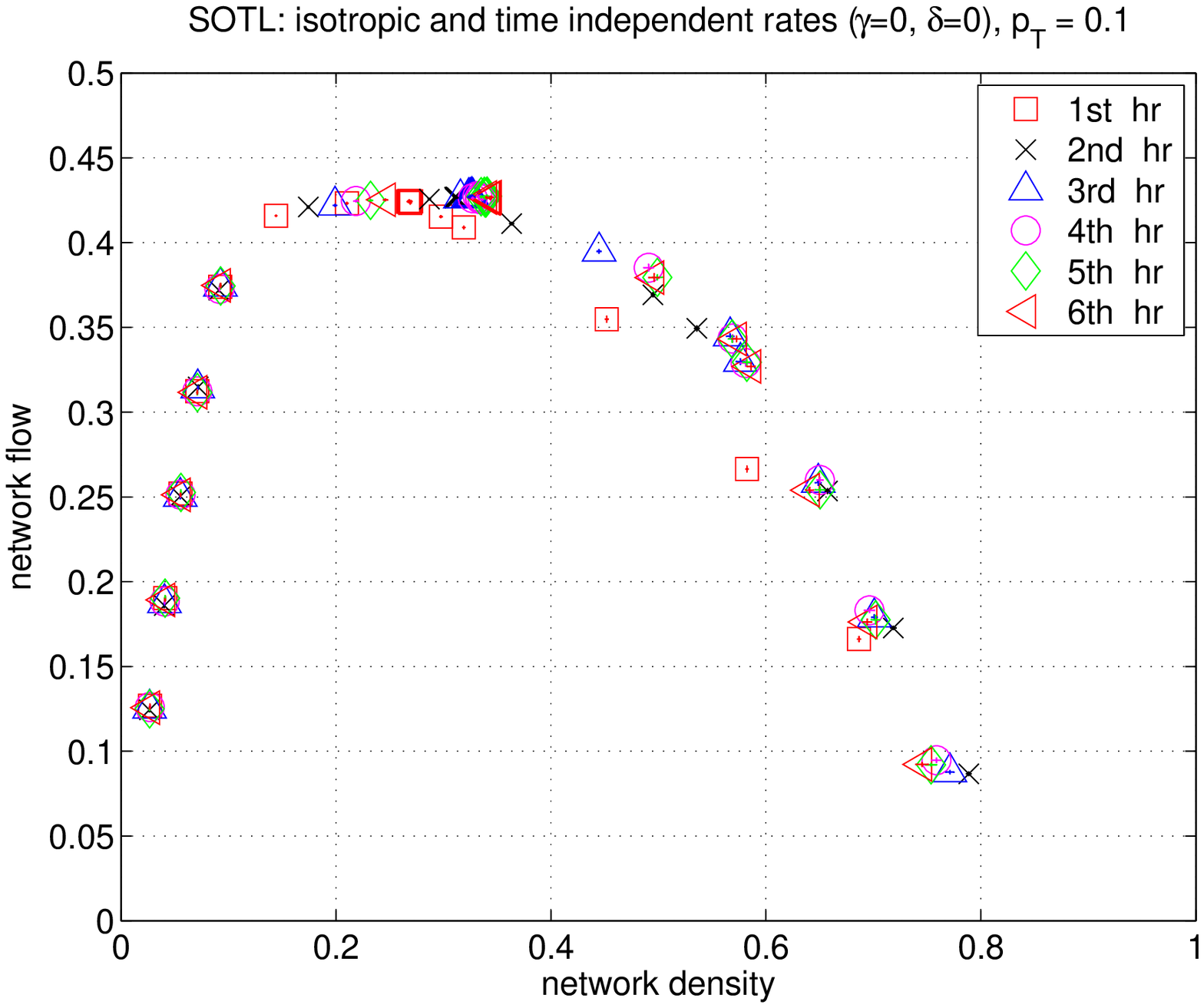}}
\def\fduniformturnten		{\includegraphics[scale=\oneup]{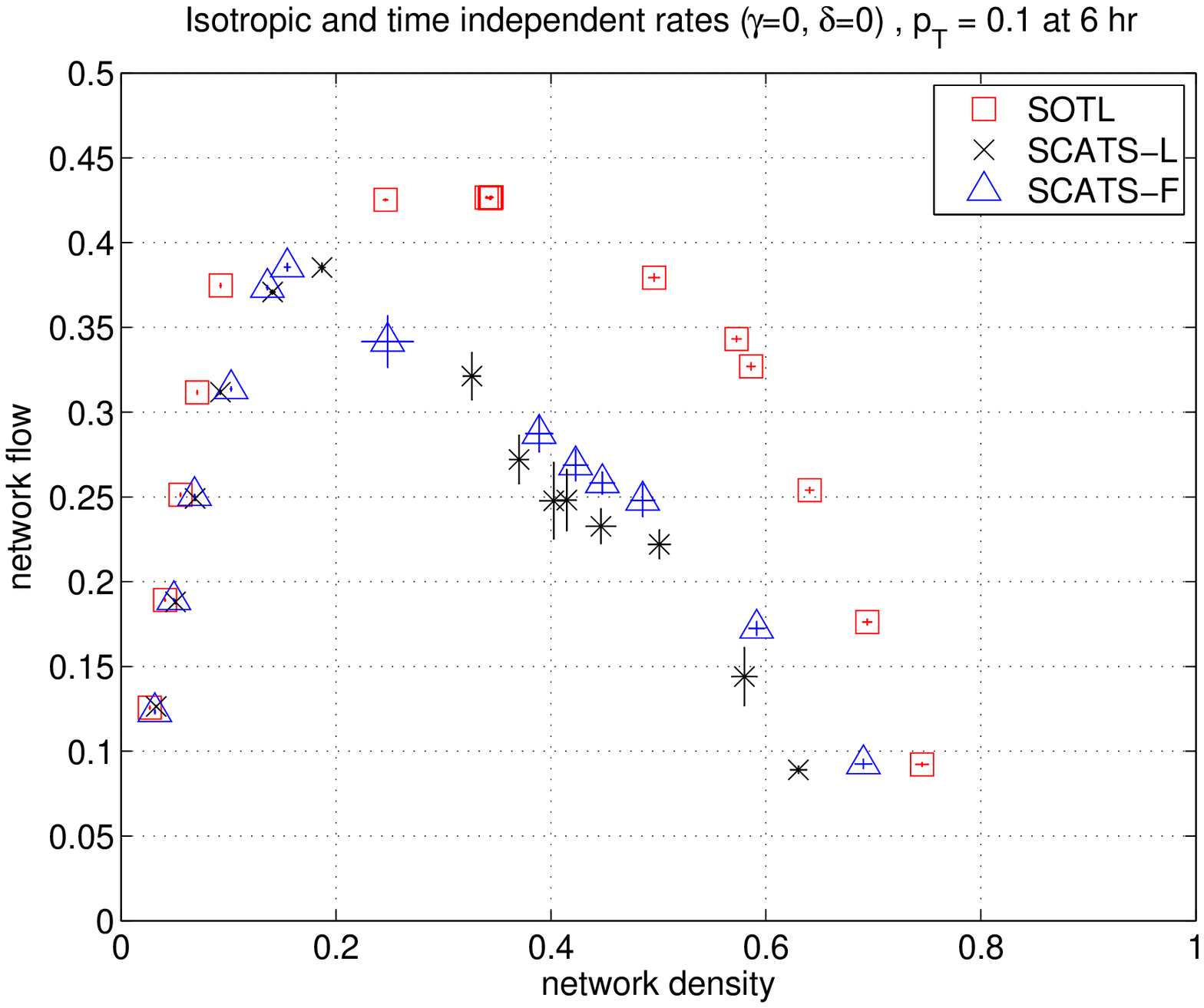}}

\def\fdheteroRhoscatslhoursturnten {\includegraphics[scale=\oneup]{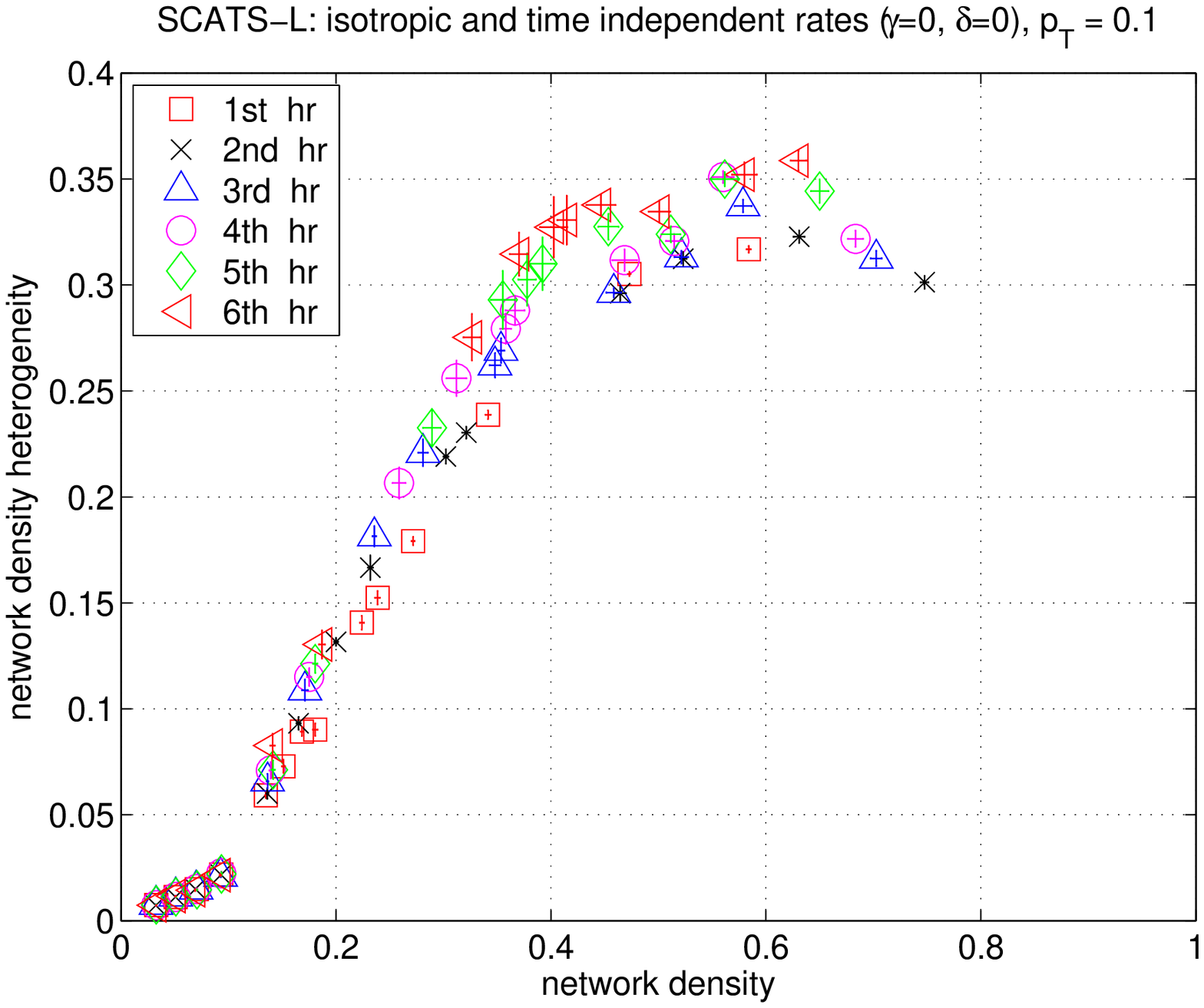}}
\def\fdheteroJscatslhoursturnten {\includegraphics[scale=\oneup]{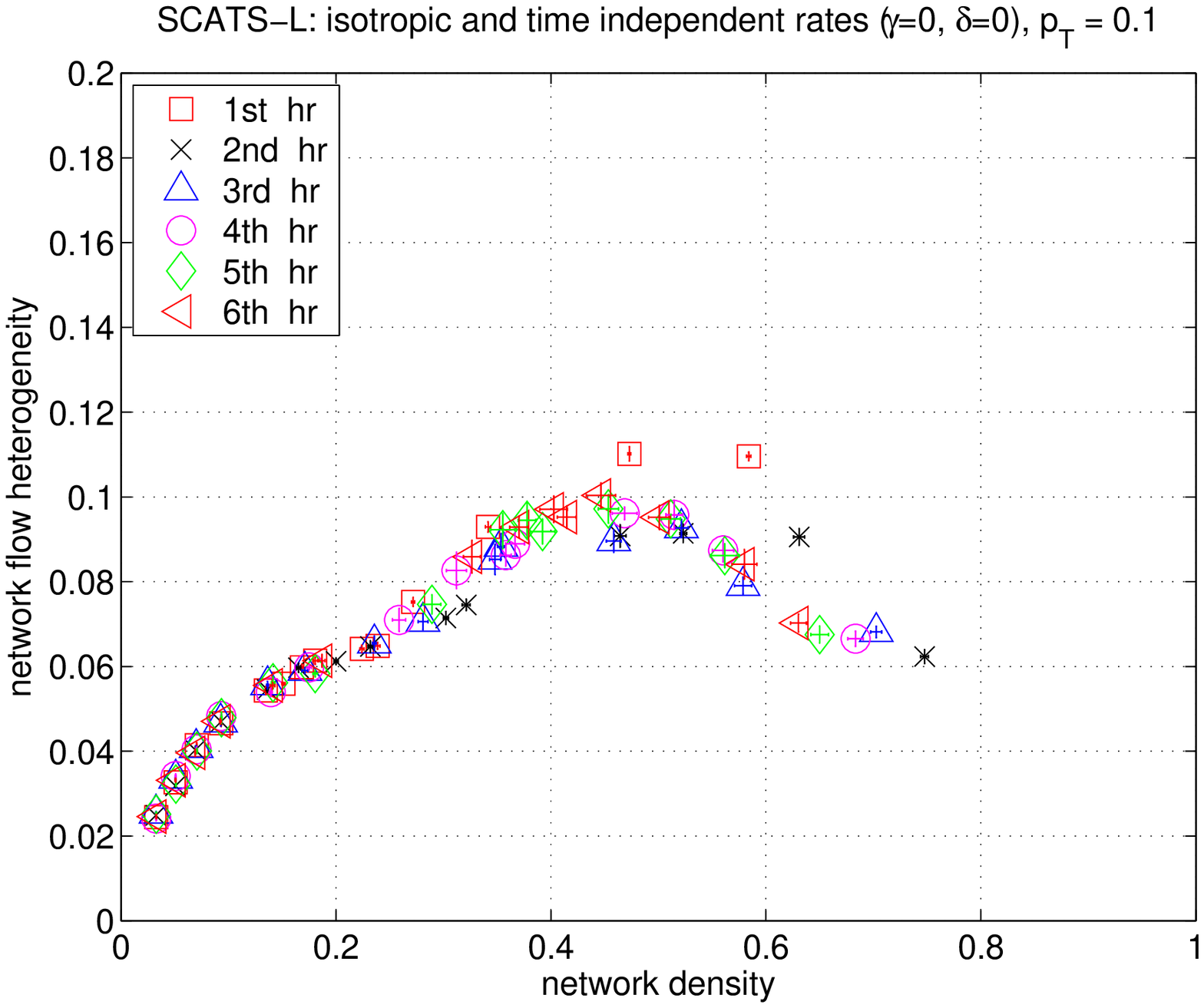}}
\def\fdheteroRhosotlhoursturnten {\includegraphics[scale=\oneup]{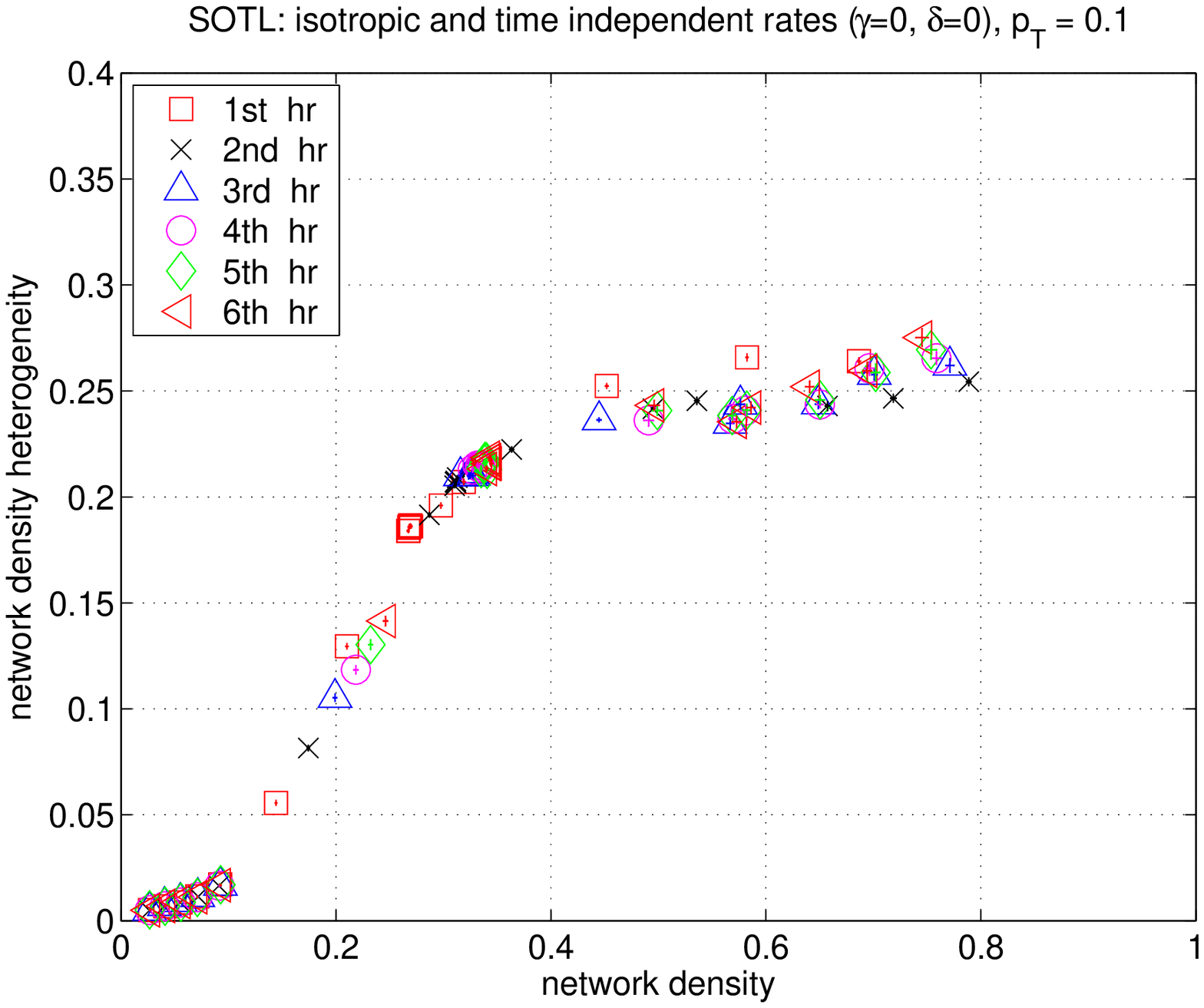}}
\def\fdheteroJsotlhoursturnten	 {\includegraphics[scale=\oneup]{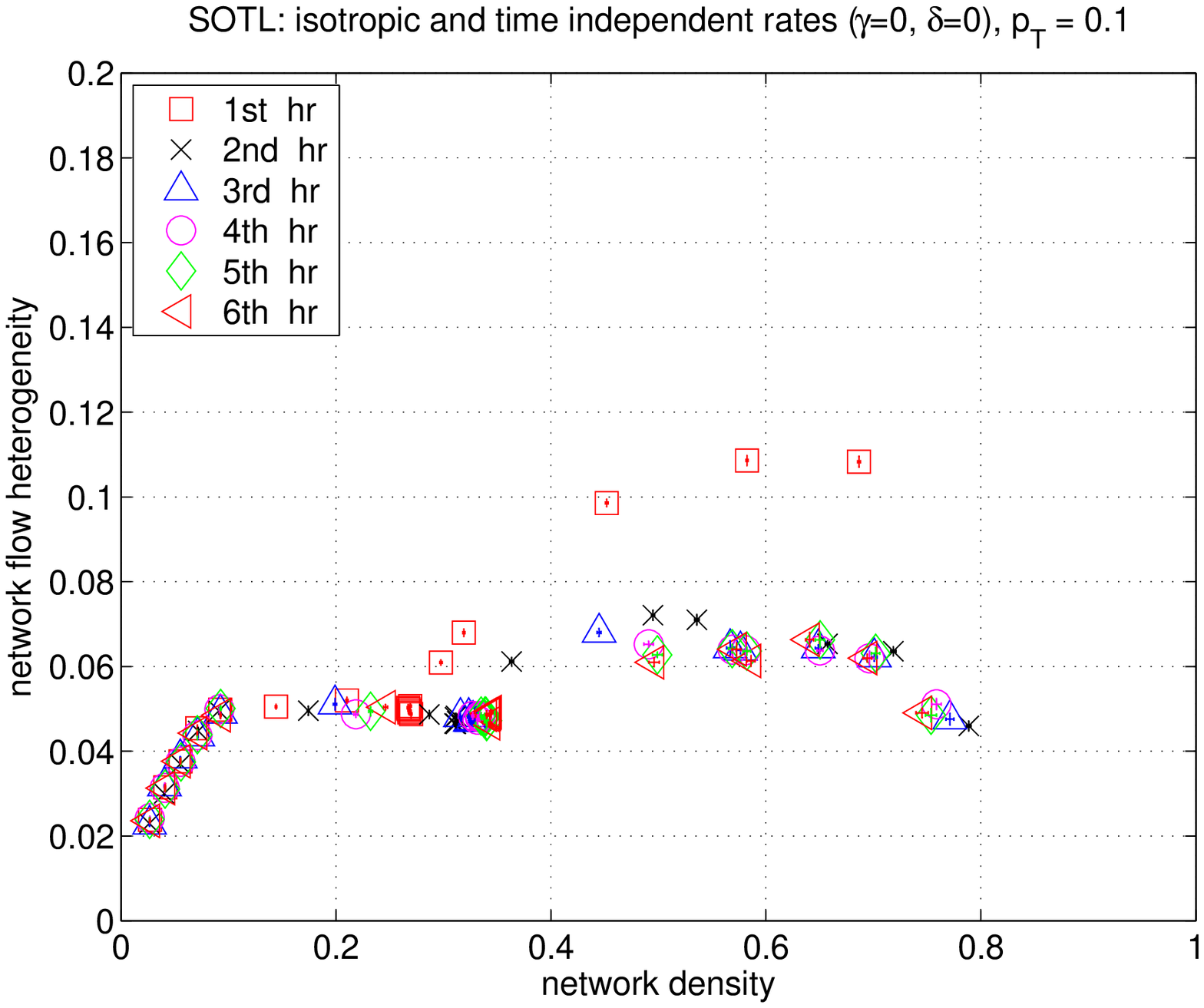}}

\def\fdcuscatsldifgdturnten	{\includegraphics[scale=\oneup]{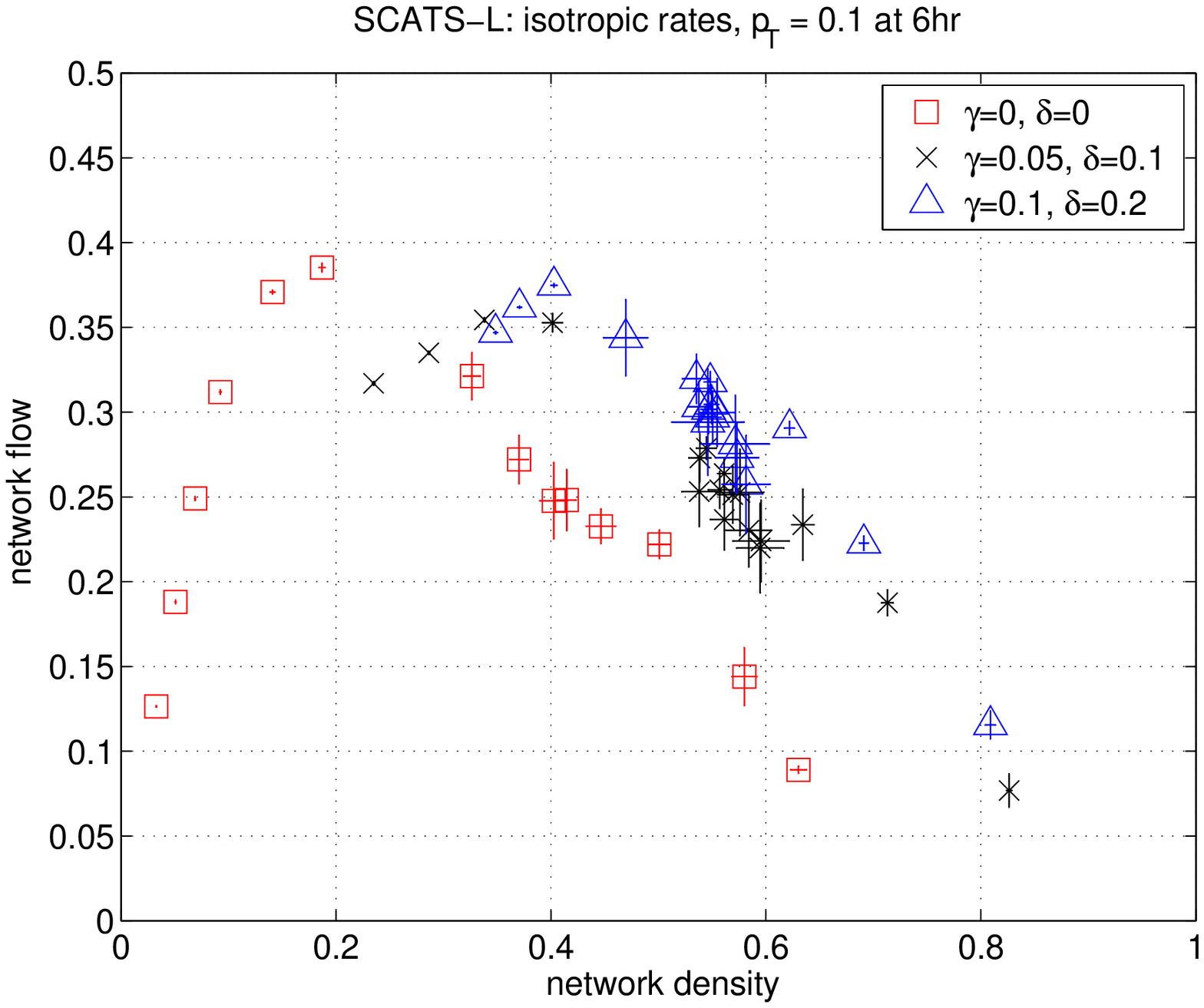}}
\def\fdcusotldifgdturnten		{\includegraphics[scale=\oneup]{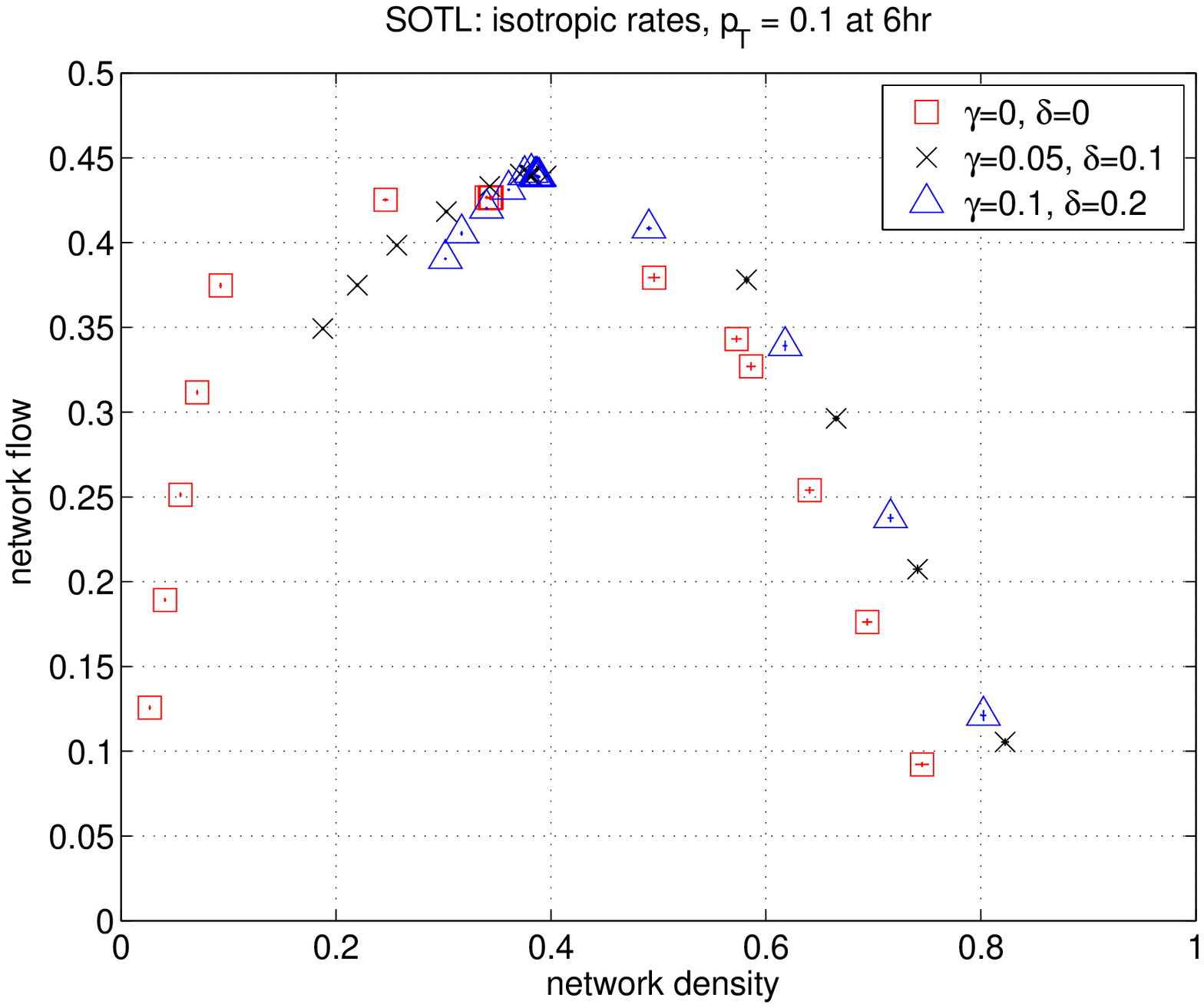}}

\def\fdcuscatsfturns			{\includegraphics[scale=\oneup]{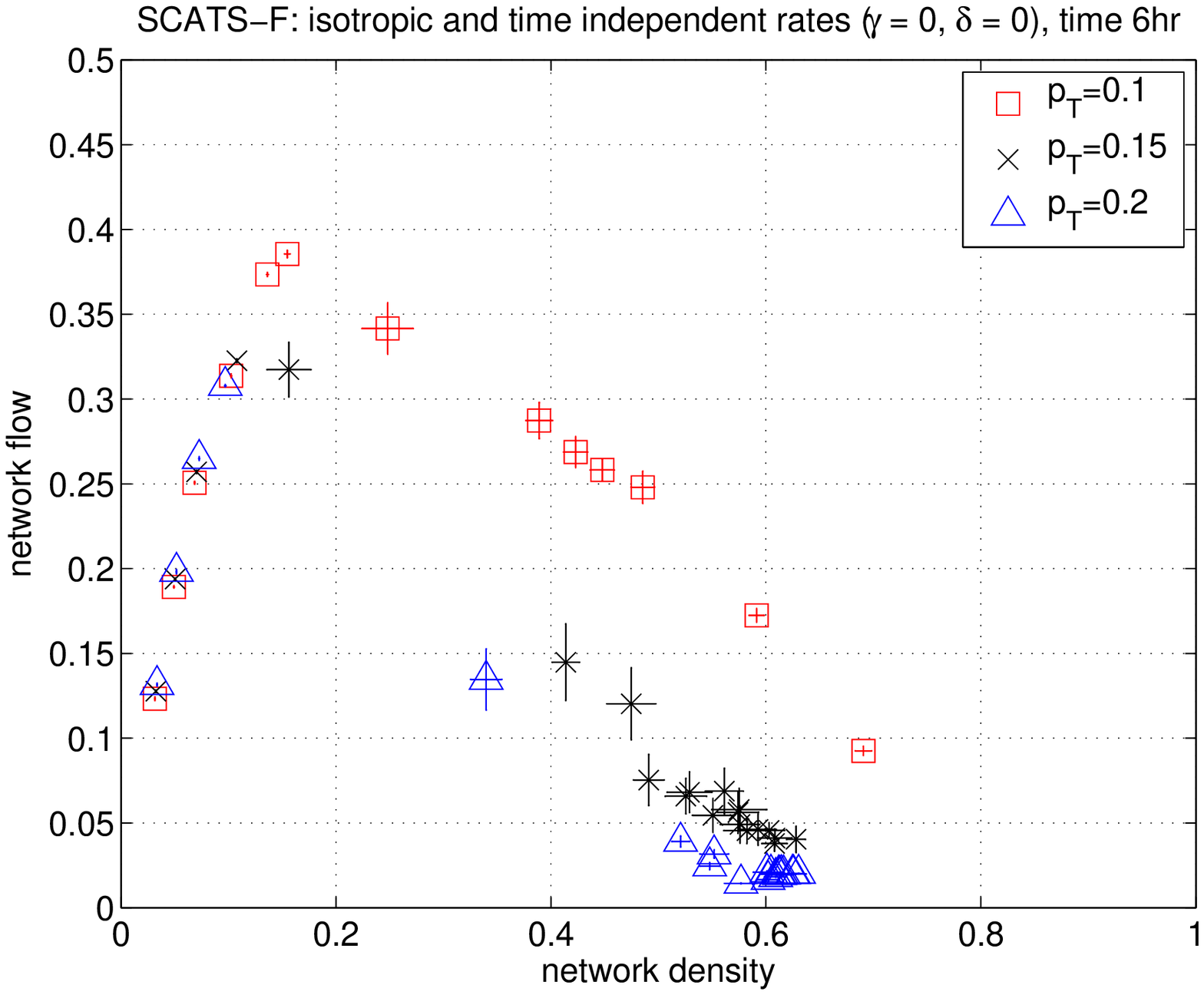}}
\def\fdcuscatslturns			{\includegraphics[scale=\oneup]{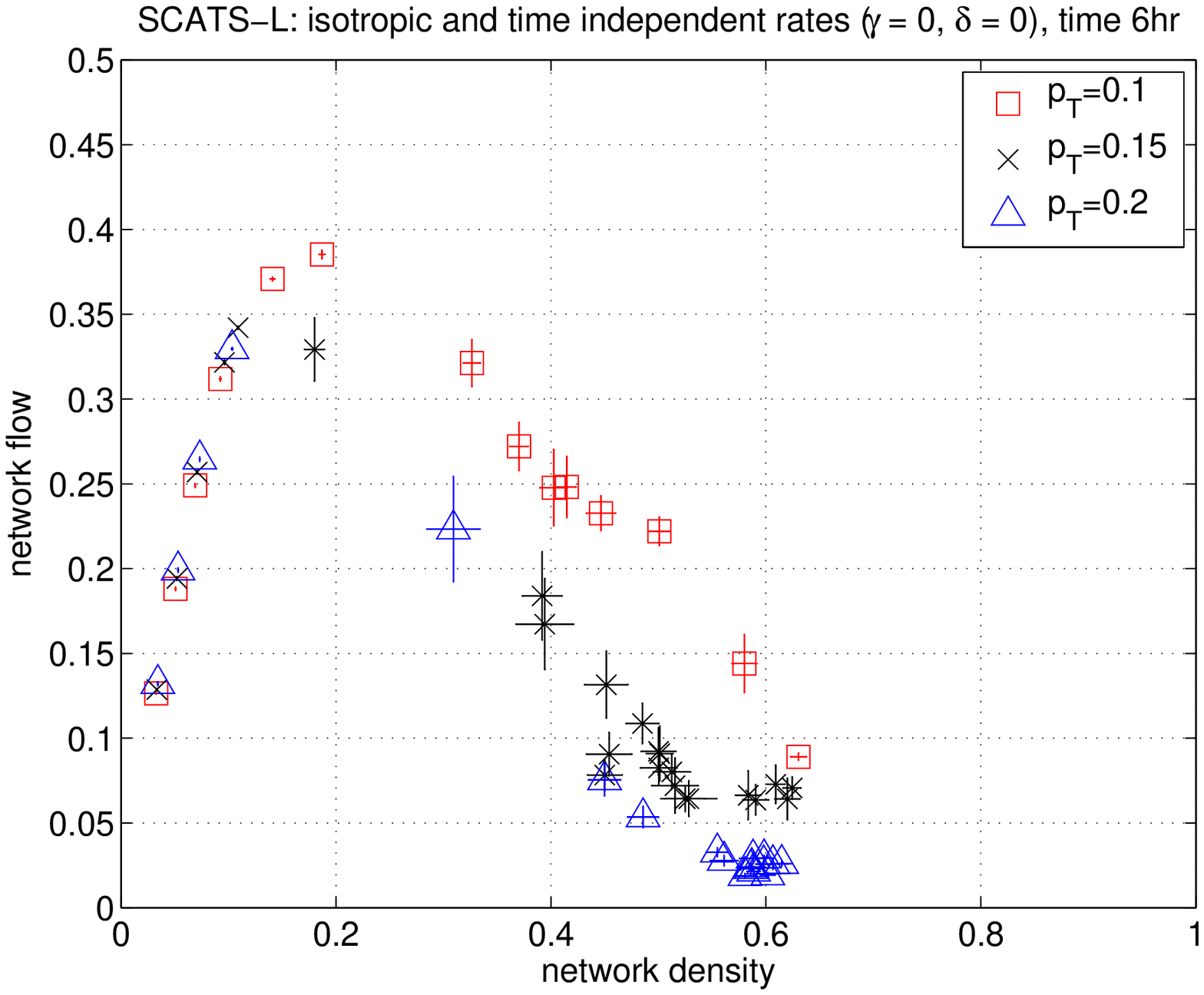}}
\def\fdcusotlturns			{\includegraphics[scale=\oneup]{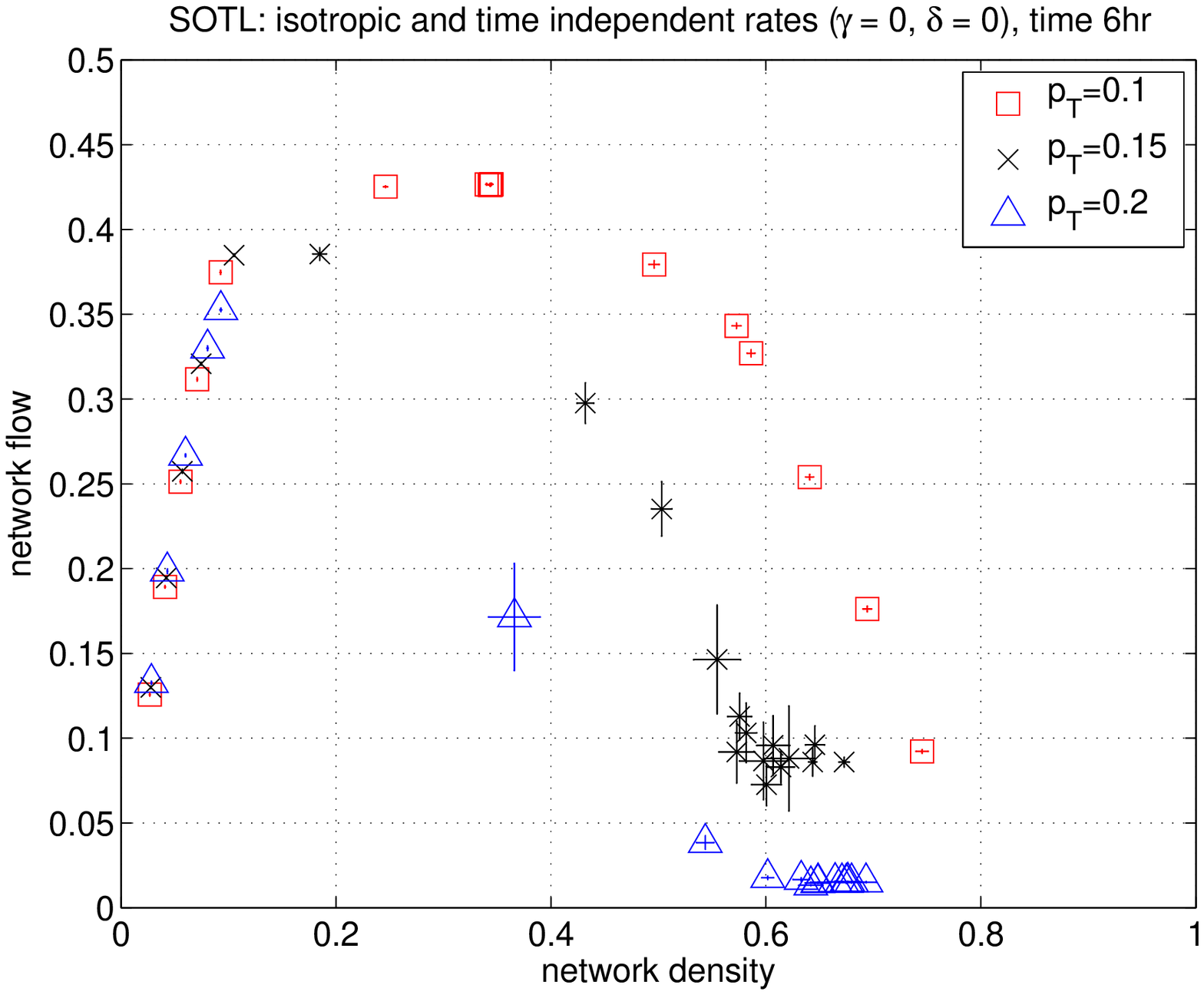}}
\def\fdscatsfhoursturns		{\includegraphics[scale=\oneup]{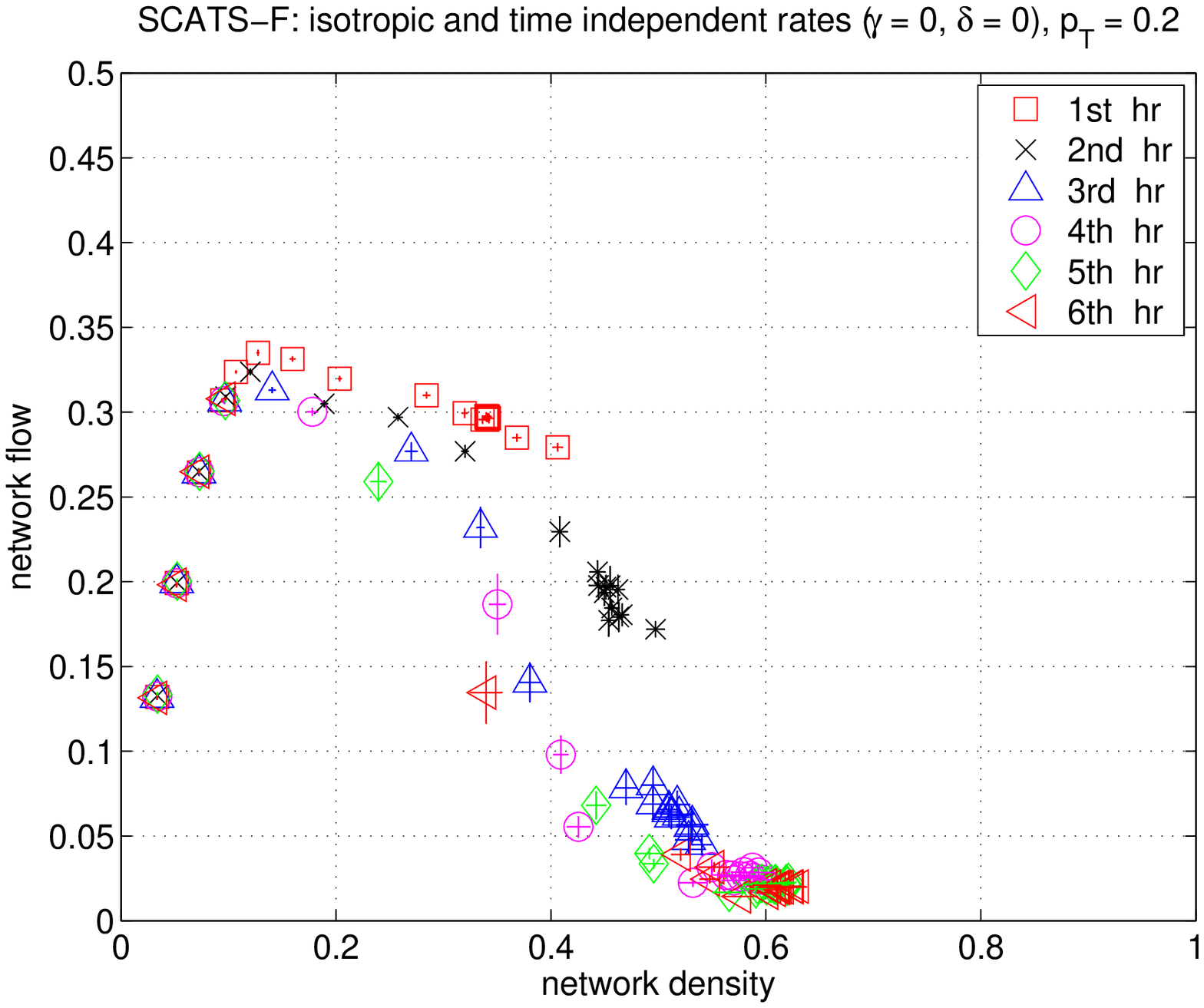}}

\def\fdlongturnssotl		{\includegraphics[scale=\oneup]{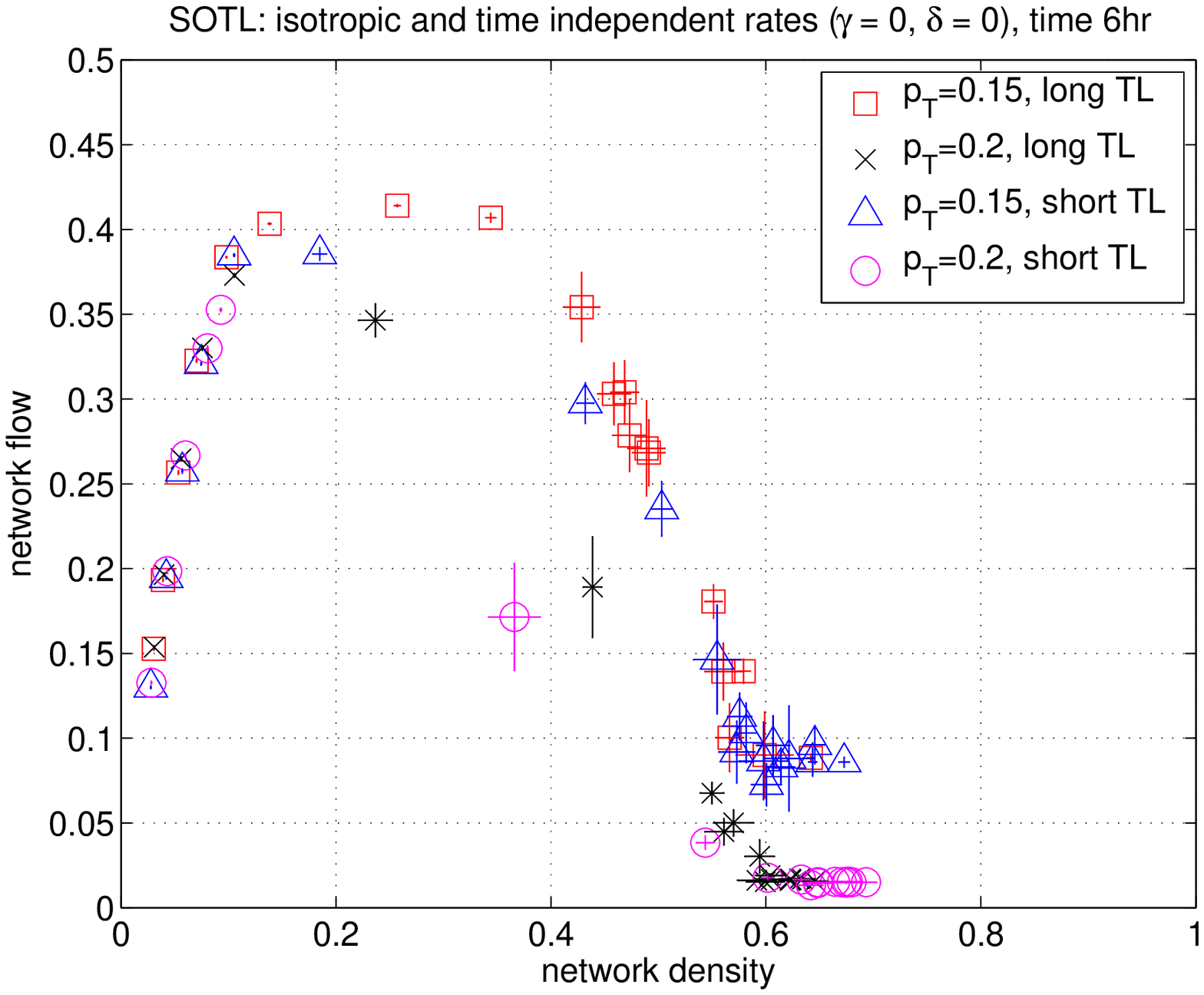}}
\def\fdshortlinks			{\includegraphics[scale=\oneup]{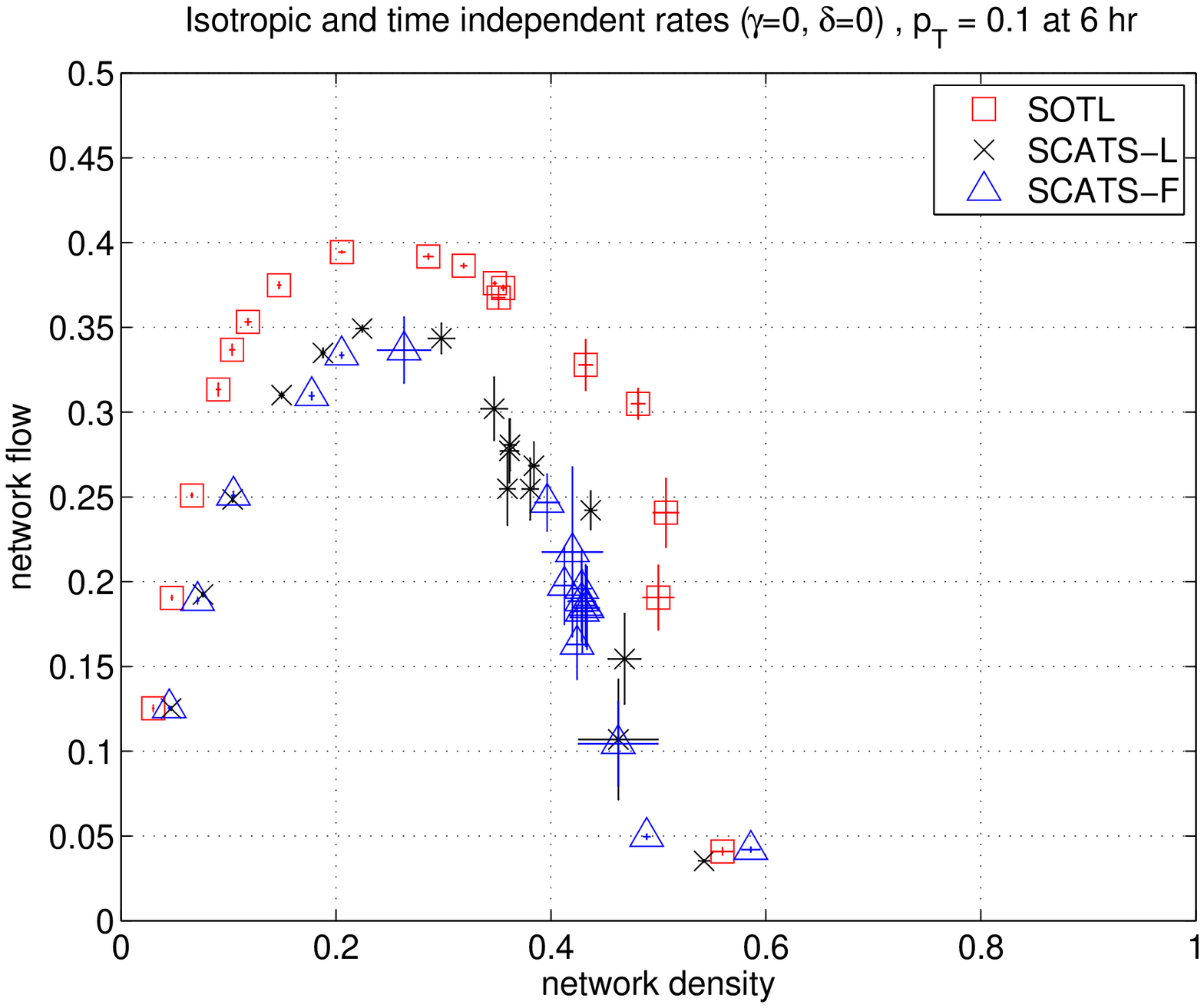}}

\def\fdbiasedturnten			{\includegraphics[scale=\oneup]{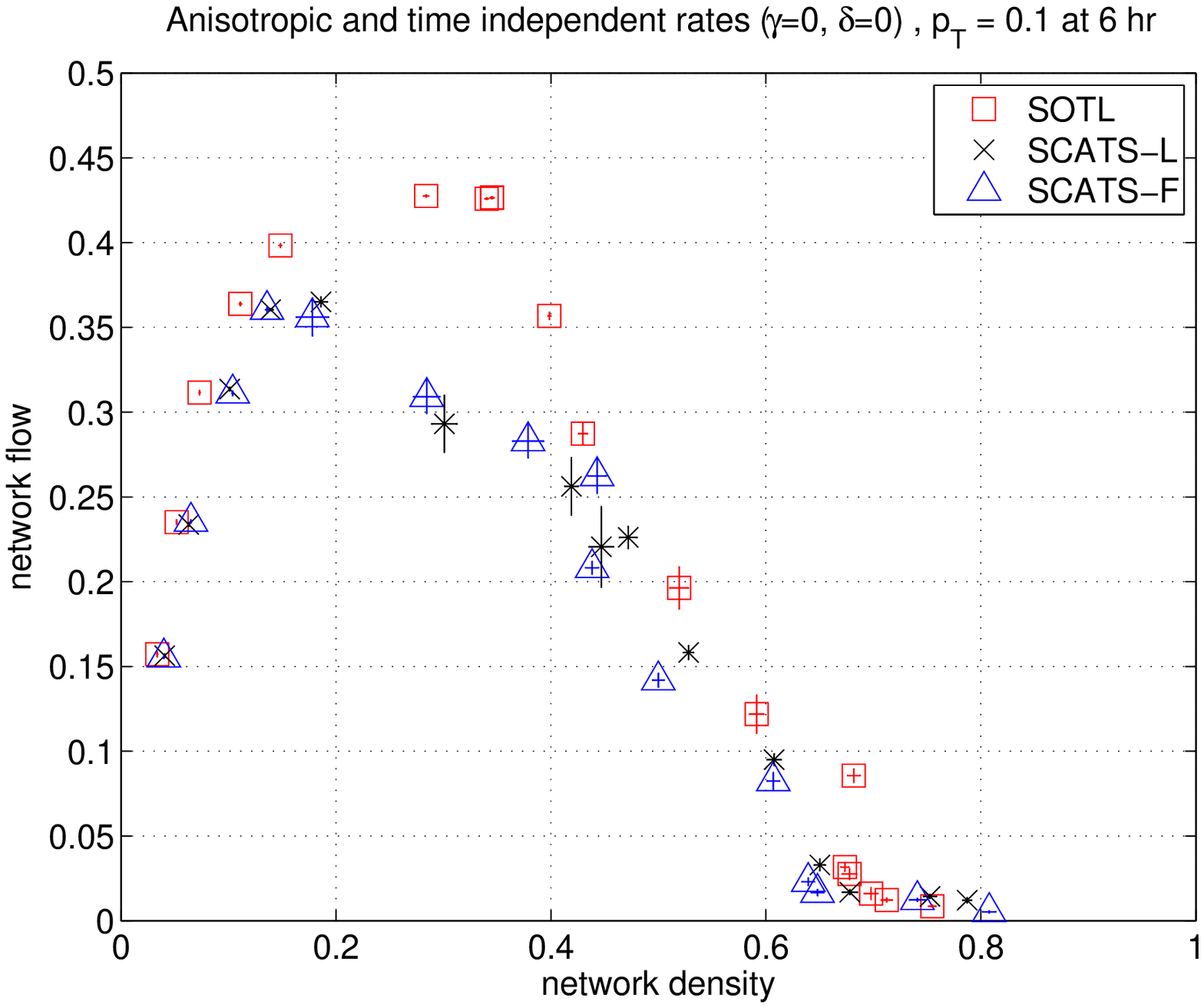}}
\def\fdbiasedRhoturnten 		{\includegraphics[scale=\oneup]{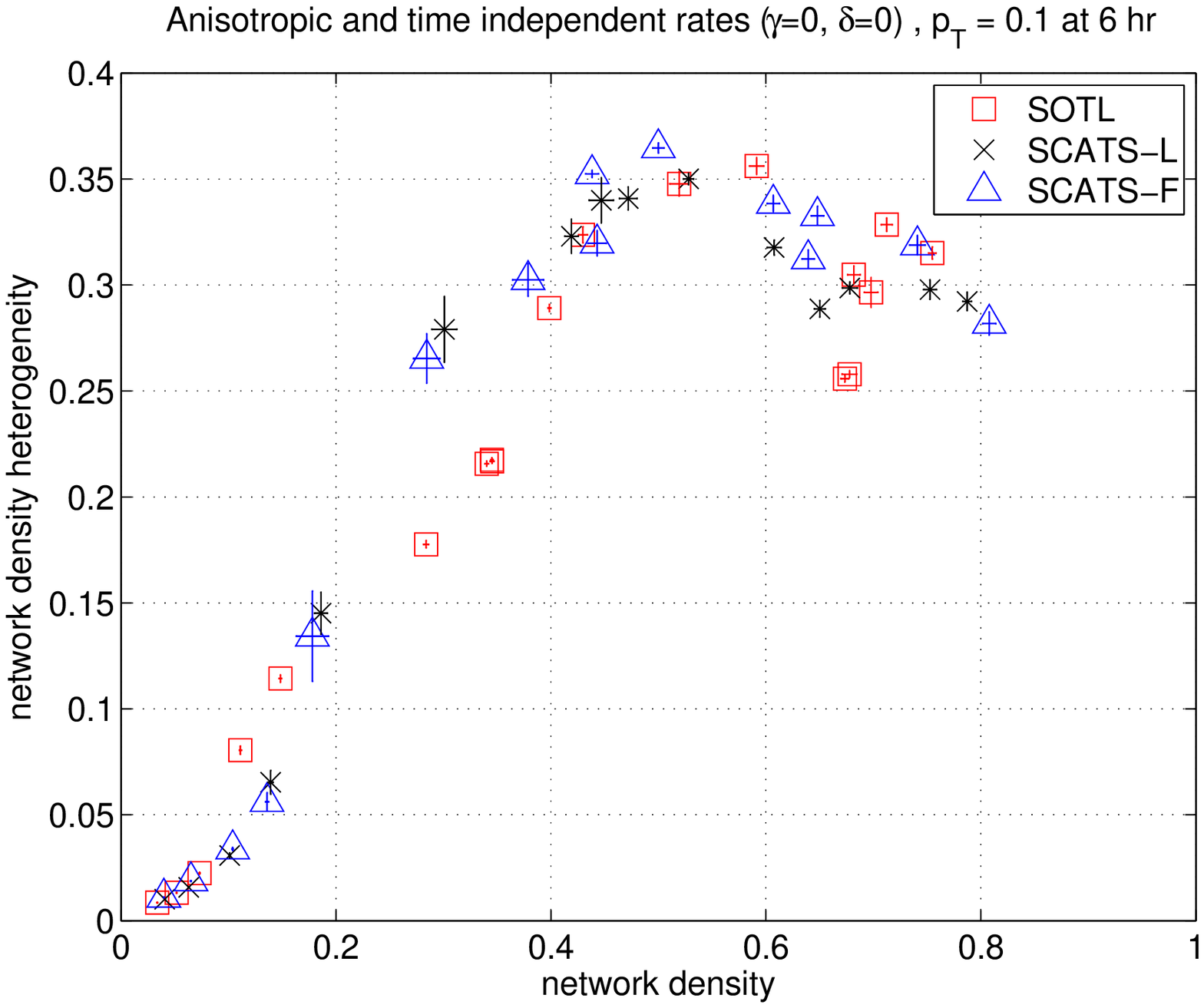}}

\def\densityflowtdscatslturntenboundary 	{\includegraphics[scale=\smaller]{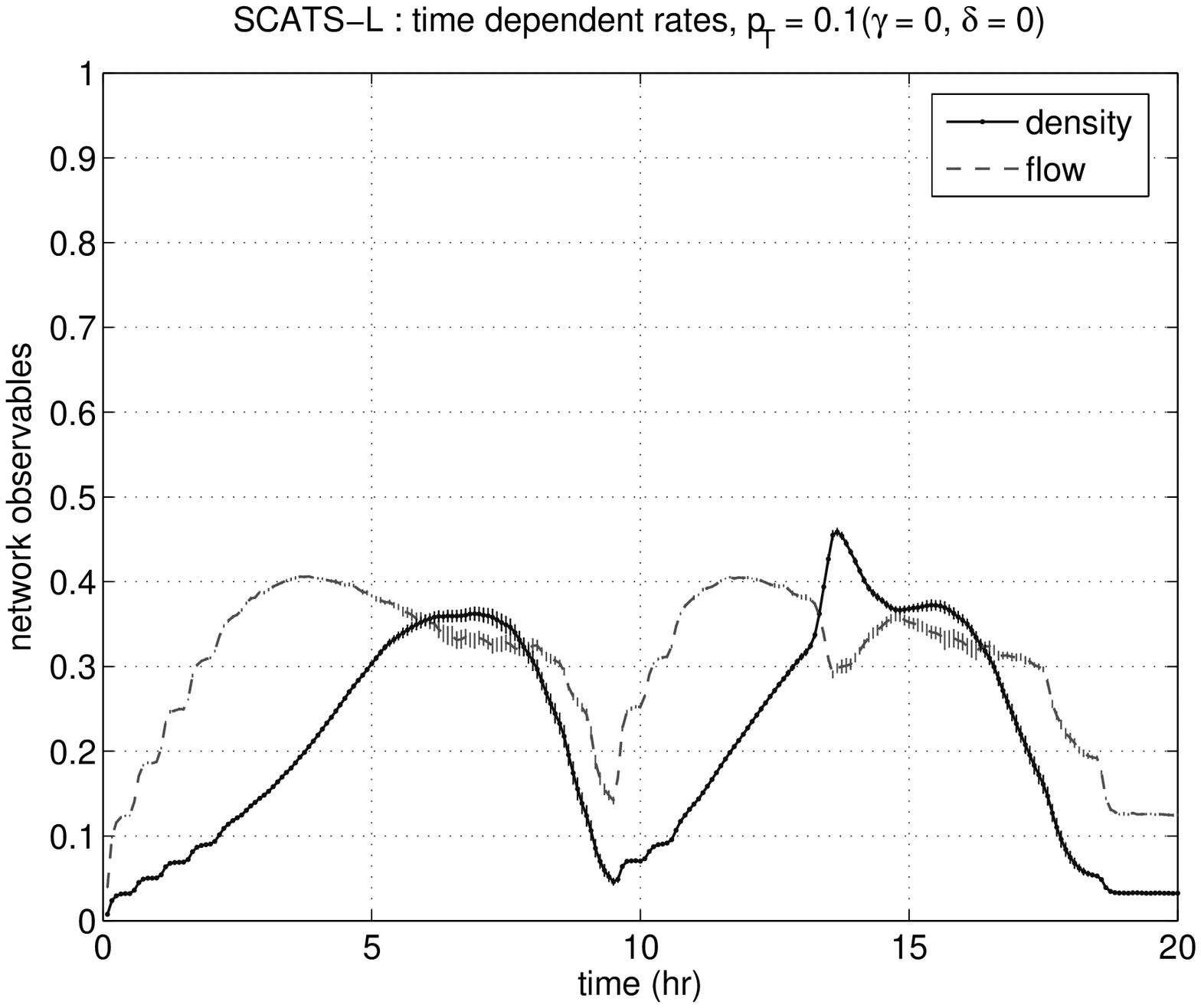}}
\def\densityflowtdscatslturntenuniform	{\includegraphics[scale=\smaller]{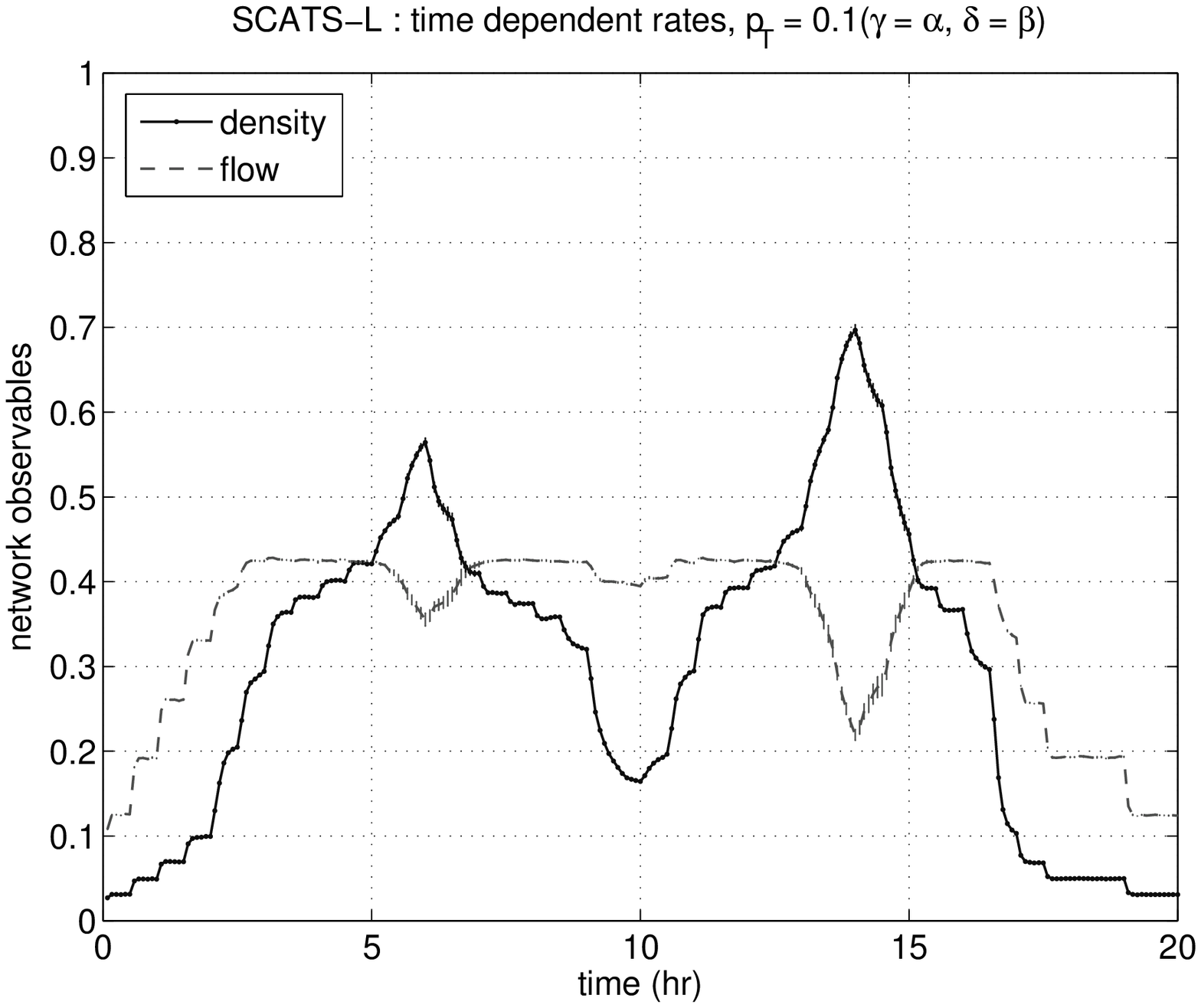}}

\def\fdtdscatsfturntenbulkIO		  		{\includegraphics[scale=\threeup]{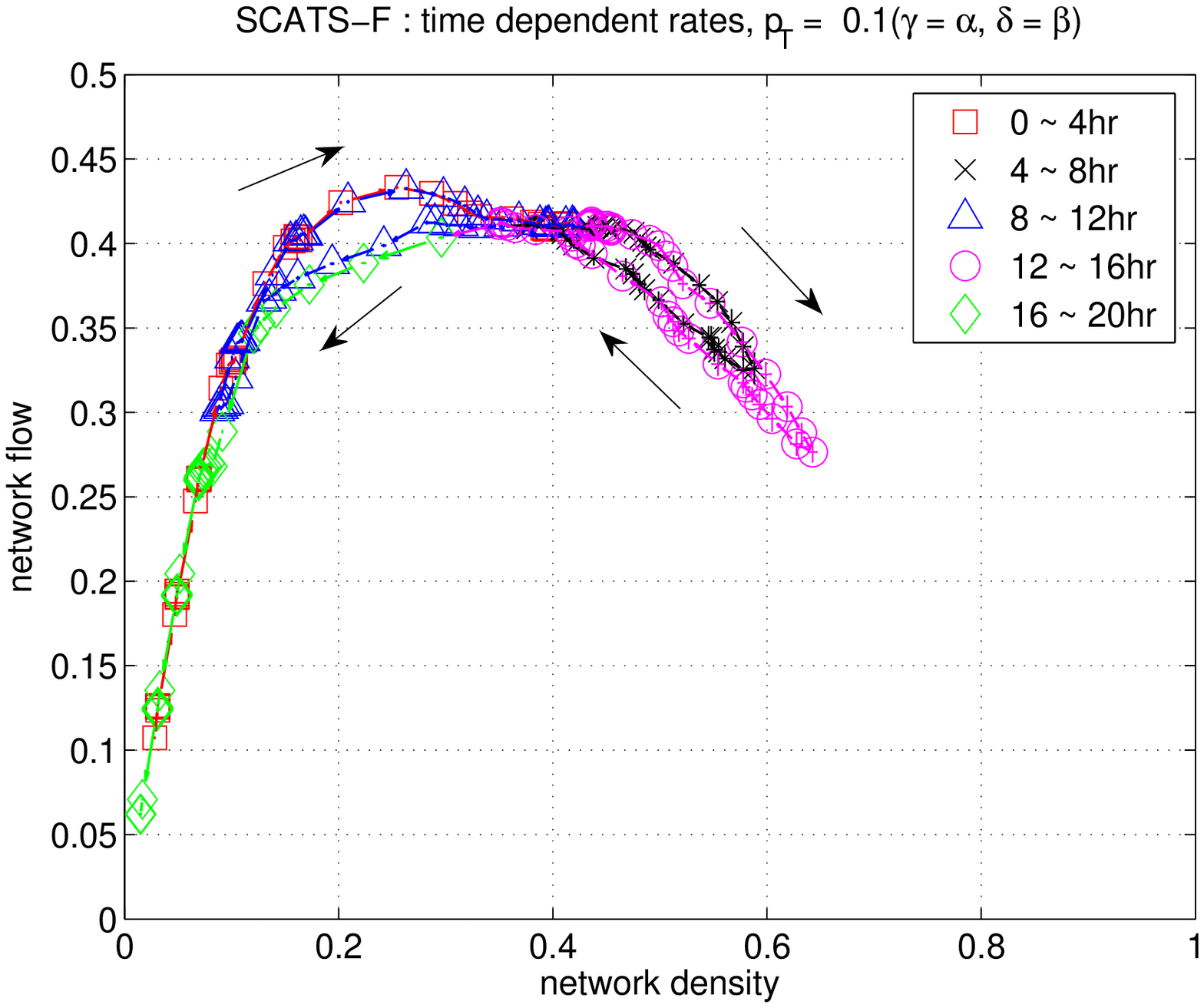}}
\def\dependentHeteroScatsfDensitybulkIO   {\includegraphics[scale=\threeup]{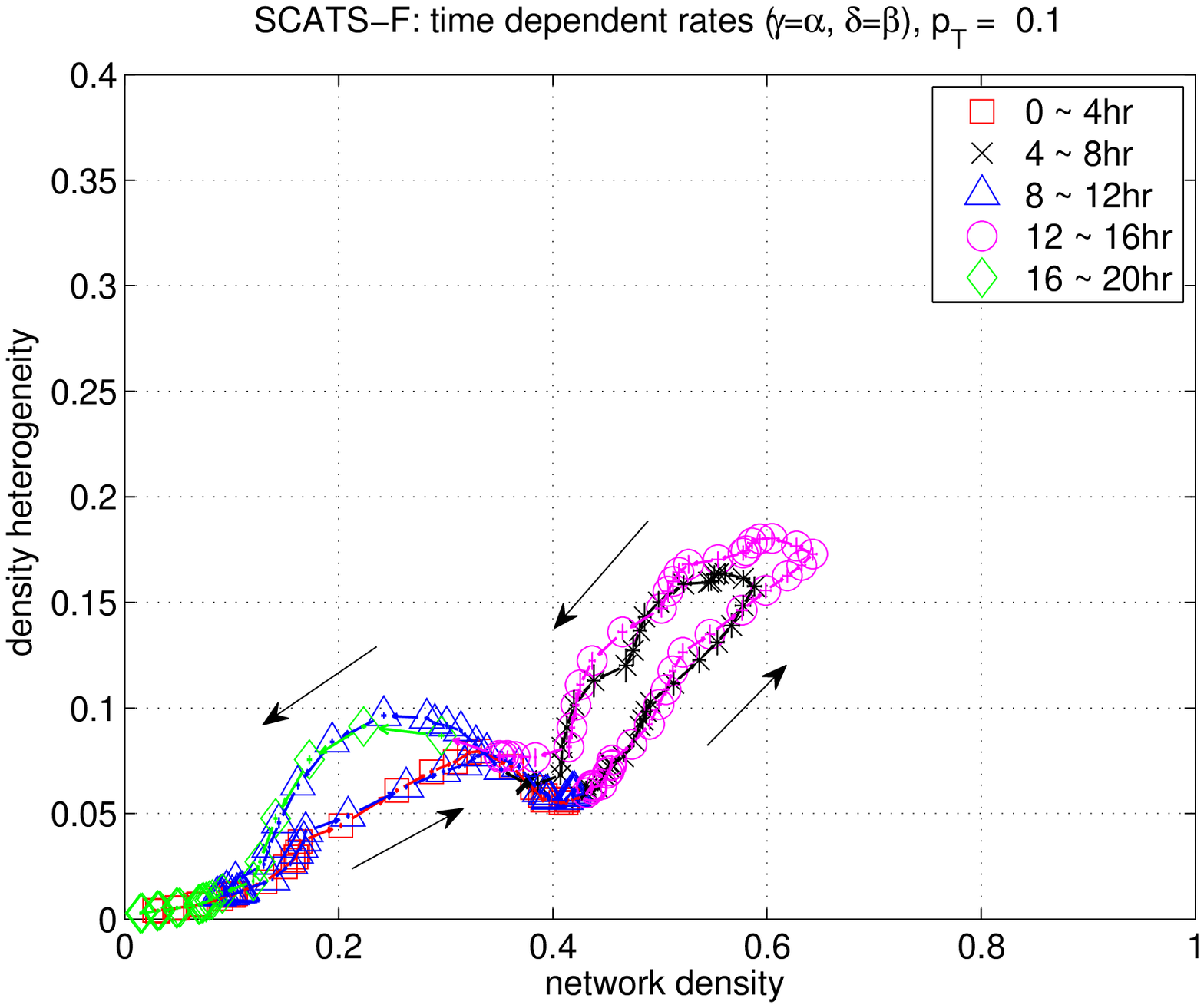}}
\def\fdtdscatslturntenbulkIO		  		{\includegraphics[scale=\threeup]{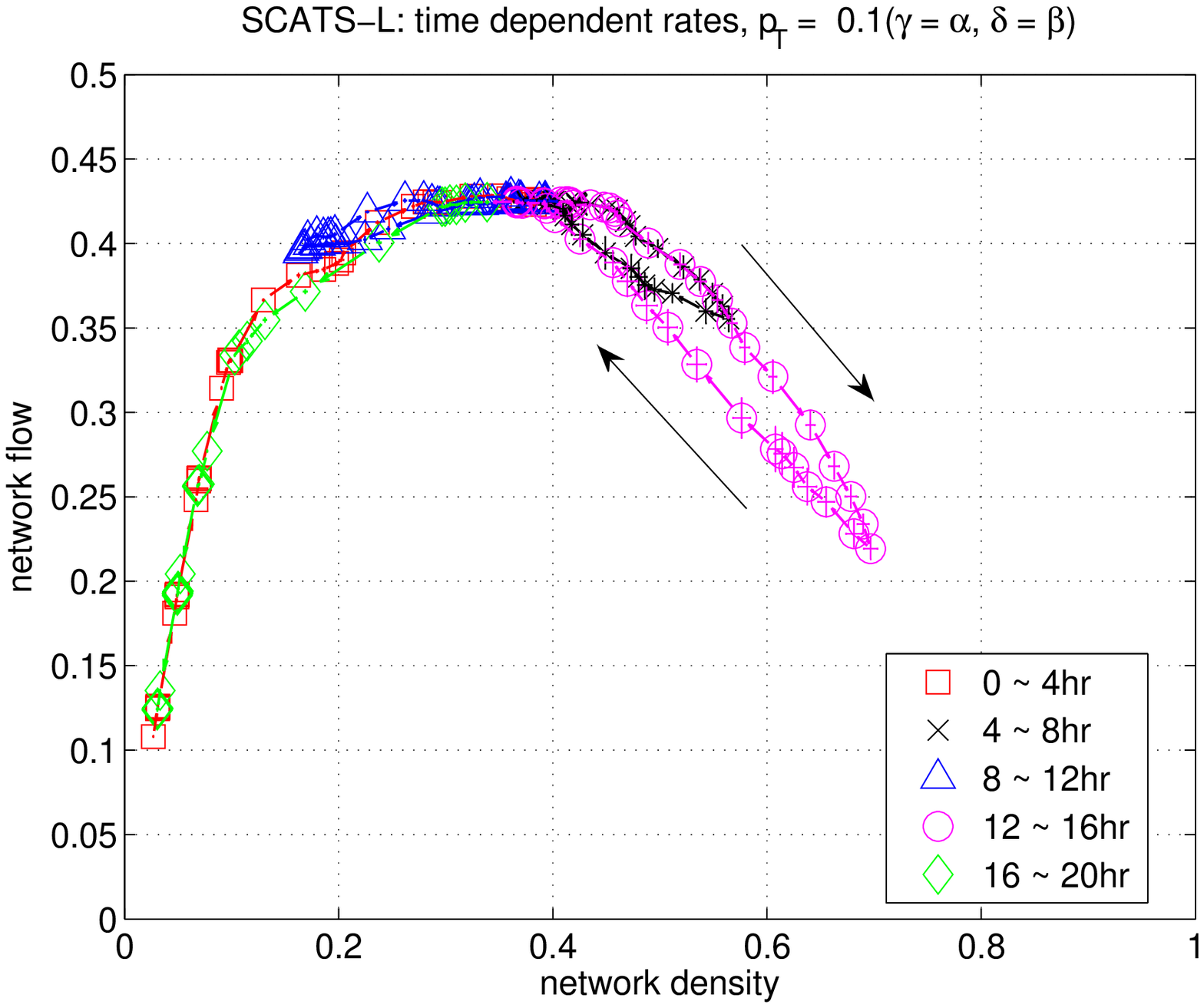}}
\def\dependentHeteroScatslDensitybulkIO   {\includegraphics[scale=\threeup]{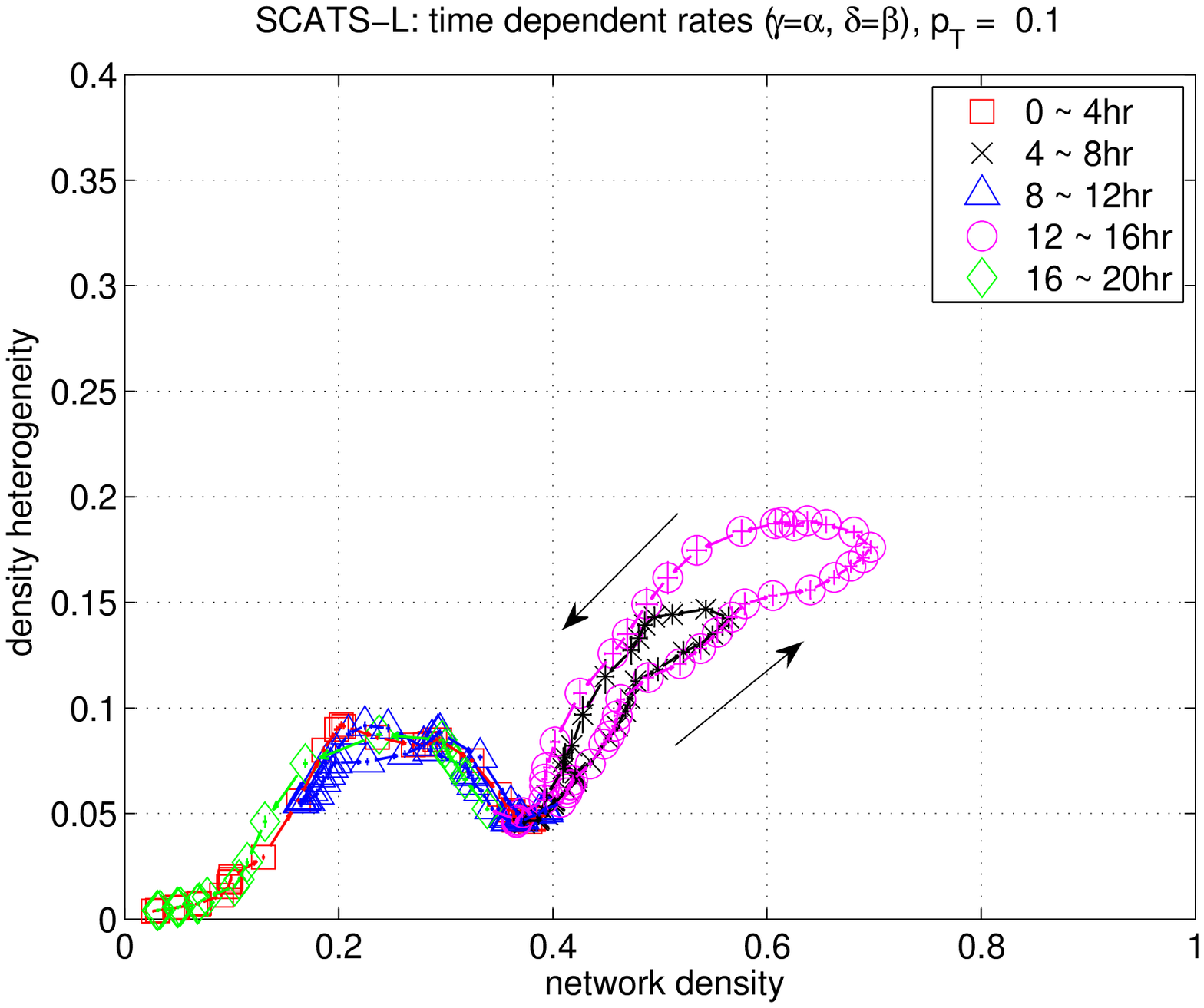}}
\def\fdtdsotlturntenbulkIO	          	{\includegraphics[scale=\threeup]{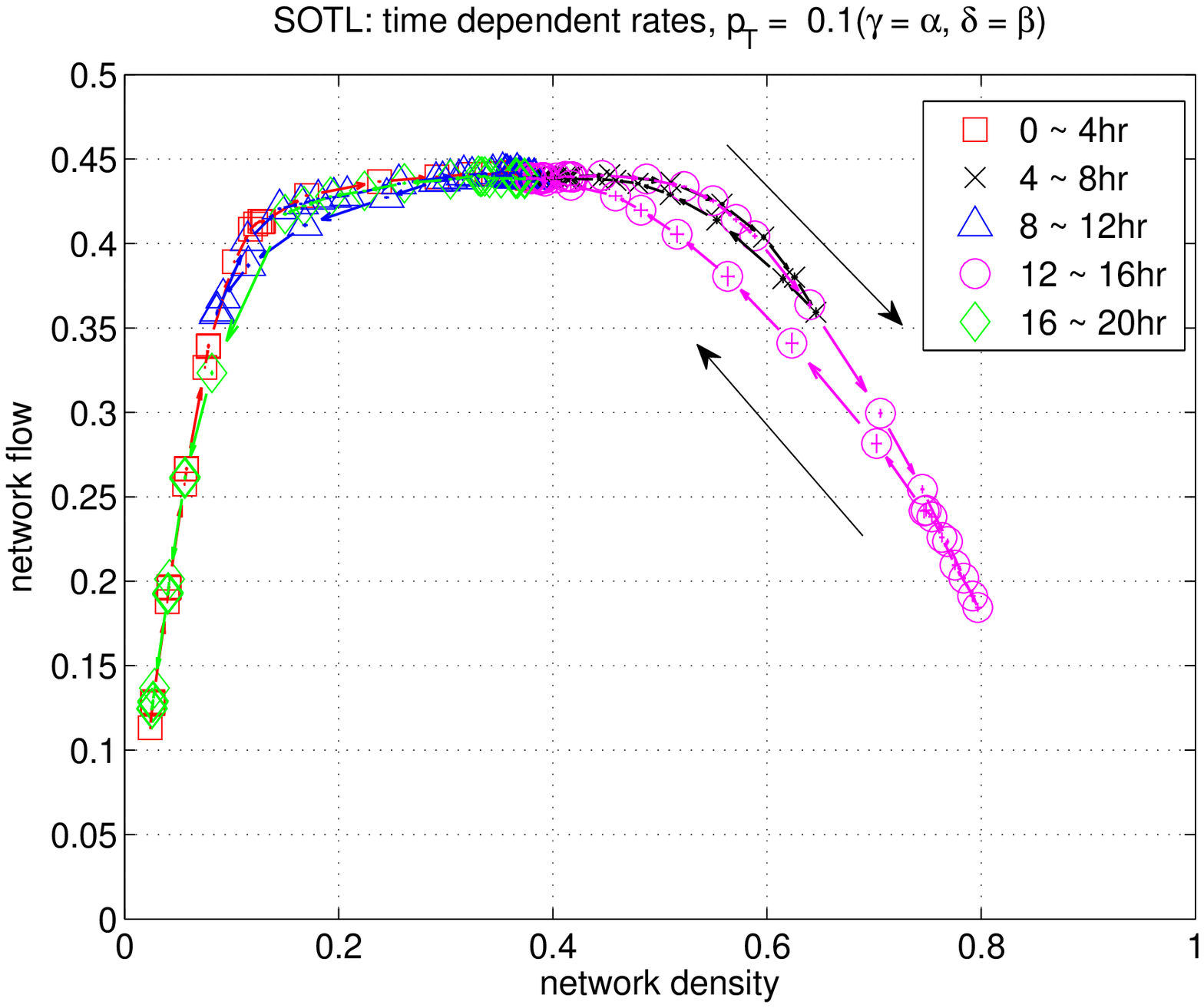}}
\def\dependentHeteroSotlDensitybulkIO     {\includegraphics[scale=\threeup]{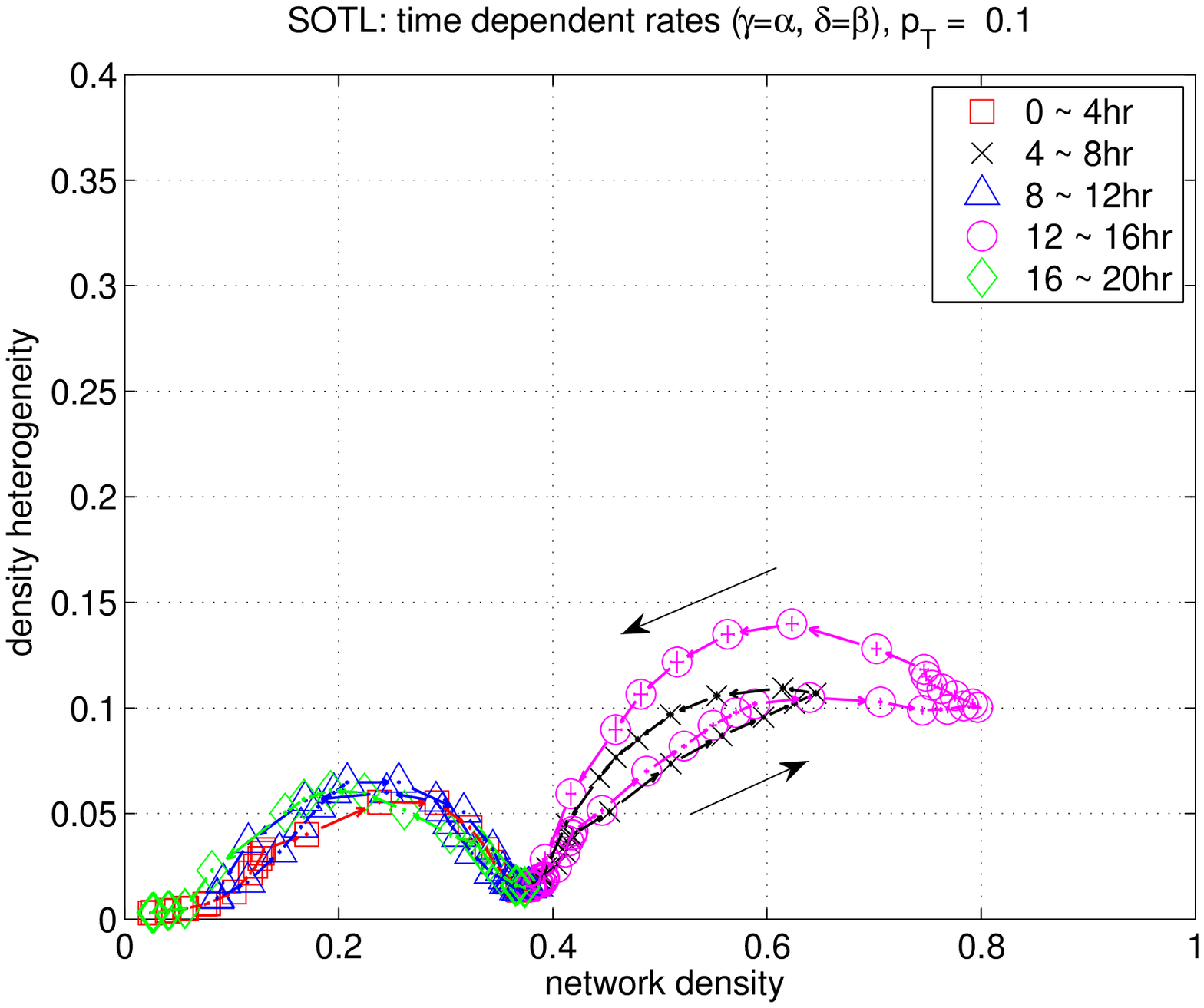}}
\def\fdtdscatsfturnten		  			{\includegraphics[scale=\threeup]{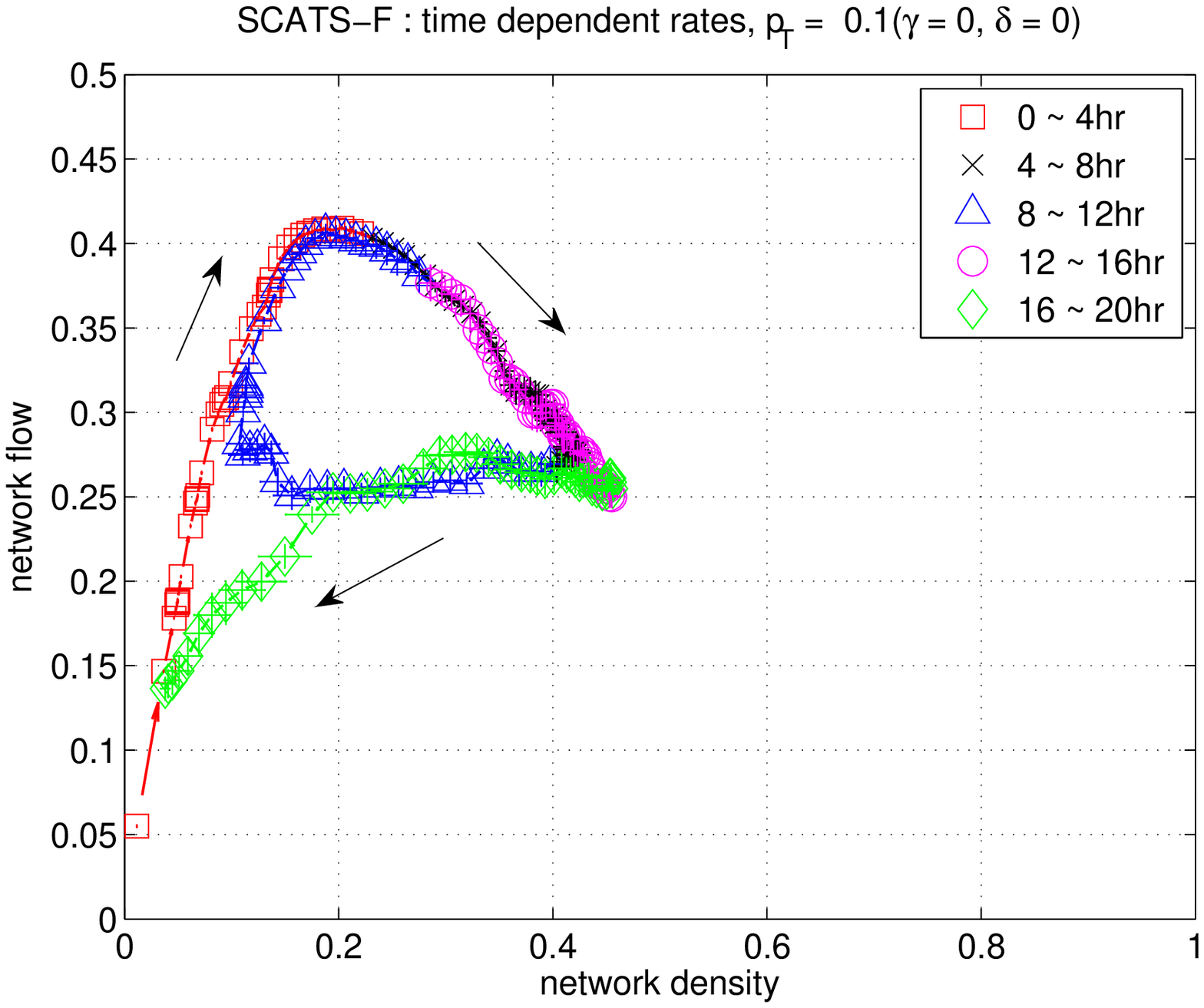}}
\def\dependentHeteroScatsfDensity 		{\includegraphics[scale=\threeup]{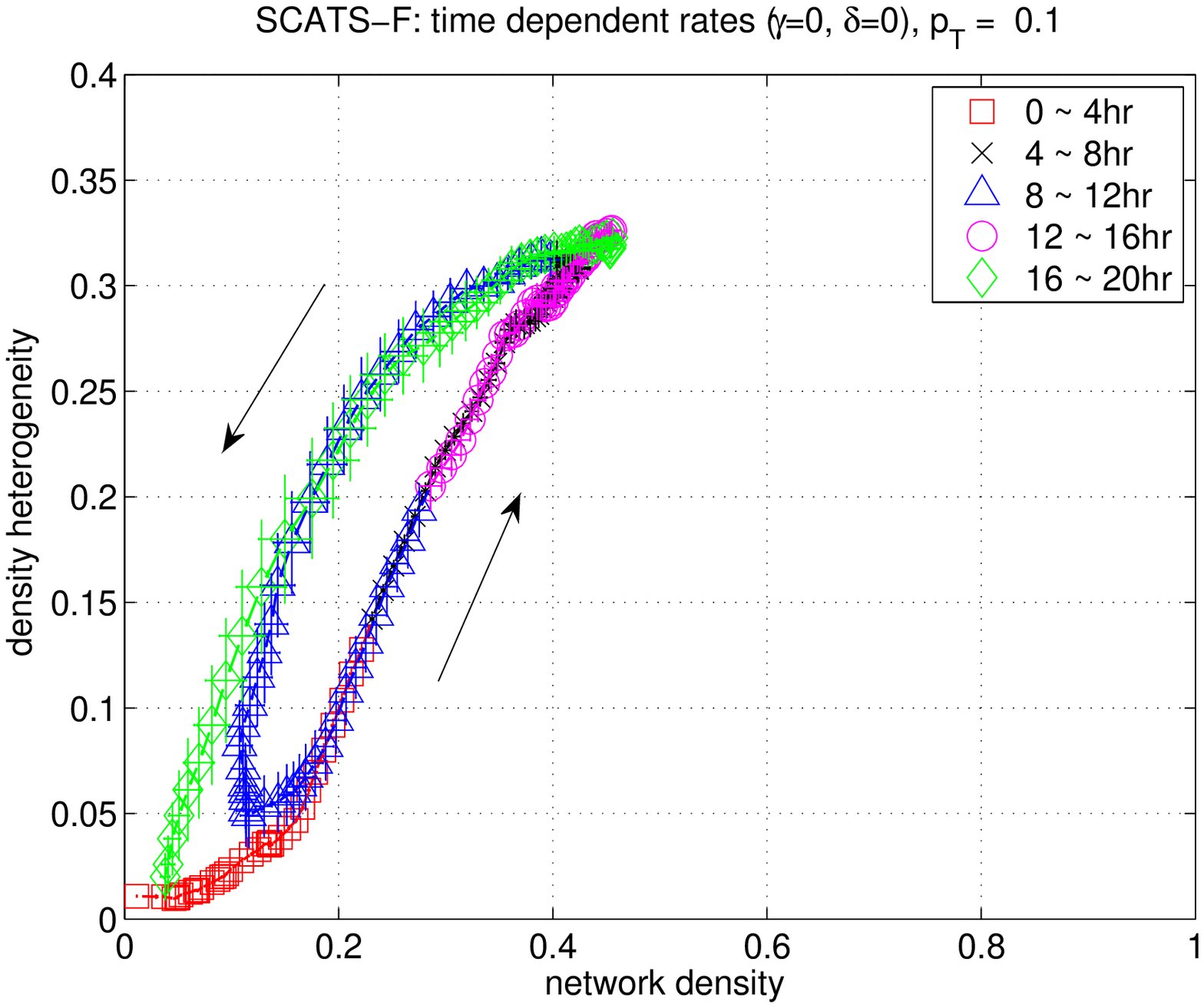}}
\def\fdtdscatslturnten		  			{\includegraphics[scale=\threeup]{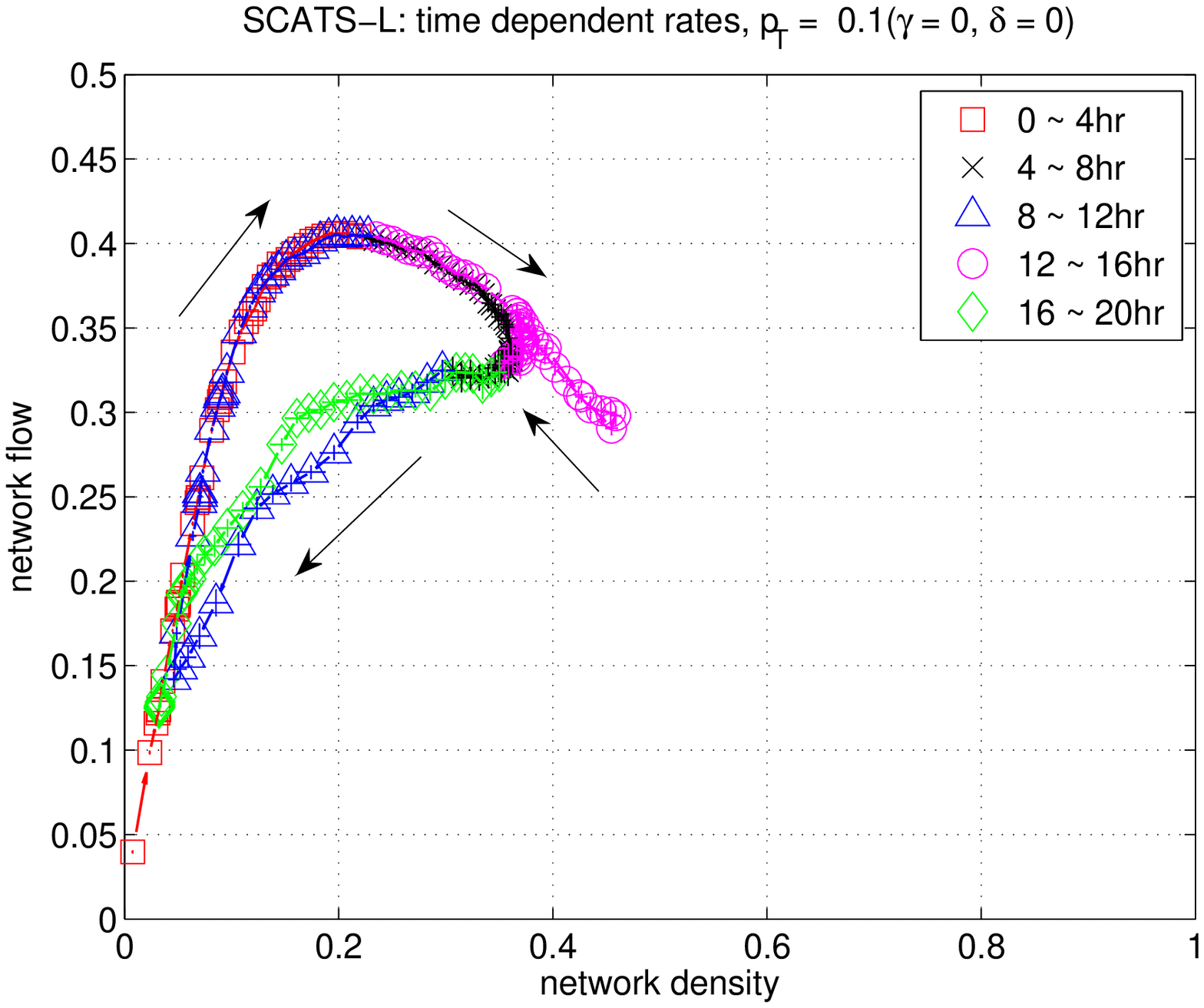}}
\def\dependentHeteroScatslDensity 		{\includegraphics[scale=\threeup]{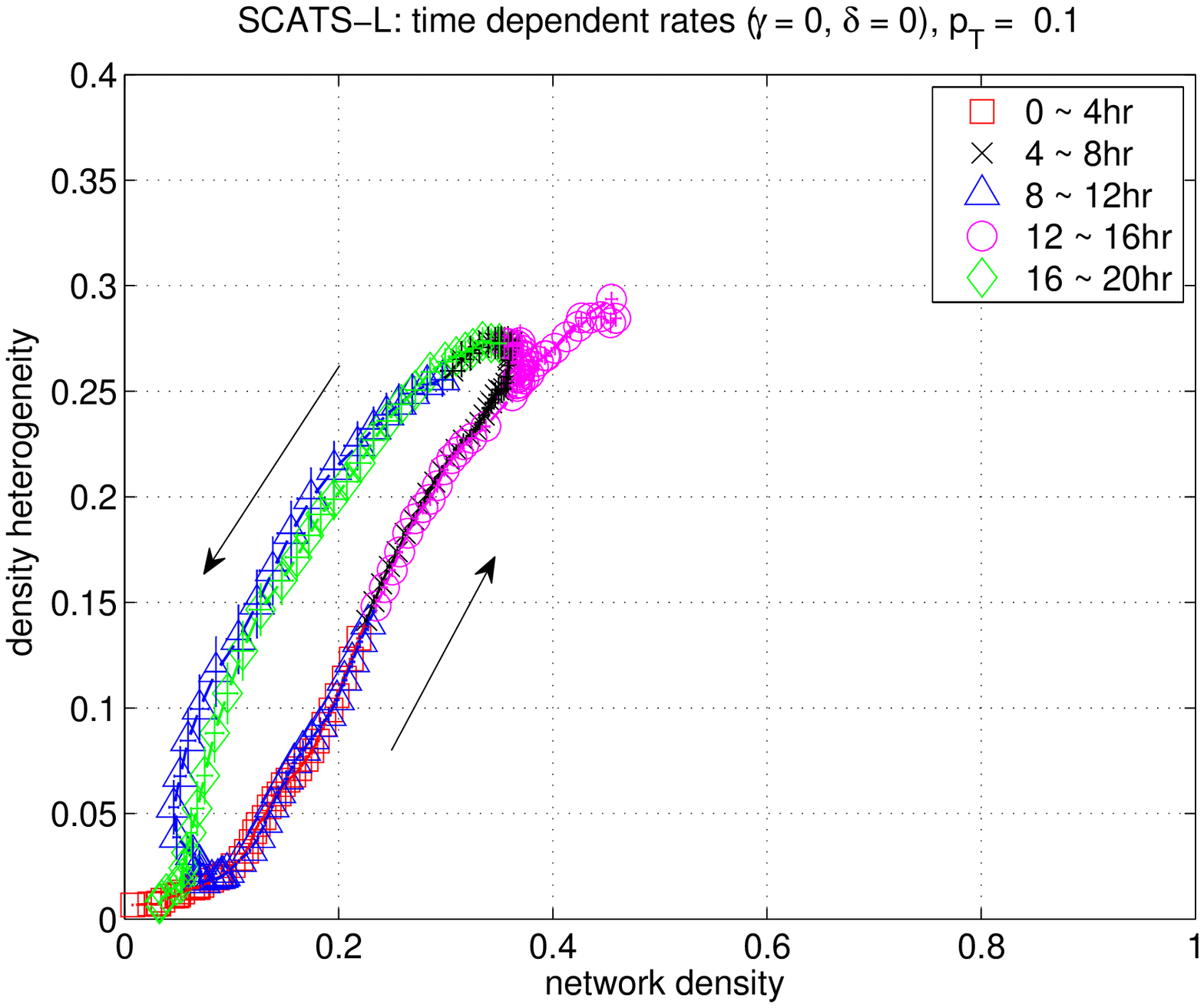}}
\def\fdtdsotlturnten		  				{\includegraphics[scale=\threeup]{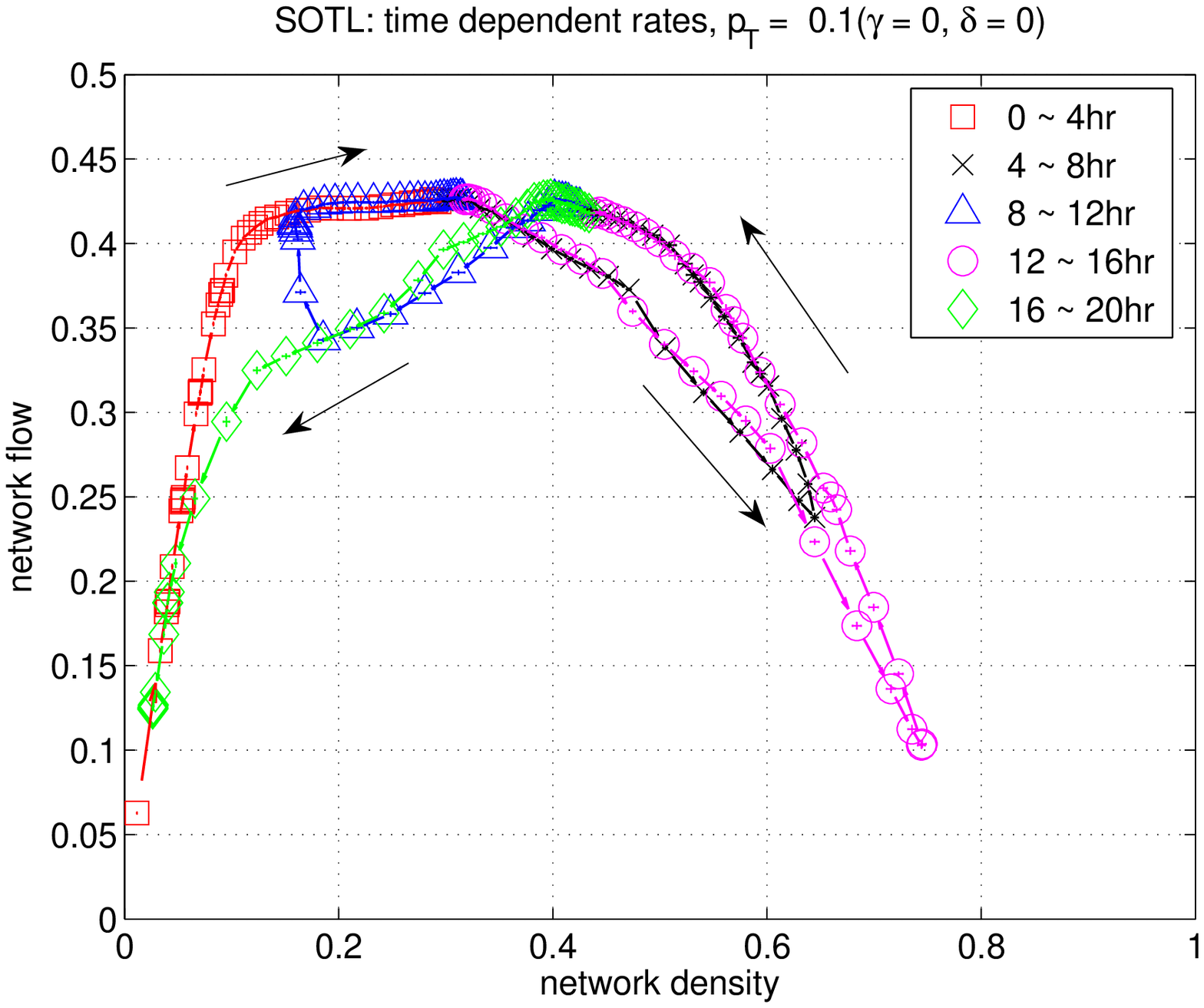}}
\def\dependentHeteroSotlDensity   		{\includegraphics[scale=\threeup]{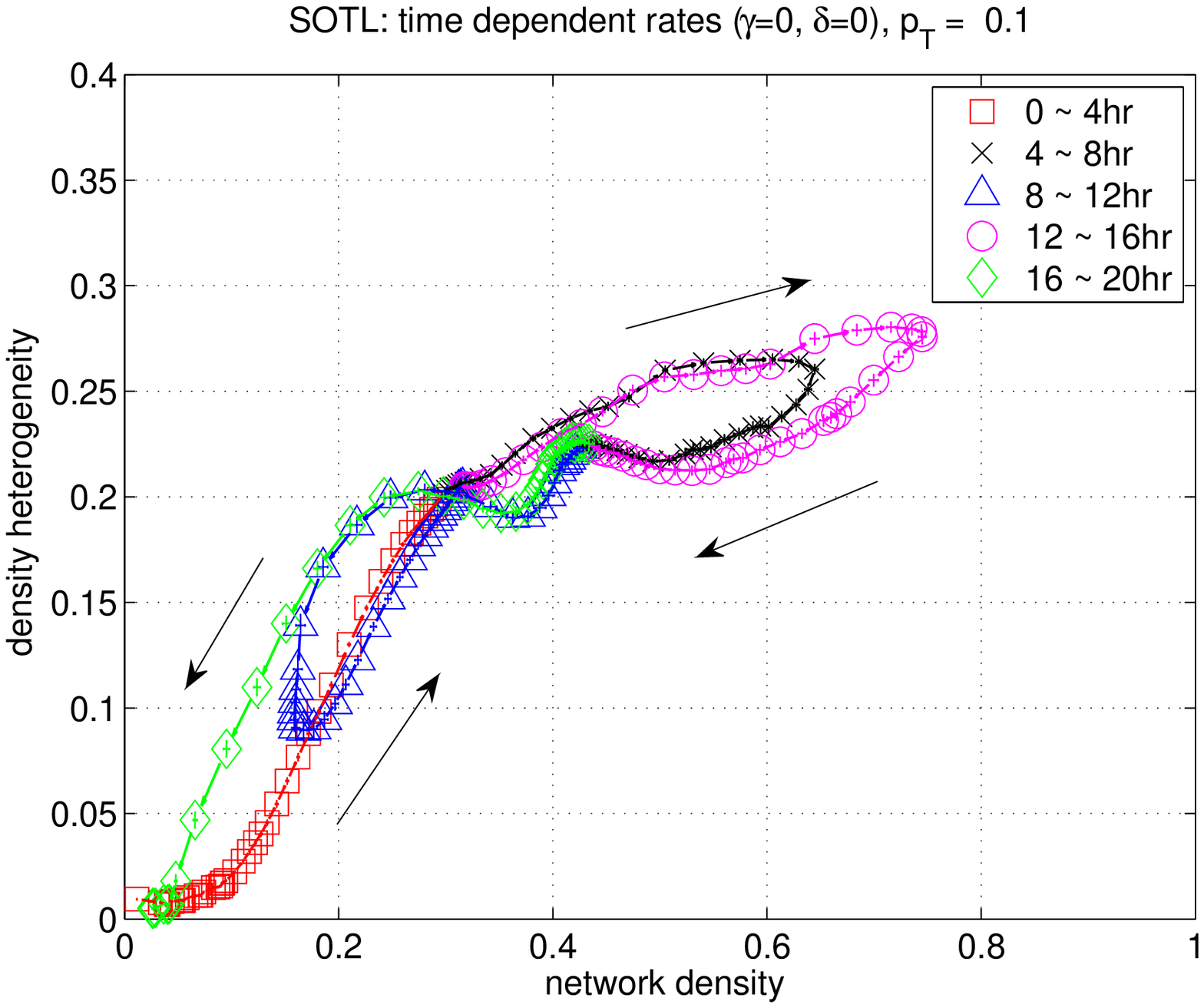}}

\def\densityDistributionLoading {\includegraphics[scale=\fourup,trim= 0 20 0 0, clip]{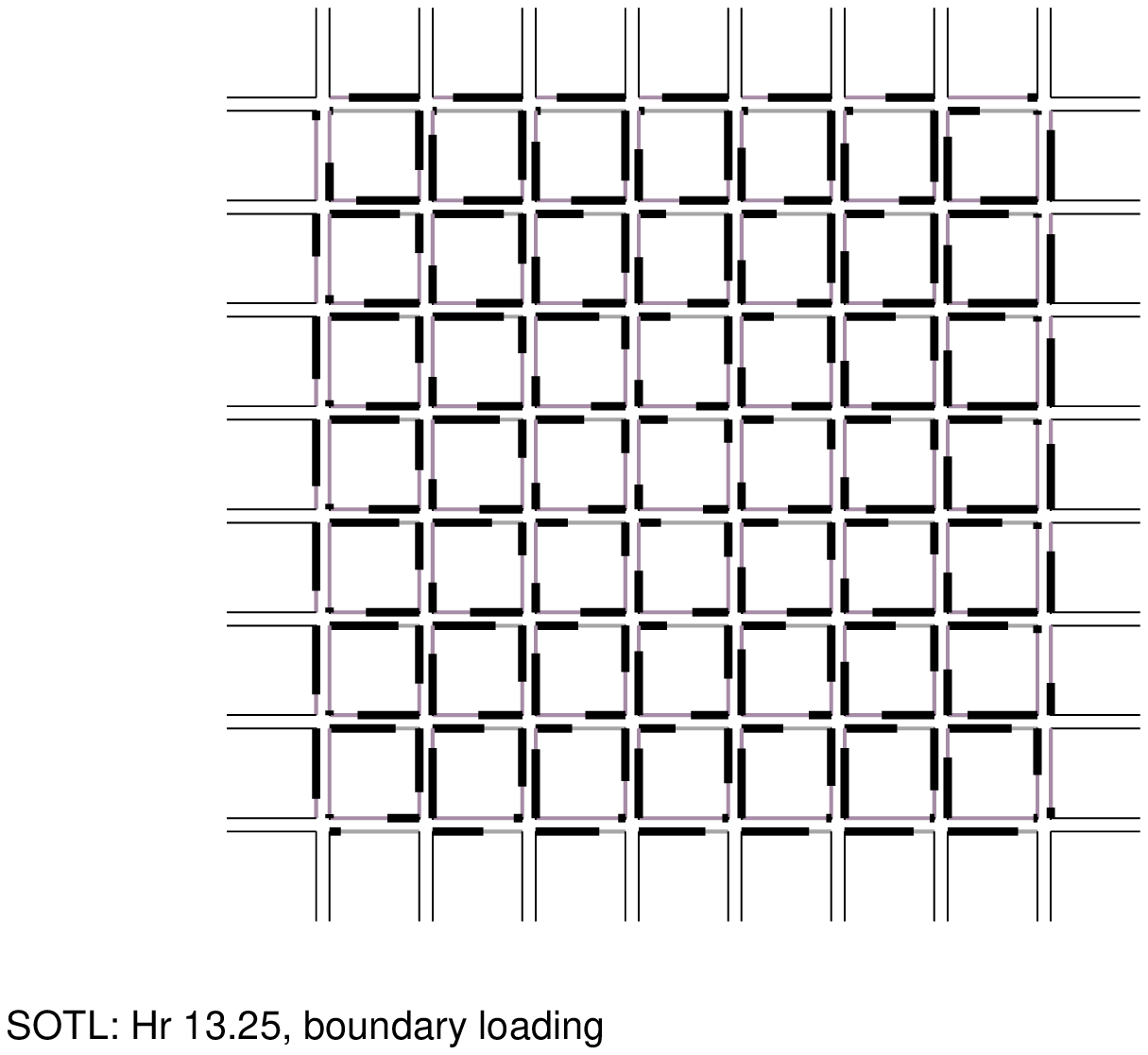}}
\def\densityDistributionRecovery {\includegraphics[scale=\fourup,trim= 0 20 0 0, clip]{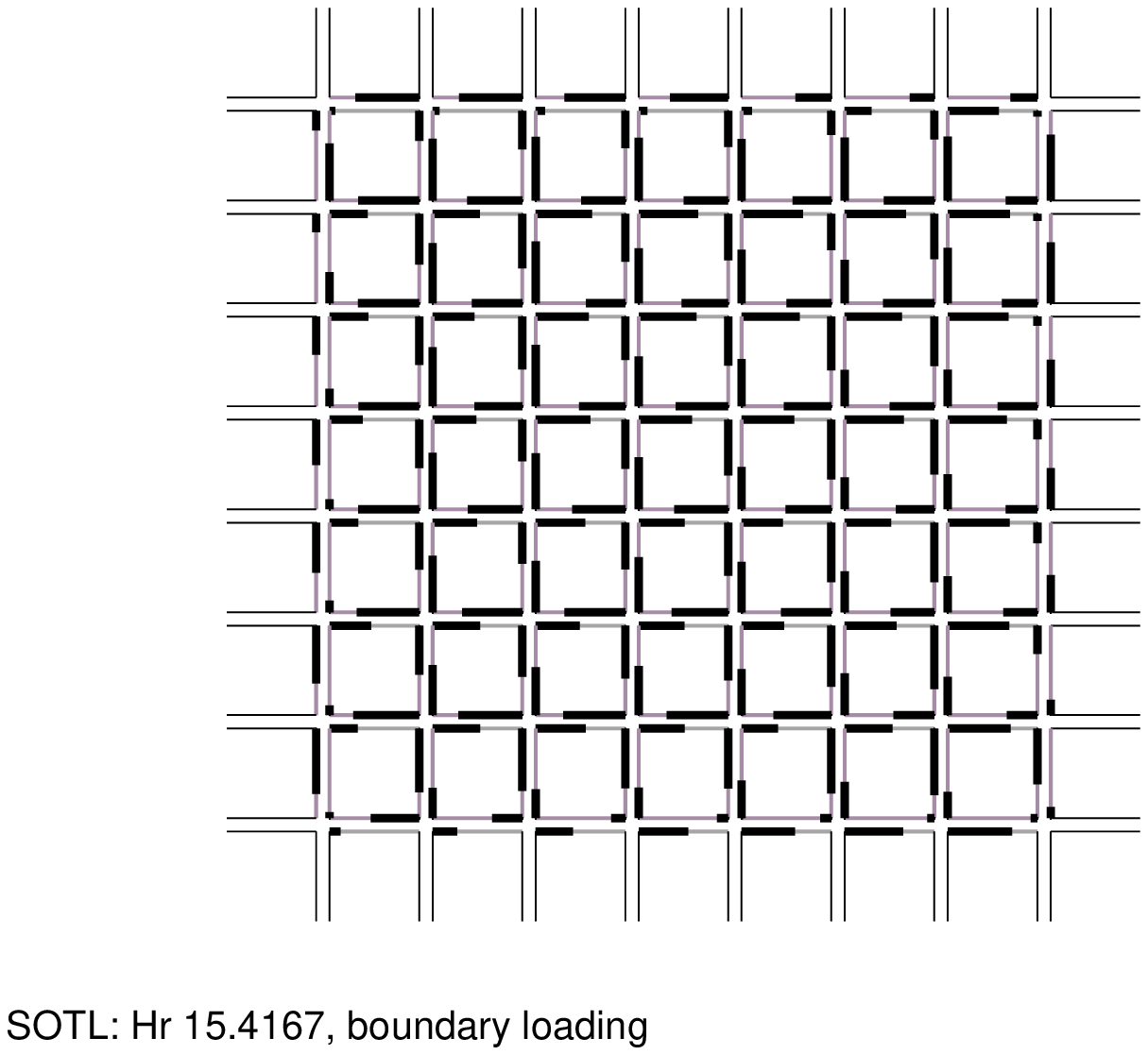}}

\def\twobinLoadingA			{\includegraphics[scale=\oneup]{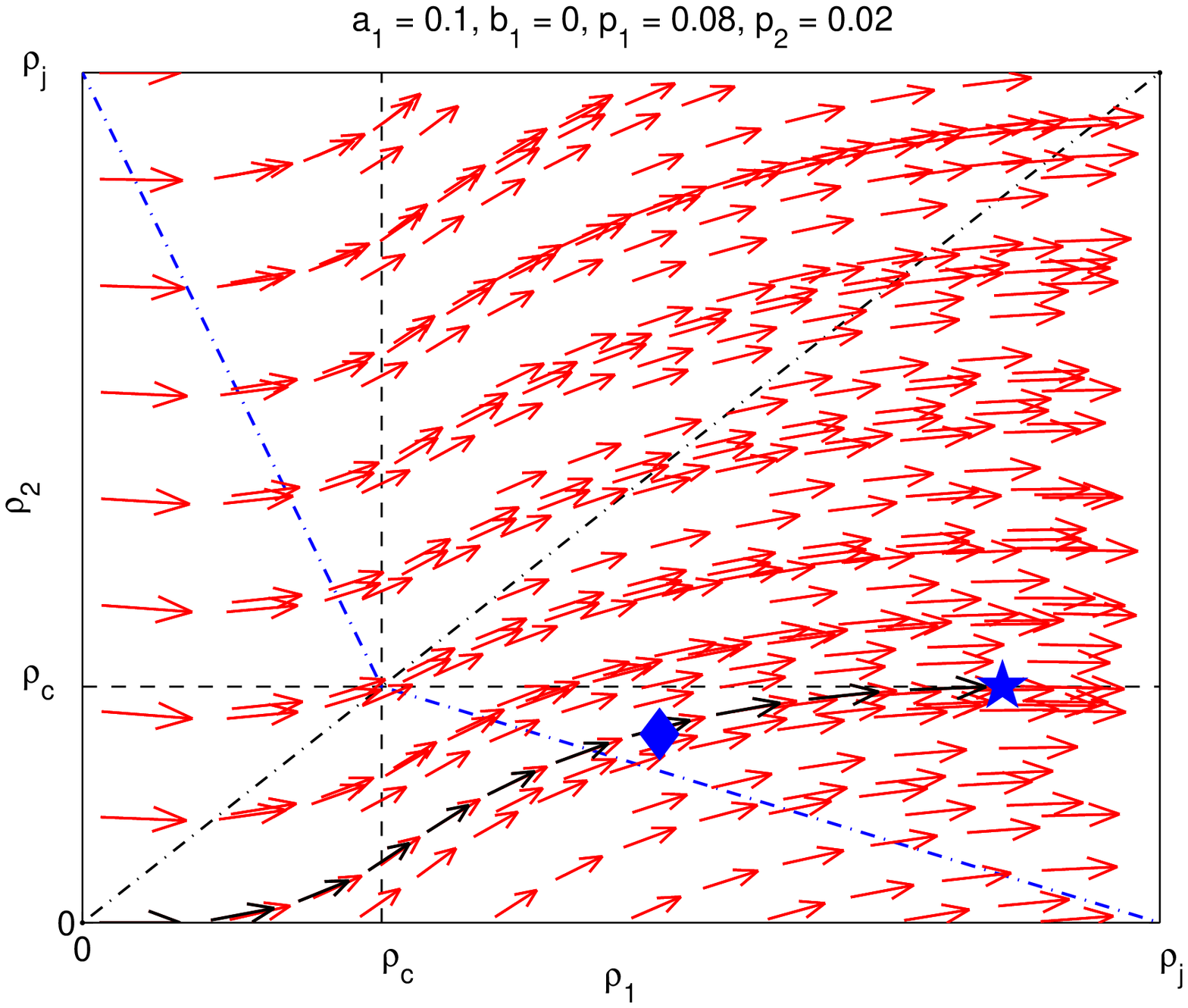}}
\def\twobinUnloading			{\includegraphics[scale=\oneup]{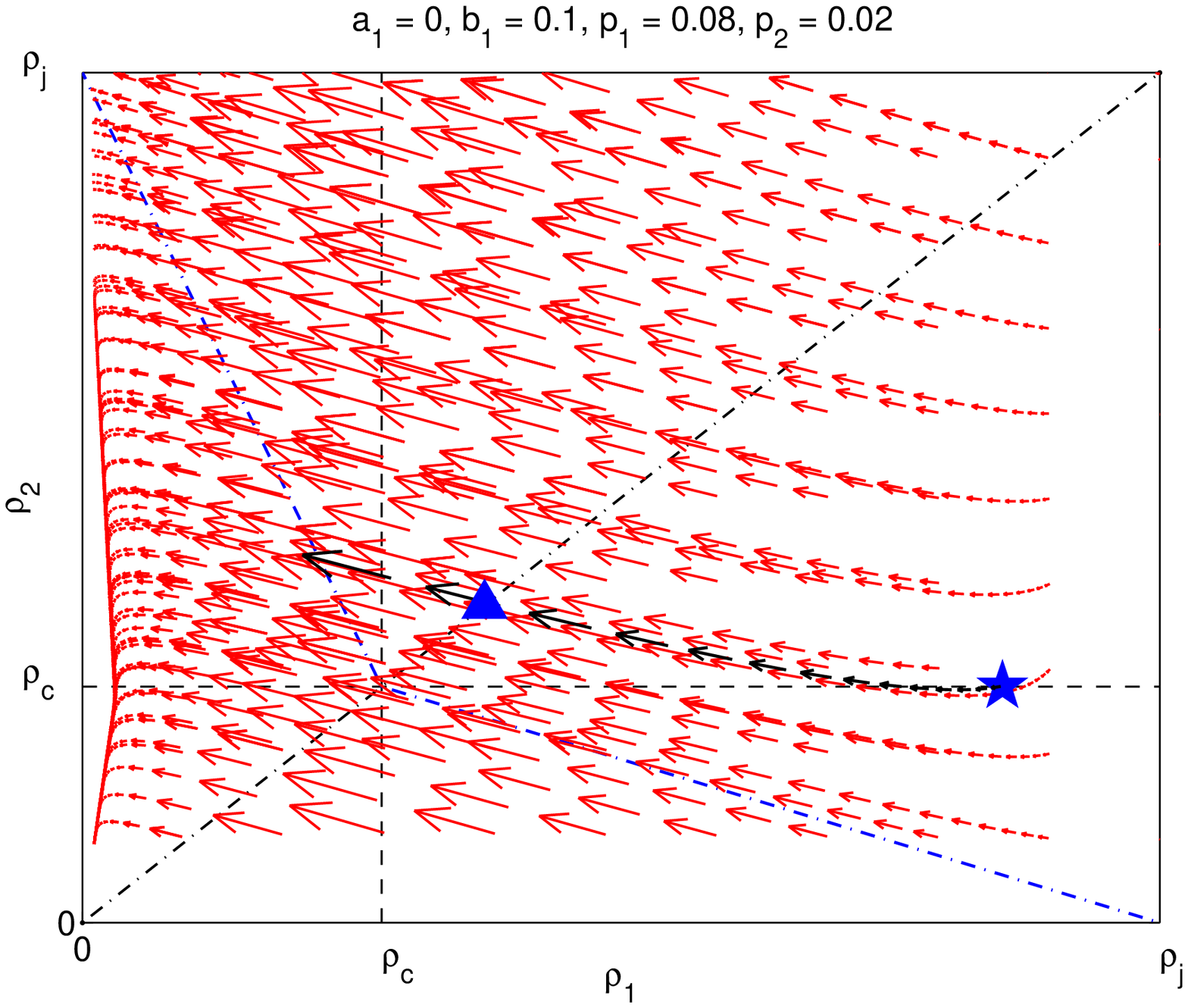}}
\def\twobinMFD				{\includegraphics[scale=\oneup]{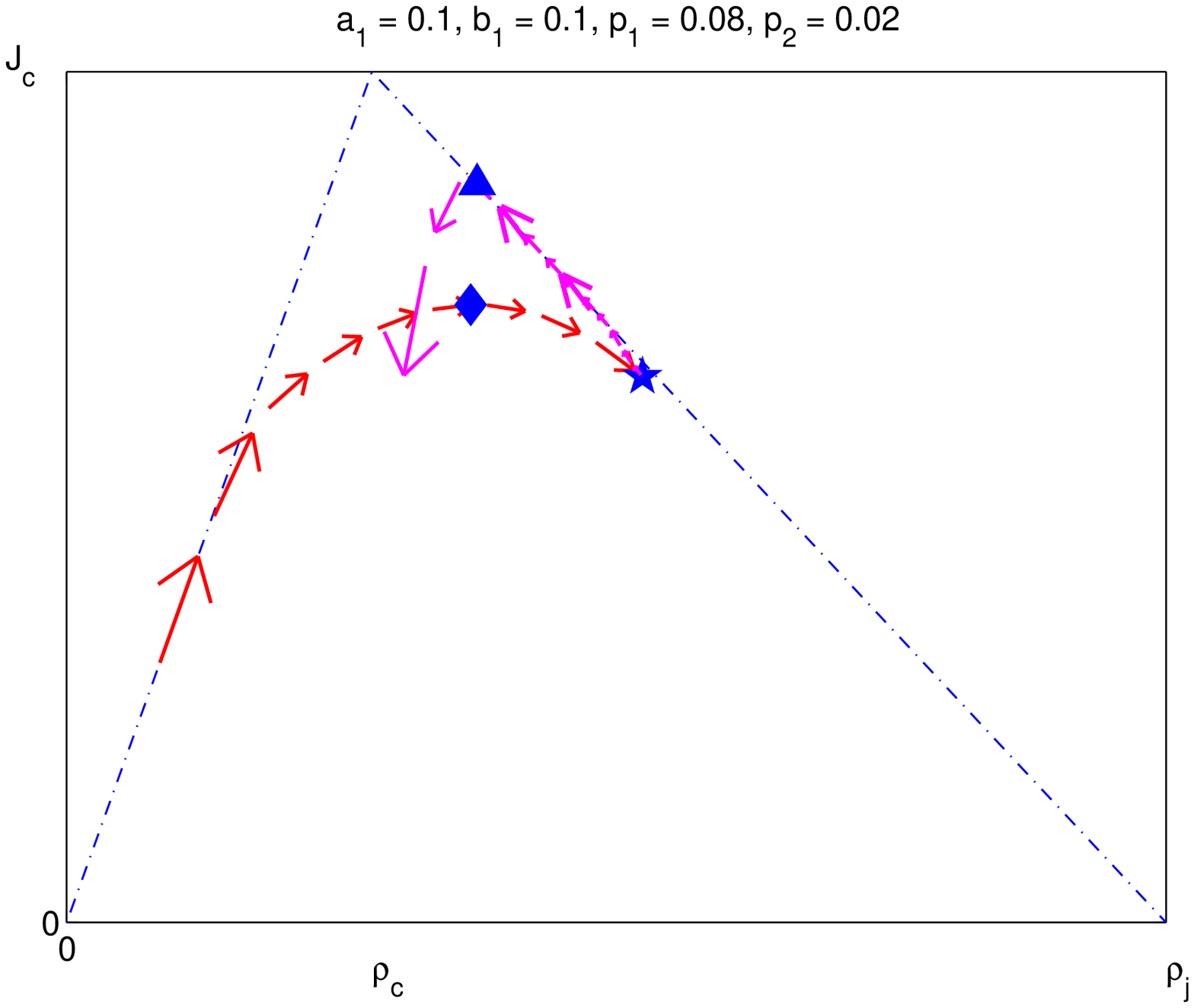}}
\def\twobinLoadingB			{\includegraphics[scale=\oneup]{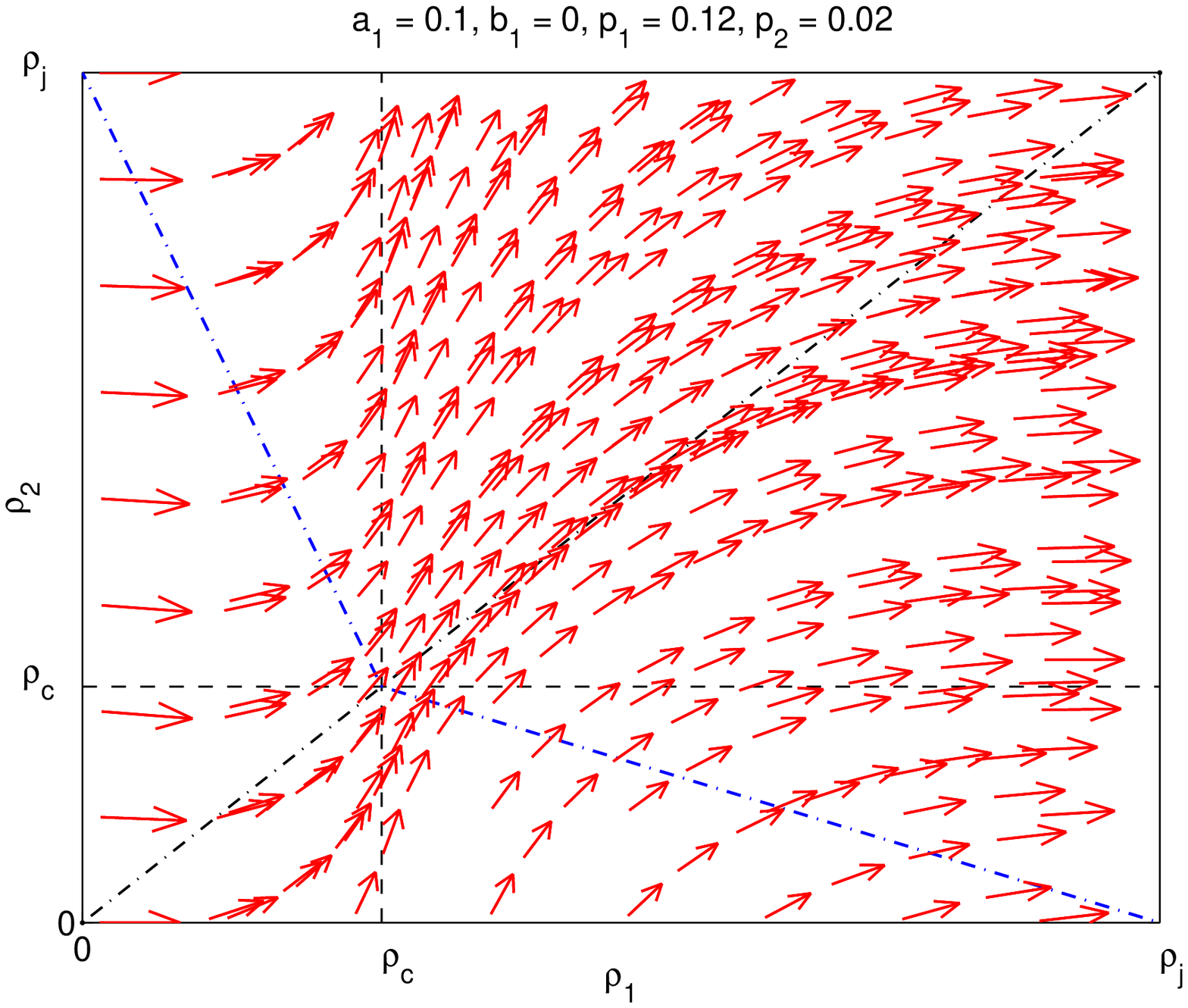}}
\def\scatsclselection		  	{\includegraphics[scale=\twoup]{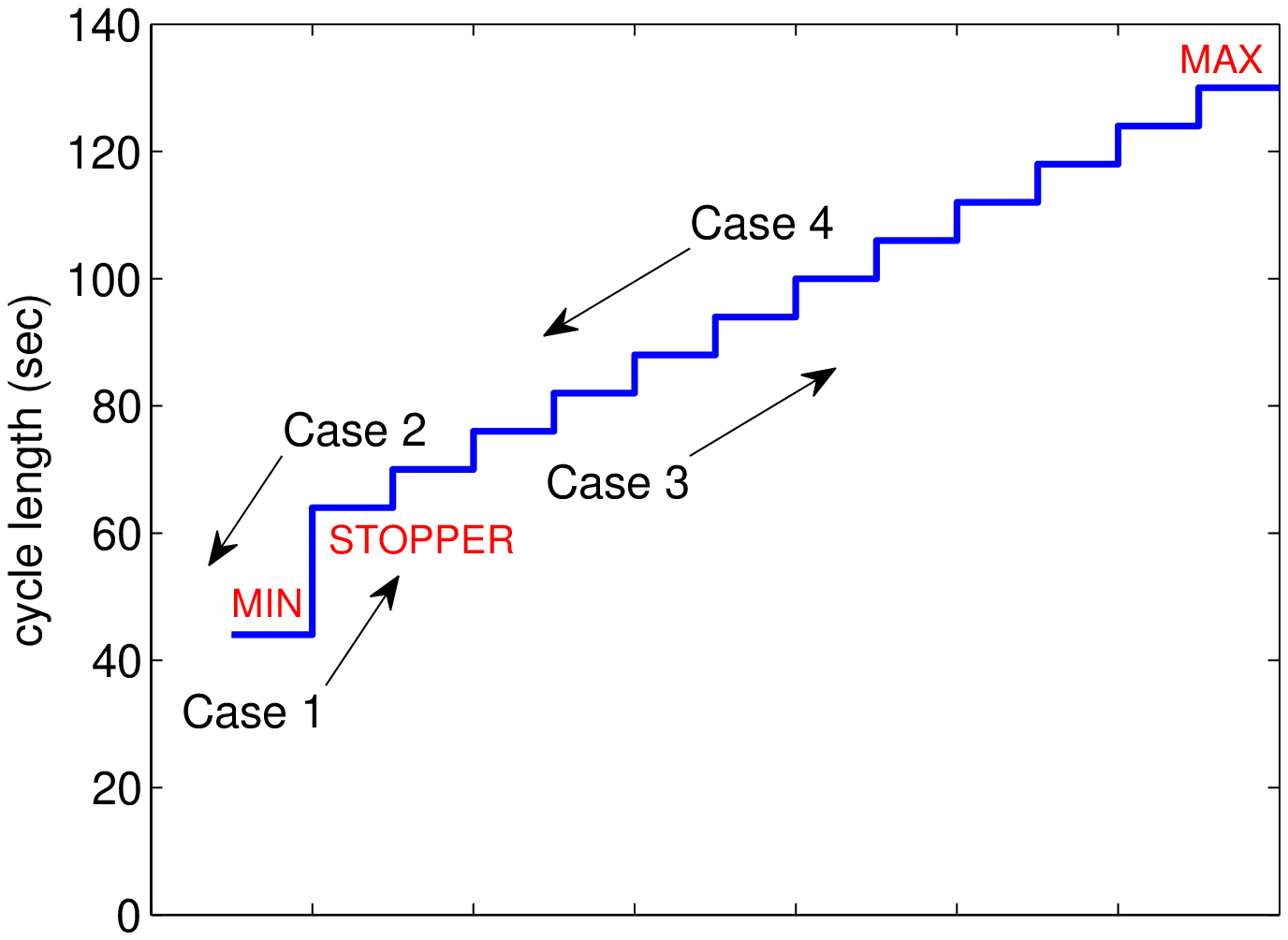}}

\begin{document}

\begin{frontmatter}
\title{A comparative study of Macroscopic~Fundamental~Diagrams of arterial road networks governed by adaptive traffic signal systems}

\author[mascos,monash]{Lele Zhang}
\ead{lele.zhang@monash.edu}
\author[monash]{Timothy M Garoni\corref{cor1}}
\ead{tim.garoni@monash.edu}
\author[unimelb]{Jan de Gier}
\ead{jdgier@unimelb.edu.au}
\cortext[cor1]{Corresponding author}
\address[mascos]{ARC Centre of Excellence for Mathematics and Statistics of Complex Systems, Department of Mathematics and Statistics, University of Melbourne, Victoria~3010, Australia}
\address[monash]{School of Mathematical Sciences, Monash University, Clayton, Victoria~3800, Australia}
\address[unimelb]{Department of Mathematics and Statistics, The University of Melbourne, Victoria~3010, Australia}

\begin{abstract}
Using a stochastic cellular automaton model for urban traffic flow,
we study and compare Macroscopic Fundamental Diagrams (MFDs) of arterial road networks governed by different types of adaptive traffic signal systems, under various boundary conditions.
In particular, we simulate realistic signal systems that include signal linking and adaptive cycle times,
and compare their performance against a highly adaptive system of self-organizing traffic signals which is designed to uniformly distribute the network density.
We find that for networks with time-independent boundary conditions, well-defined stationary MFDs are observed, whose shape depends on the particular signal system used, 
and also on the level of heterogeneity in the system.
We find that the spatial heterogeneity of both density and flow provide important indicators of network performance.
We also study networks with time-dependent boundary conditions, containing morning and afternoon peaks. 
In this case, intricate hysteresis loops are observed in the MFDs which are strongly correlated with the density heterogeneity.
Our results show that the MFD of the self-organizing traffic signals lies above the MFD for the realistic systems,
suggesting that by adaptively homogenizing the network density, overall better performance and higher capacity can be achieved.
\end{abstract}

\end{frontmatter}

\section{Introduction}\label{sec:intro}
A central goal of traffic science is the formulation of appropriate macroscopic variables characterizing and relating demand and performance of road infrastructure.
On the level of a single street (or freeway), the {\em fundamental diagram} (FD), introduced by~\cite{Greenshields35}, expresses flow as a function of density.
Fundamental diagrams for a single link are generically unimodal, describing a free-flow regime at low densities and a congested regime at high densities\footnote{Even in this simplest of cases, 
however, it is typically found experimentally that significant scatter is observed in flow-density relations of congested links; see~\cite{Kerner98}.}.
It is far from clear, however, to what extent such simple relations should extend to more complex systems such as urban road networks.
Early studies in this direction date back at least to~\cite{Godfrey69}. The first convincing empirical evidence that 
congested urban networks can display simple relationships between network-aggregated demand and performance was presented in \cite{GeroliminisDaganzo07,GeroliminisDaganzo08}.
These works clearly indicate the existence of a {\em Macroscopic Fundamental Diagram} (MFD) in the city of Yokohama, relating network-aggregated production and accumulation\footnote{The production and 
accumulation are surrogates for the flow and density, and are more readily measured in empirical trials.}.
Analytical theories attempting to explain the existence of MFDs have been developed by \cite{DaganzoGeroliminis08} and \cite{Helbing09}, and match the Yokohama data quite well.
The discussion in~\cite{DaganzoGeroliminis08} analyzes the effects on MFDs of applying different signal timings, by treating traffic signals as exogenous capacity constraints.
One of the main aims of the current work is to study and compare MFDs in networks governed by different types of {\em adaptive} traffic signal systems.

Given the existence of MFDs, it is natural to ask under what conditions should they be observed? 
In previous work on MFDs, \cite{DaganzoGeroliminis08} postulate sufficient regularity conditions under which MFDs should be expected to exist, including slow-varying and distributed demand, 
and homogeneous network infrastructure.
\cite{Helbing09} argues that the details of MFD curves should be expected to depend not only on the aggregated density, but also on the spatial density distribution.
Taking these observations further, \cite{MazloumianGeroliminisHelbing10} argue that the aggregated flow should in fact be a function of both the aggregated density and also its spatial variation.
\cite{GeroliminisSun11TRB} demonstrate using empirical data that while strict homogeneity of traffic states is not necessary to observe a well-defined MFD, 
the spatial distribution of density is indeed a key quantity.
In this work we study the spatial heterogeneity of both density and flow, and demonstrate how their behavior can be used as predictors of network performance.
 
In particular, we discuss the relationship between the time evolution of spatial heterogeneity and {\em hysteresis}.
Hysteresis has been observed and studied in freeway networks in \cite{GeroliminisSun11TRA}.
While hysteresis was not observed in~\cite{GeroliminisDaganzo08}, a careful empirical study of MFDs in the city of Toulouse was undertaken by \cite{BuissonLadier09},
in which hysteresis effects were clearly observed. 
A theoretical explanation of the clockwise hysteresis loops found in~\cite{BuissonLadier09} was presented in~\cite{GayahDaganzo11}, 
by treating the network as a dynamical system and performing a stability analysis. 
This analysis, which assumed homogeneous loading of the network, suggested that clockwise hysteresis loops should be typical, while anticlockwise hysteresis loops should be rare.
If the constraint of homogeneous loading is relaxed however, different behaviour can arise.
For example, for networks corresponding to {\em commuter corridors}, 
i.e. portions of an arterial network wedged between a strong source (e.g. a residential area) and a strong sink (e.g. the central business district), 
the input/output rates on the boundary may be significantly higher than in the bulk of the network.
Other situations where such behavior might arise include arterial networks in the presence of perimeter control.
We observe from our simulations that for such systems both clockwise and anticlockwise hysteresis loops can generically appear,
and we show that these observations can be explained using a modified version of the model presented in~\cite{GayahDaganzo11}.

Our simulations utilize a stochastic cellular automaton (CA) model, introduced by~\cite{deGierGaroniRojas11}.
This model is mesoscopic, in the sense that although individual vehicles are modeled, fine-grained details of individual driver behavior are deliberately treated in a course-grained, statistical, manner.
While details of vehicular motion through intersections are deliberately ignored, realistic signal phasing at intersections is included in the model.
In fact, the model was specifically designed to provide a simple and fast way to study arbitrary traffic signal systems, on arbitrary networks.
Using this CA model we study the existence and shape of MFDs for three specific traffic signal systems, using both time-dependent and time-independent boundary conditions.
In particular, we simulate variants of the SCATS\footnote{Sydney Coordinated Adaptive Traffic System} traffic signal system,
which is currently employed by numerous road authorities worldwide, including in Sydney and Melbourne. 
In order to study the effect of increased adaptivity on MFDs,
we then compare these results for SCATS with the highly-adaptive (idealized) {\em self-organizing traffic lights} (\sotl) system, 
originally introduced for Manhattan lattices by~\cite{Gershenson05}, and then generalized to arbitrary networks in~\cite{deGierGaroniRojas11}.
In particular, the version of \sotl\ that we study is specifically designed to minimize the spatial heterogeneity of the density within the network.

\cite{JiGeroliminis12} have recently addressed the question of how to decompose road networks into subnetworks so that each subnetwork has a well-defined MFD,
while \cite{Daganzo07} and \cite{HaddadGeroliminis12} have studied how to optimize the control of flows between such subnetworks.
\sotl\ can be viewed as a possible mechanism for adaptively homogenizing the density distributions within subnetworks.

The remainder of this paper is organized as follows.
In Section \ref{sec:camodel}, we describe the CA model introduced in~\cite{deGierGaroniRojas11}, and define the network parameters we use for the simulations in this paper.
Section~\ref{sec:observables} then defines the macroscopic quantities of interest in terms of observables of the CA model.
In Section \ref{sec:schemes}, we describe the three traffic signal systems that we study in our simulations, each of which has a different level of adaptivity.
Then Sections~\ref{sec:independent} and~\ref{sec:dependent} respectively describe the results of our simulations using time-independent and time-dependent boundary conditions.
Section~\ref{sec:twobin} discusses a modified version of the two-bin model in which the two bins are no longer assumed to be identical.
Finally, Section~\ref{sec:concl} concludes with a discussion.

\section{Cellular Automata Model}\label{sec:camodel}
We briefly outline the cellular automata model used in our simulations, which we refer to as the {\em NetNaSch} model. For a comprehensive description of the model see \cite{deGierGaroniRojas11}.

Cellular automata (CA) are models which are discrete in time, space and state variables, whose dynamical rules are local. 
The NetNaSch model represents a road network by a directed graph, in which the nodes represent intersections and the links represent streets.
With each link is associated an ordered list of lanes, and each lane is a simple one-dimensional stochastic CA obeying a (slight generalization of) the Nagel-Schreckenberg (NaSch) dynamics;
see~\cite{NagelSchreckenberg92}.
In addition, vehicles may move between neighboring lanes via simple lane-changing rules.
Thus, the dynamics along each given link is essentially a standard CA freeway model, albeit with input and output rates that are determined dynamically by the rest of the network.
The NetNaSch model intentionally avoids modeling the detailed motion of vehicles as they move through intersections;
the underlying assumption being that the actual time a vehicle physically spends in an intersection is unimportant compared to the time spent on the inbound link waiting to traverse the intersection.
This course-grained approach allows the model to be easily applied to networks of arbitrary topology, using any choice of desired signal phasing.

In order to mimic origin-destination behavior, the NetNaSch model demands that each vehicle makes a random decision about which link it wants to turn into at the approaching intersection.
More precisely, for each node $n$, we assign to each ordered pair $(l,l')$, where $l$ is an inlink and $l'$ an outlink of $n$, the probability $p_{\text T}(l \to l')$ that a vehicle on $l$
wants to turn into $l'$ when it reaches $n$.
The turning decision is made when the vehicle first enters $l$, since its choice of which link to turn into at the approaching intersection should influence its dynamics as it travels along $l$.
In particular, it influences the vehicle's choice of when to change lanes.
In order to guarantee the robustness of the model, however, we do allow for drivers to adaptively change their turning decisions when faced with very high levels of congestion. 
This enables the model to avoid becoming {\em frozen} in pathological states of gridlock caused by drivers adhering strictly to their turning decisions.
Specifically, suppose that a vehicle is queued at an intersection in a lane that is consistent with its current turning decision, but that the vehicle has waited through more than 
$n_G$ green signals without being able 
to clear the intersection, due to spillback on the link onto which it wishes to turn. In this instance,
the NetNaSch model allows the vehicle to remake its random turning decision. We set $n_G=6$ in the simulations in this work.

The NetNaSch model can be used with a variety of boundary conditions. In this paper we use open boundary conditions, and so the density in the network is not controlled directly.
Instead, at each time step, vehicles enter and exit the network stochastically, according to prescribed input/output rates. 
We call in- and output at the boundary of the network {\em exogenous}, while internal sources/sinks are called {\em endogenous}, for example representing parking garages.
In general, both exogenous and endogenous input and output are allowed.

A {\em boundary link} is a link which has one of its two endpoints within the network, and one external to the network.
Boundary links are classified as either boundary {\em inlinks}, if their to-node belongs to the network, or boundary {\em outlinks}, if their from-node belongs to the network. 
A {\em bulk link} is a link whose endpoints are both contained in the chosen network.
In the NetNaSch model, each lane $\lambda$ of each boundary inlink is assigned an input probability $\alpha_{\lambda}$:
at each discrete time step a new vehicle is inserted into the first cell of lane $\lambda$ with probability $\alpha_{\lambda}$.
Likewise, each lane $\lambda$ of each boundary outlink is assigned output probability $\beta_{\lambda}$, 
which determines the probability that a vehicle wishing to exit the network from the last cell of lane $\lambda$ at a given time step actually be allowed to do so.

The collections $\{\alpha_{\lambda}\}$, $\{\beta_{\lambda}\}$ therefore specify the exogenous input/output of the network, i.e. they describe the level of demand imposed on the network by its environment.
Intuitively, one can view $\alpha_{\lambda}$ as being the density of an external reservoir of vehicles being fed into boundary lane $\lambda$.
And likewise, one can view $(1-\beta_{\lambda})$ as being the density of an external sink being fed vehicles by boundary out-lane $\lambda$.
Indeed, for the 1-lane NaSch freeway model, where the reservoir and sink can be thought of as on- and off-ramps, these interpretations are quite realistic.

As discussed in Section~\ref{sec:observables}, the boundary links are not considered to be part of our network, in the sense that we do not include their densities and flows in our network aggregated values.
Instead, the boundary links are simply viewed as buffers allowing a realistic way to couple the bulk network to its external environment. 
A practical issue that must be decided upon is how long to make the boundary links. 
There is no unique best answer to this question; a detailed discussion of the pros and cons of different possibilities is presented in~\cite{deGierGaroniRojas11}.
For the simulations discussed in the present work, we simply set the length of the boundary links equal to the length of the bulk links. 
One advantage of this approach is that spillback caused by over-saturation on boundary outlinks is modelled dynamically, in a realistic way.

In addition to these exogenous inputs/outputs, each lane of each bulk link is assigned an input probability $\gamma_{\lambda}$, and an output probability $\delta_{\lambda}$,
which determine the rate of input and output from internal sources and sinks.
Intuitively, one can view these internal sources and sinks as parking garages, for example.
Then $\gamma_\lambda$ is the probability that at a given time step, a vehicle will leave the parking garage and enter the network, while
$\delta_\lambda$ is the probability that a vehicle passing such a parking garage will leave the network to enter the parking garage.
For the simulations discussed in the present work, the internal sources and sinks were located near the middle of the lane, with the sink occurring before the source.

For simplicity, we refer to the collection of all exogenous and endogenous inflow and outflow rates as the ``boundary conditions'' for the network, 
despite the fact that the endogenous rates are actually properties of the bulk.
In principle then, the boundary conditions for a network are specified by the collections
$\{\alpha_{\lambda}:\lambda\text{ is a lane of an inlink}\}$, $\{\beta_{\lambda}:\lambda\text{ is a lane of an outlink}\}$,
$\{\gamma_{\lambda}:\lambda\text{ is a lane of a bulk link}\}$ and $\{\delta_{\lambda}:\lambda\text{ is a lane of a bulk link}\}$.
For a given network, one could conceive of varying all of these parameters independently, from 0 to 1, and studying the resulting distributions of flow and density.
In order to meaningfully investigate MFDs however, we instead vary the $\alpha_{\lambda}$, $\beta_{\lambda}$, $\gamma_{\lambda}$ and $\delta_{\lambda}$ in a given systematic manner, 
corresponding to a reasonable demand scenario for an arterial network.
We discuss several such scenarios in Section~\ref{ssec:boundary}.
We emphasize that the values of $\alpha_{\lambda}$, $\beta_{\lambda}$, $\gamma_{\lambda}$ and $\delta_{\lambda}$ can vary with time.

We now summarize the details of the specific network and input parameters simulated in the present study.

\subsection{Links and lanes}\label{ssec:links}
According to the NaSch model, the speed $v$ of each vehicle can take one of $v_{\max} + 1$ allowed integer values $v = 0, 1, 2,\ldots, v_{\max}$.
Taking the length of a cell to be 7.5m, corresponding to the typical space occupied by each vehicle in a jam, and the duration of each time step to be 1s, 
suggests $v_{\max} = 3$ is a reasonable choice for an urban network.
I.e., each vehicle can move 0, 1, 2 or 3 cells per time step in such a CA model, depending on local traffic conditions. 
These are the values used in our simulations.
In addition, the NaSch model (and consequently the NetNaSch model) includes, at each time step and for each vehicle,
a random unit deceleration which is applied with probability $p_{\rm noise}$. By setting $p_{\rm noise}$ so that it is $0.5$ when the current vehicle speed is $v_{\max}$, and $0.2$ otherwise
we obtain an average free-flow speed of approximately 60 km/hr, which is typical of an arterial network. See the discussion in~\cite{deGierGaroniRojas11} for further details.

\begin{figure}
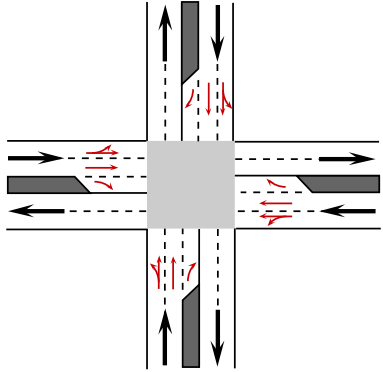

\centering
{\intersectionturnlanes}
\caption{Illustration of a typical node in the simulated network.}
\label{fig:intersectionturnlanes}
\end{figure}

The particular network we simulated in this study consists of a regular $8 \times 8$ square grid.
Each link in the network has two lanes plus an additional right-turning lane\footnote{Vehicles drive on the left side of the road in Australia.}.
Fig.~\ref{fig:intersectionturnlanes} shows a typical intersection in detail.
With the exception of Section~\ref{sssec:shortlinks}, the length of each bulk and boundary link was set to $750$m, corresponding to $100$ cells, and the length of each turning lane was set to $120$m.
This choice of link length corresponds to the distance between {\em signalized} intersections in an arterial network, 
and is typical of arterial road networks in predominantly suburban cities such as Melbourne.

\begin{figure}
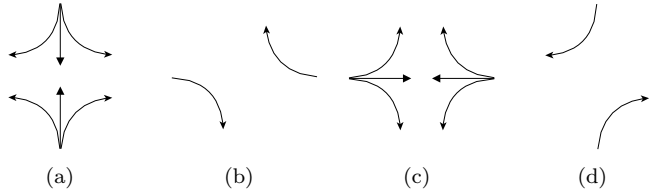

\centering
\subfigure[]{{\phasea}}\quad\subfigure[]{{\phaseb}}\quad\subfigure[]{{\phasec}}\quad\subfigure[]{{\phased}}
\caption{The four phases used at each node of the simulated network.}
\label{fig:phases}
\end{figure}

\subsection{Phases}\label{ssec:phases}

Each node was given the same four phases: a north/south phase, an east/west turning phase, an east/west phase and a north/south turning phase.
See Fig.~\ref{fig:phases}. This fixed ordering of phases was applied to our simulations of \scats.
Note that the phase in Fig.~\ref{fig:phases}-(a) is not necessarily the first phase of the cycle; 
for SCATS with signal linking this is determined by the linking protocol as described in Section~\ref{ssec:scats}. 

\subsection{Turning probabilities}\label{ssec:turn}
In all our simulations, each link was assigned the same turning probability of $p_{\text T}$ for left and right turns, implying a probability $1-2\,p_{\text T}$ of continuing straight ahead. 
The probability $p_{\text T}$ was set to $0.1$ for the majority of our simulations. For comparison, in Section~\ref{sssec:turning} we discuss simulations using $p_{\text T}=0.15,\ 0.2$.

\subsection{Boundary conditions}\label{ssec:boundary}
We consider a number of representative scenarios for the boundary conditions, which we summarize as follows.
\begin{enumerate}[label=\Roman{*}.]
\item\label{enum:Tind} {\em Time-independent.} All input and output rates are constant in time. Two main variations were studied.
\begin{enumerate}
\item\label{subenum:isotropic} {\em Isotropic.} The same value of $\alpha$ is applied to all inlinks, the same value of $\beta$ to all outlinks, 
  and the same values of $\gamma$ and $\delta$ to all bulk links. To obtain MFDs, the values of $\alpha$ and $\beta$ were varied
  while the values of $\gamma$ and $\delta$ were held constant.
\item\label{subenum:anisotropic} {\em Anisotropic}. The value of $\alpha$ on the west boundary is twice as large as on the other three boundaries,
  and likewise $\beta$ is only half as large on the east boundary. This sets up a west-to-east anisotropy in the demand imposed on the network. The endogenous sources and sinks were ignored in this case,
  $(\gamma,\delta)=(0,0)$.
\end{enumerate}
\item\label{enum:time-dependent} {\em Time-dependent}. In this case the boundary conditions are isotropic, but the values of $\alpha$, $\beta$, $\gamma$ and $\delta$ are time-dependent.

  The values of $\alpha$ and $\beta$ were changed every 30 minutes, and the network was simulated for 20 hours.
  For each of the three signal systems, the profiles of $\alpha$ and $\beta$ were reverse-engineered to produce time series for the density profiles that resemble the empirical data for
  Yokohama presented in~\cite{GeroliminisDaganzo08}.
  In each case, the peak densities were chosen so as to drive the MFD into the high-density regime during loading, and then allow it to relax back to the stationary low-density curve during recovery.
  Two extreme cases were studied.
\begin{enumerate}
\item\label{subenum:boundary-loading} {\em Boundary Loading}. Vehicles can enter and exit the network only via boundary links. I.e. $\gamma$ and $\delta$ are strictly zero.
\item\label{subenum:bulk-loading} {\em Uniform Loading}. We allow vehicles to be loaded and unloaded uniformly throughout the network by setting $\gamma=\alpha$ and $\delta=\beta$.
\end{enumerate}
\end{enumerate}

In the time-independent cases we compute MFDs by varying the exogenous demand, for a number of fixed choices of endogenous demand.
This corresponds in some sense to viewing the endogenous demand as an inherent part of the network, comparable to the choice of signal system etc,
and studying how the network responds to different levels of external demand.
Systems with low endogenous demand correspond to ``commuter corridors''; portions of the arterial network wedged between a strong source (e.g. a residential area) and a strong sink 
(e.g. the central business district). 
Higher values of $(\gamma,\delta)$ correspond to introducing shopping centers and parking garages etc into the bulk of the network.

In the time-dependent case, in addition to the two extreme cases of boundary loading and uniform loading, we also studied a number of intermediate scenarios, 
in which the strength of the endogenous demand was non-zero but less than the exogenous demand.
The resulting behavior of these systems was intermediate between the two extremes (\ref{subenum:boundary-loading}) 
and (\ref{subenum:bulk-loading}) and so for the purposes of illustration it suffices to focus on boundary and bulk loading.

Finally, we note that \cite{BuissonLadier09} present a careful discussion of possible sources of network heterogeneity in empirical studies of MFDs, 
including the position of detectors, types of roads, and traffic signals. We deliberately study symmetric square-lattice networks for which all links are equivalent and all nodes are equivalent. 

\section{Observables}\label{sec:observables}
We define the density, $\rho_l(k)$, of link $l$ at the $k$th time step of a simulation to be the fraction of all cells on $l$ which are occupied at that instant.
The flow, $J_{\lambda}(k)$, of lane $\lambda$ during the $k$th time step is simply the indicator for the event that a vehicle crosses the boundary between 
a fixed pair of neighboring cells during the $k$th update\footnote{In our simulations, the flow is measured $2v_{\max}$ cells from the upstream node of the link.}.
The flow $J_{l}(k)$ on link $l$ at the $k$th time step is then simply the sum of the $J_{\lambda}(k)$ over all lanes $\lambda$ in link $l$.
We emphasize that since our model is stochastic, the observables $\rho_l(k)$ and $J_l(k)$ are {\em random variables}.

Since we are interested in the dynamics on the order of traffic cycles, rather than iterations of our model, 
we {\em bin} the instantaneous link flow and density into bins of size $b$, using $b=5$ minutes in our simulations. In a slight abuse of notation, we define
\be
\rho_l(t) := \frac1b \sum_{k=(t-1)b+1}^{kb} \rho_l(k)\quad \text{and}\quad J_l(t)=\frac1b \sum_{k=(t-1)b+1}^{kb} J_l(k),
\label{link observables}
\ee
where the physical time $t$ is measured in intervals of $b=5$ minutes.

Let us denote the set of all {\em bulk links} in the network by $\Lambda$.
We emphasize that the {\em boundary} links in the NetNaSch model are not considered to be part of the network, and serve only as an effective means of connecting the bulk to the external environment.
From the link observables~\eqref{link observables}, we then define the following macroscopic network-aggregated observables:
\begin{equation}
\begin{split}
\rho(t) &:= \frac{1}{|\Lambda|} \sum_{l\in\Lambda}\rho_l(t),\\
h_{\rho}(t) &:= \sqrt{\frac{1}{|\Lambda|} \sum_{l\in\Lambda}[\rho_l(t) - \rho(t)]^2},\\
J(t) &:= \frac{1}{|\Lambda|} \sum_{l\in\Lambda}J_l(t),\\
h_{J}(t) &:= \sqrt{\frac{1}{|\Lambda|} \sum_{l\in\Lambda}[J_l(t) - J(t)]^2}.
\label{network observables}
\end{split}
\end{equation}
Again, we emphasize that these macroscopic observables are random variables in our model, although by aggregating the data over both time and space the fluctuations of these macroscopic observables
will be significantly suppressed relative to the original instantaneous link observables. 

The quantities $J(t)$ and $\rho(t)$ are the network-aggregated flow and density.
We refer to the quantities $h_{\rho}(t)$ and $h_{J}(t)$ as the {\em heterogeneity} (spatial variability) of the density and flow respectively, 
since they give a measure of the extent to which the spatial distribution of the link-level observables differ from the corresponding network-aggregated values. 
Note that the heterogeneities achieve their lower bound of 0 only when the corresponding link observables are all equal.

A fundamental question to be studied via our simulations is the extent to which $\rho(t)$ and/or $h_{\rho}(t)$ determine the value of $J(t)$. 
The statement that an invariant MFD exists implies that $J(t)$ should be a function of $\rho(t)$ alone. 
However, recent work by \cite{MazloumianGeroliminisHelbing10} and \cite{GeroliminisSun11TRB} suggest that $h_{\rho}(t)$ is also an important indicator of network performance. 
Our simulations confirm this. In fact, we find that both $h_{\rho}(t)$ and $h_{J}(t)$ provide valuable indicators of network performance.

\subsection{Statistics}
For each distinct choice of traffic signal system and boundary conditions, we performed $n$ independent simulations (with $n$ ranging between $10$ and $30$),
in order to estimate the expected values of the network-aggregated quantities 
defined in~\eqref{network observables}. For a given observable $X(t)$, if we denote its realization in the $i$th run by $X^{(i)}(t)$ then we compute
\begin{align}
\overline{X(t)} &=\frac1n\sum_{i=1}^n X^{(i)}(t),\\
\text{err}(\overline{X(t)}) &= \sqrt{\frac1{n(n-1)}\sum_{i=1}^{n}[X^{(i)}(t)-\overline{X(t)}]^2},
\end{align}
where $\overline{X(t)}$ is the natural estimator for the expected value of $X(t)$ and $\text{err}(\overline{X(t)})$ is its standard error.

\subsection{Physical Units}\label{subsec:units}
The density, $\rho_l$, gives the fractional occupation of the link $l$, relative to the maximum physically possible density, $\varrho_{\max}$.
Consequently, $\rho_l$ is a dimensionless quantity lying in the interval $[0,1]$.
Taking the length of a cell in the NetNaSch model to be 7.5m, as discussed in Section~\ref{ssec:links}, gives a maximum density of $\varrho_{\max}=400/3\approx133 \text{veh/km}$.
To obtain densities in physical units of veh/km, therefore, the dimensionless densities $\rho_l$ can be multiplied by $\varrho_{\max}=133$.
The network aggregated density, $\rho$, and its heterogeneity, $h_{\rho}$, have the same dimensions as $\rho_l$ and the same conversions apply.

Taking the duration of each time step in the NetNaSch model to be 1s, as discussed in Section~\ref{ssec:links}, implies that the flow $J_{l}$ on link $l$ is measured in units of veh/s.
We emphasize that the link flows we measure are not normalized by the number of lanes, which in all cases is 2.
The network aggregated flow, $J$, and its heterogeneity, $h_{J}$, have the same dimensions as $J_l$.

\section{Traffic Signal Systems}\label{sec:schemes}
We simulate and study the existence of MFDs in networks using three distinct traffic signal systems:
\begin{description}
\item[\scatsl:] A model of \scats\ with linking and adaptive cycle lengths.
\item[\scatsf:] A ``free'' version of \scatsl,\ with no signal linking.
\item[\sotla:] Self-organizing traffic lights. 
\end{description}

The \scats\ traffic signal system, which controls the traffic signals in numerous cities around the world,
uses knowledge of the recent state of traffic to choose appropriate values of three key signal parameters: cycle length, split time, and linking offset.
At each intersection it can adaptively adjust both the total cycle length, and the fraction ({\em split}) of the cycle given to each particular phase.
In addition, it can coordinate ({\em link}) the traffic signals of several consecutive nodes along a predetermined route
by introducing fixed {\em offsets} between the starting times of specific phases, thereby creating a green wave.
Both the \scatsl\ and \scatsf\ models are special cases of our general \scats\ model, which we outline in Section~\ref{ssec:scats}

As a benchmark with which to compare the \scats-like traffic signal systems, we also considered the highly-adaptive {\em self-organizing traffic lights} (\sotl) system (see~\cite{deGierGaroniRojas11}).
The SOTL system is based on the simple principle that each node (intersection) should choose its current phase to be the phase which currently has the highest demand.
Unlike SCATS, no direct coordination is enforced between the signals at neighboring nodes, however such coordination is often seen to arise via {\em self-organization}, since
neighboring nodes do indirectly communicate with each other via the levels of traffic that they accept and release.
The particular version of \sotl\ that we study here uses density as the demand metric, and it therefore strives to adaptively minimize the network's density heterogeneity.
The details of the SOTL system are summarized in Section~\ref{ssec:sotl}.

\subsection{\scats\ }\label{ssec:scats}
The \scats\ control system adaptively controls three key signal parameters: linking offset, cycle length and split time. We discuss below how our model of SCATS chooses each of these parameters.

\subsubsection{Linking}
\begin{figure}
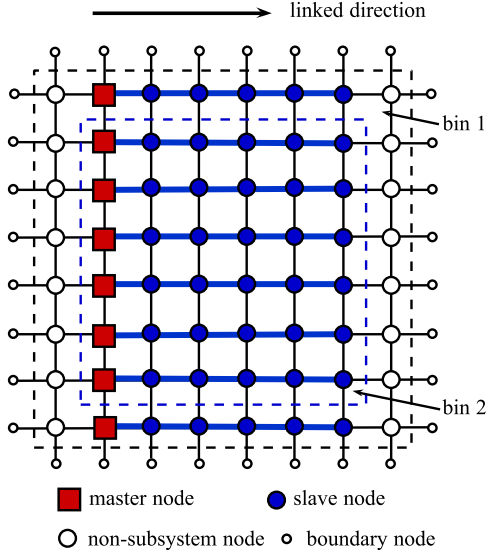

\centering
\latticeeighta
\caption{An 8 by 8 lattice network with 8 linearly linked subsystems.}
\label{fig:lattice8a}
\end{figure}

A {\em subsystem} in a SCATS network is a group of nodes which all share a common cycle length.
If a node does not belong to any subsystem, we call it a {\em non-subsystem node}.
Within each subsystem, we appoint a unique {\em master} node $m$ and a number of {\em slave} nodes $s$.
Fig.~\ref{fig:lattice8a} illustrates an 8 by 8 network with eight subsystems, each consisting of one master node and six slave nodes.
To implement linking, each node is assigned a special phase $\mp^*$, which is its {\em linked phase}.
If the linked phase $\mp^*_m$ of the master node $m$ starts at time $t$ then $\mp^*_s$ of the slave node starts at time $t+T_s$, where $T_s$ is the {\em offset}.
Ideally, the linking offset should be chosen based on the distance $L$ between $m$ and $s$, and the instantaneous local space-mean speed.
In practice, actual implementations of \scats\ tend to operate with fixed offsets during a given period of the day (for example morning peak hour). 
In our simulations we therefore use a fixed {\em linking speed} $\overline{v}=54$km/hr, which is just slightly less than the average free-flow speed of about 60km/hr, see Section~\ref{ssec:links}.

\subsubsection{Cycle length and split plan}
\scats\ chooses the unique cycle length within a subsystem based on the local traffic conditions in the neighborhood of the master node, as quantified by the {\em Degree of Saturation} (DoS).
Every time a master node is about to restart its cycle, the cycle length is adjusted adaptively based on recent measured values of the DoS.
In our model of \scats, the cycle length is selected based on the {\em volume ratio}.
For an inlink $\ml$ and phase $\mp$ the volume ratio is defined to be 
\be
R(\ml,\mp) = \frac{1}{N(\ml,\mp)}\frac{\mv(l,\mp)}{\ms(\mp)},
\label{volume ratio definition}
\ee
where $\mv(l,\mp)$ is the measured traffic volume out of inlink $\ml$ during phase $\mp$, and $\ms(\mp)$ is $\mp$'s current split time.
The quantity $N(\ml,\mp)$ denotes a fixed benchmark volume for the link and phase,
measured in vehicles per second\footnote{In our simulations $N(\ml,\mp)$ was set to 1 veh/sec for all phases.}.
If the volume ratio was large during the previous cycle, then the cycle length is increased by a fixed amount.
Conversely, if green time was wasted during the previous cycle, the cycle length is decreased.
The key underlying strategy is to attempt to keep the volume ratio within the range $[0.85,0.95]$.
Once the cycle length is determined, the split of a phase is taken to be proportional to its traffic volume during the previous cycle.
The specific details of cycle length and split time selection are discussed more fully in~\ref{asec:scats}.

\subsubsection{Versions of \scats\ }
Based on the model of \scats\ outlined above, we considered two variants: {\scatsl\ and \scatsf}. The first variant, \scatsl, operates on the linked network shown in Fig.~\ref{fig:lattice8a}.
By contrast, \scatsf\ operates on the network where no subsystems or linking are imposed, so that each node chooses its own cycle length and split time plan according to its local traffic state,
independently of its neighbors.

\subsection{SOTL}\label{ssec:sotl}
The third signal system we study is the self-organizing traffic lights (SOTL) system described in~\cite{deGierGaroniRojas11}.
While \scatsl\ and \scatsf\ are adaptive in their cycle length and split time selections, they both maintain a fixed cyclic ordering of each node's phases. By contrast, the
\sotla\ system is acyclic, and is designed so that at each phase change, the phase which currently has highest demand is selected. This procedure is applied independently to each node.

Suppose we agree on a suitable demand function $d(\mp)$ which quantifies the demand of each phase $\mp$ of each given node.
Phases with large values of $d(\mp)$ should be candidates for being the next choice of the active phase.
However, one should also keep track of the time $\tau(\mp)$ that each phase has been idle, since we do not want a given phase to remain idle for too long, unless it has strictly zero demand.
The key idea behind SOTL is to compute a threshold function, $\kappa(\mp)$, for each phase $\mp$, which depends on both the phase's idle time and demand function, 
and when $\kappa(\mp)$ reaches a predetermined threshold value,
$$
\kappa(\mp)>\theta,
$$
we consider making $\mp$ the active phase. For a detailed general discussion of the SOTL methodology, see~\cite{deGierGaroniRojas11}.

In the simulations performed in the current work, the demand $d(\mp)$ of phase $\mp$ was simply chosen to be the total number of vehicles over all its inlinks, and the threshold function was
\be
\kappa(\mp)=\frac{d(\mp)\tau(\mp)}{\sum_{\mp'} d(\mp')},
\ee
This particular choice for the demand function implies that \sotl\ attempts, at each instant of time, to adaptively minimize the network's density heterogeneity.
We used a threshold value of $\theta=5$.

A precise algorithmic description of \sotla\ is given in Algorithm~\ref{alg:sotl} in ~\ref{asec:sotl}.

\section{Simulations: Time-independent boundary conditions}\label{sec:independent}
We simulated the network described in Section~\ref{sec:camodel}, using the three different traffic signal systems described in Section~\ref{sec:schemes},
and measured the macroscopic observables defined in~\eqref{network observables}.
In this section, we present the results for the time-independent boundary conditions defined in Section~\ref{ssec:boundary}.
We simulated each system for 10 hours, which ensured that stationarity was always reached.

\subsection{Isotropic boundary conditions}\label{ssec:uniform}
In this section, we present our results for simulations using the isotropic Boundary Condition~(\ref{subenum:isotropic}).
In this case, we have two free parameters, $\alpha$ and $\beta$, where $\alpha$ is the input probability on each boundary inlink, and $\beta$ the output probability on each boundary outlink.
Fixed values for the bulk input and output probabilities $\gamma$ and $\delta$ were applied to all bulk links. 
Three different choices for the fixed values of $(\gamma,\delta)$ were simulated: $(\gamma,\delta)=(0,0), (0.05,0.1), (0.1,0.2)$.

We note that in the isotropic case, the \scatsl\ system is somewhat artificial, since there is no motivation for imposing linking if the demand is isotropic; we include \scatsl\ here to enable comparisons with 
the results for the anisotropic network discussed in Section~\ref{ssec:biased}.

\subsubsection{Zero $\gamma$ and $\delta$}\label{sssec:zerogammadelta}
Fixing $\gamma=\delta=0$,
we simulated the network using a number of different values of $\alpha$ and $\beta$, in order to obtain a range of values of the aggregated network density, ranging from very low to very high.
We observed that, in all cases, the flow and density reach approximately stationary values by hours 5 or 6. 
For a given choice of traffic signals, the longest relaxation times were observed for flows close to capacity.
For \sotl, stationarity was always achieved much faster than for \scats\ (at comparable densities), typically by hours 3 or 4,
which implies that the relaxation time of the network using \sotl\ is much smaller than when using \scats.

In Figs.~\ref{subfig:uniform-scatsf-mfd}, \subref{subfig:uniform-scatsl-mfd} and \subref{subfig:uniform-sotl-mfd} we plot $\overline{J}$ against $\overline{\rho}$ for \scatsf, \scatsl, and \sotl\ respectively,
at hours $1, 2, \dots, 6$ of the simulations.
For each signal system, the low-density branch of the curves are essentially time-independent, as is the highly-oversaturated region of the congested branch.
For intermediate values of density, approximately in the range $[0.3,0.8]$, we see a time dependence in the early hours of the simulation, 
however we also clearly see that the $\overline{J}$ vs $\overline{\rho}$ curves are converging to a well-defined stationary MFD as time increases. 
After approximately 6 hours of simulation, the $\overline{J}$ vs $\overline{\rho}$ curves at all later times are essentially indistinguishable. 
We note that \cite{MazloumianGeroliminisHelbing10} observed similar time-dependent behavior during three-hour simulations of their model,
and they concluded that such time-dependence would likely persist at all later times.
Fig.~\ref{fig:fdhoursT10} would suggest however that such time-dependence is in fact transient, and, for all practical purposes, ceases to be observable after some finite time.
It is conceivable that the behavior observed in \cite{MazloumianGeroliminisHelbing10} is an consequence of their use of periodic boundary conditions.

\begin{figure}
\centering
\subfigure[Time evolution of MFDs: \scatsf]{\label{subfig:uniform-scatsf-mfd}\fdcuscatsfhoursturnten}
\subfigure[Time evolution of MFDs: \scatsl]{\label{subfig:uniform-scatsl-mfd}\fdcuscatslhoursturnten}
\subfigure[Time evolution of MFDs: \sotl]{\label{subfig:uniform-sotl-mfd}\fdcusotlhoursturnten}
\subfigure[Stationary MFDs]{\label{subfig:uniform-comparison-mfd}\fduniformturnten}
\caption{Figs.~\subref{subfig:uniform-scatsf-mfd}, \subref{subfig:uniform-scatsl-mfd}, and~\subref{subfig:uniform-sotl-mfd} show MFDs of \scatsf, \scatsl, and \sotl,
at hours 1,2,\ldots 6, for a network with isotropic and time-independent boundary conditions, and no internal sources/sinks.
Fig.~\subref{subfig:uniform-comparison-mfd} shows a comparison of the stationary MFDs for the three signal systems.
Error bars corresponding to one standard deviation are shown, but are often smaller than the symbol size of the data point.
}
\label{fig:fdhoursT10}
\end{figure}

From Figs.~\ref{subfig:uniform-scatsf-mfd} and~\ref{subfig:uniform-scatsl-mfd}, we can see that for both \scatsf\ and \scatsl,
the instantaneous MFD curves appear to decrease monotonically with time towards their stationary limits. 
From Fig.~\ref{subfig:uniform-sotl-mfd} however, we notice that for \sotl\ the MFD at hour 1 lies below the limiting stationary curve.
We shall return to a discussion of this transient behavior in Section~\ref{sssec:stationary_heterogeneity}.

Fig.~\ref{subfig:uniform-comparison-mfd} shows a comparison of the stationary MFDs for the three traffic signal systems, \scatsf, \scatsl, and \sotl.
In the low density regime, the performance of each system is quite similar.
However, \sotl\ clearly allows the network to reach a higher capacity:
\sotla\ achieves a maximum flow of approximately $0.43$ at a density around $0.34$, while \scatsf\ and \scatsl\ obtain maxima of approximately $0.39$ at densities around $0.19$.
This represents a 10\% increase in network capacity by using \sotl, compared with \scats.
We note that \scatsf\ performs better than \scatsl\ in the high density regime. 
This is in fact to be expected, since for an isotropic network, linking should at best be merely unhelpful, and at worst it will be counterproductive because it will reduce the system's adaptivity.
We return to the comparison between \scatsf\ and \scatsl\ in Section~\ref{ssec:biased}.

\subsubsection{Heterogeneity}\label{sssec:stationary_heterogeneity}
In order to gain insight into the underlying cause of the transient behavior displayed by the MFDs, we consider how the corresponding heterogeneity curves evolve with time.
We begin by considering the density heterogeneity.

\begin{figure}
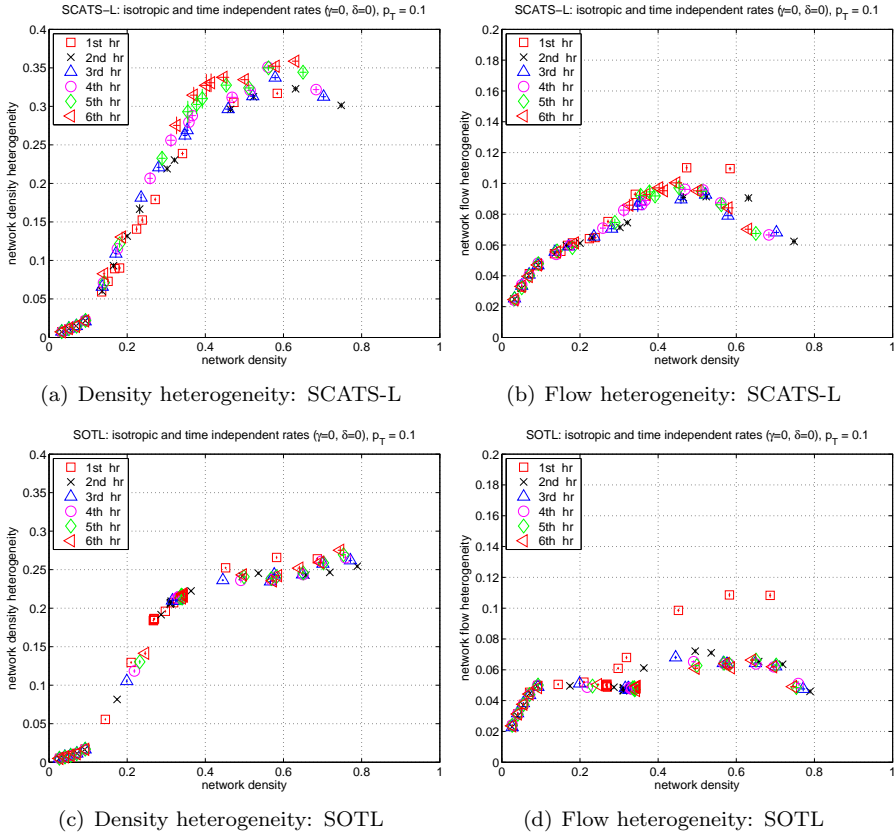

\centering
\subfigure[Density heterogeneity: \scatsl\ ]{\label{subfig:uniform-scatsl-hetero-rho}\fdheteroRhoscatslhoursturnten}
\subfigure[Flow heterogeneity: \scatsl\ ]{\label{subfig:uniform-scatsl-hetero-J}\fdheteroJscatslhoursturnten}
\subfigure[Density heterogeneity: \sotl\ ]{\label{subfig:uniform-sotl-hetero-rho}\fdheteroRhosotlhoursturnten}
\subfigure[Flow heterogeneity: \sotl\ ]{\label{subfig:uniform-sotl-hetero-J}\fdheteroJsotlhoursturnten}
\caption{Heterogeneity versus density at hours 1,2,\ldots,6 for network governed by \scatsl\ and \sotl, with isotropic and time-independent boundary conditions.
Error bars corresponding to one standard deviation are shown, but are often smaller than the symbol size of the data point.
}
\label{fig:uniform_heteroT10}
\end{figure}
In Fig.~\ref{subfig:uniform-scatsl-hetero-rho} we plot $\overline{h}_{\rho}$ for \scatsl\ at hours 1,2,\ldots 6.
\cite{GeroliminisSun11TRB} find empirically that for a given network density, the flow should be lower when the density heterogeneity is higher.
Comparison of Fig.~\ref{subfig:uniform-scatsl-mfd} with Fig.~\ref{subfig:uniform-scatsl-hetero-rho} confirms this;
if one observes the time evolutions of the MFD curve and $\overline{h}_{\rho}$ vs $\overline{\rho}$ curve in the neighborhood of a given value of $\overline{\rho}$,
there is a clear anticorrelation between the flow and the density heterogeneity.
At moderate values of the density, $\overline{h}_{\rho}$ starts at low values early in the simulation, and then increases to a stationary curve at around hour 5 or 6.
Conversely, the flow starts at relatively high values and decreases to its stationary value, again reaching stationarity at around hour 5 or 6.
Similar behavior is observed for \scatsf.

If one considers the analogous plots for \sotl, Figs.~\ref{subfig:uniform-sotl-mfd} and~\ref{subfig:uniform-sotl-hetero-rho}, at hours 3,\ldots, 6, precisely the same behavior can be observed,
although it is less pronounced.
The most obvious feature in Figs.~\ref{subfig:uniform-sotl-mfd} and~\ref{subfig:uniform-sotl-hetero-rho}, however,
is the behavior of the instantaneous curves at hour 1 (and to a lesser extent, hour 2): 
for moderate values of density, the flow is below its stationary limit, while $\overline{h}_{\rho}$ is above its stationary limit.
This again demonstrates the anticorrelation between the network flow and the density heterogeneity.
In fact, the same type of behavior displayed by \sotl\ at hour 1 is also present in \scatsl\ and \scatsf, however it largely occurs before hour 1 in these cases.
This type of behavior is presumably related to the fact that the simulations are started from a completely empty network, with vehicles only entering via the boundary.

The general conclusion to be drawn from these observations is that while the transient behavior of the flow at a given density might be subtle (e.g. non-monotonic in time), 
at all times there is a strong anticorrelation between flow and density heterogeneity.
We shall see in Section~\ref{sec:dependent} that this relationship between the transient behavior of $J$ and $h_{\rho}$ plays an important role in understanding hysteresis in networks with
time-dependent boundary conditions.

As a practical observation, we note that $\overline{h}_{\rho}$ is considerably smaller when the network is governed by \sotl\ than when governed by \scatsl, 
which is unsurprising given that \sotl\ is specifically designed to adaptively reduce the density heterogeneity.
Given the previous observation that the \sotl\ MFD lies strictly above the \scatsl\ MFD, 
this suggests that a methodology of adaptive density homogenization can provide a highly effective means of network control.

The relationship between the transient behaviors of the flow and the flow heterogeneity is more complicated, however.
We shall present a careful time series analysis of the cross correlations between $h_J$ and $J$ elsewhere.
However, the stationary behavior of the $\overline{h}_{J}$ vs $\overline{\rho}$ curves appear to contain a considerable amount of information.
In particular, for all three signal systems studied,
there appears to be a point of inflection in the stationary $\overline{h}_J$ vs $\overline{\rho}$ curve either at, or slightly before, the density $\rho_{\text c}$ at which the network reaches capacity.
Figs.~\ref{subfig:uniform-scatsl-hetero-J} and~\ref{subfig:uniform-sotl-hetero-J} show $\overline{h}_J$ vs $\overline{\rho}$ for \scatsl\ and \sotl, respectively.
For the case of \sotl\ there is in fact a long plateau, implying that for the majority of the free-flow regime the flow heterogeneity is independent of the network density in this case.

\subsubsection{Effects of internal sources/sinks}
\begin{figure}
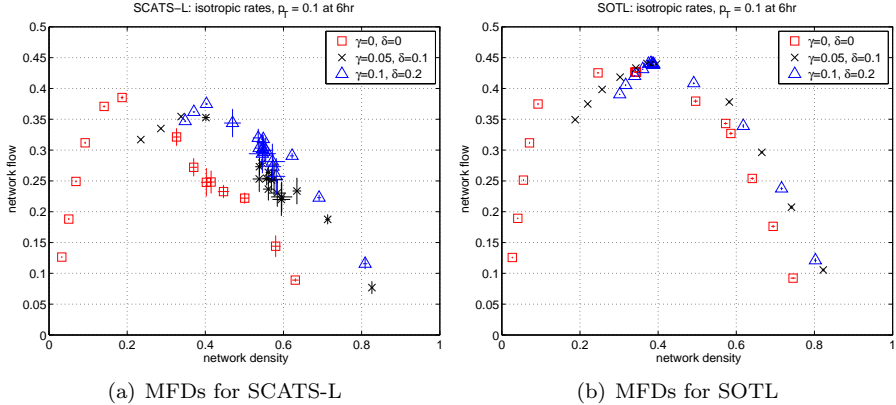

\centering
\subfigure[MFDs for \scatsl]{\label{subfig:uniform-scatsl-difgd}\fdcuscatsldifgdturnten}
\subfigure[MFDs for \sotl]{\label{subfig:uniform-sotl-difgd}\fdcusotldifgdturnten}
\caption{MFDs for \scatsl\ and \sotl\ with $\gamma = 0, 0.05, 0.1$ and $\delta = 0, 0.1, 0.2$.
Error bars corresponding to one standard deviation are shown.
}
\label{fig:fddifgd}
\end{figure}

We now consider the effect of introducing non-zero values of the internal input and output probabilities, $\gamma$ and $\delta$, which model endogenous sources/sinks such as parking garages.
Figs.~\ref{subfig:uniform-scatsl-difgd} and~\ref{subfig:uniform-sotl-difgd} show the MFDs for \scatsl\ and \sotl\
produced by varying $\alpha$ and $\beta$ for two distinct, non-zero, fixed values of $(\gamma,\delta)$, and compares them with the $(\gamma,\delta)=(0,0)$ case already 
discussed. The results for \scatsf\ are intermediate between those of \scatsl\ and \sotl.
The first observation is simply that different values of $(\gamma,\delta)$ clearly produce different MFD curves as $\alpha$ and $\beta$ are varied.
This is to be expected, since the stronger the internal sources and sinks, the more homogeneous is the network; for sufficiently strong internal sources and sinks
the effects of the boundary become essentially irrelevant.

The second observation is that for both \scatsl\ and \sotl, the MFDs are translated to the right (higher densities) as $\gamma$ and $\delta$ increase.
This is intuitively reasonable; for $(\gamma,\delta)$ strictly zero the links deep in the bulk of the network would be expected to have lower density than those near the boundary, implying that the 
aggregated network density should be lower. 
This scenario explains why the translation of the MFD is more pronounced for \scatsl\ than for \sotl, since we have already seen in Fig.~\ref{fig:uniform_heteroT10} that the 
heterogeneity of \scatsl\ is much higher than that for \sotl.

A final observation is that for \sotl\ the capacity is also marginally higher with non-zero $\gamma,\delta$ than with zero $\gamma,\delta$.
For \scatsl, there is considerably more statistical noise in the high density branch when simulating with non-zero $\gamma,\delta$, and it is not clear to what extent the capacities change, if at all.

Apart from these translations and rescalings, Figs.~\ref{subfig:uniform-scatsl-difgd} and~\ref{subfig:uniform-sotl-difgd} show that
no qualitative change in the shapes of the stationary MFDs is introduced by applying non-zero internal sources/sinks. 
By contrast, we will see in Section~\ref{sec:dependent} that when using time-dependent boundary conditions, the introduction of sufficiently strong internal sources and sinks can qualitatively affect the
behavior of the network.

\subsubsection{Turning probabilities}\label{sssec:turning}
\begin{figure}
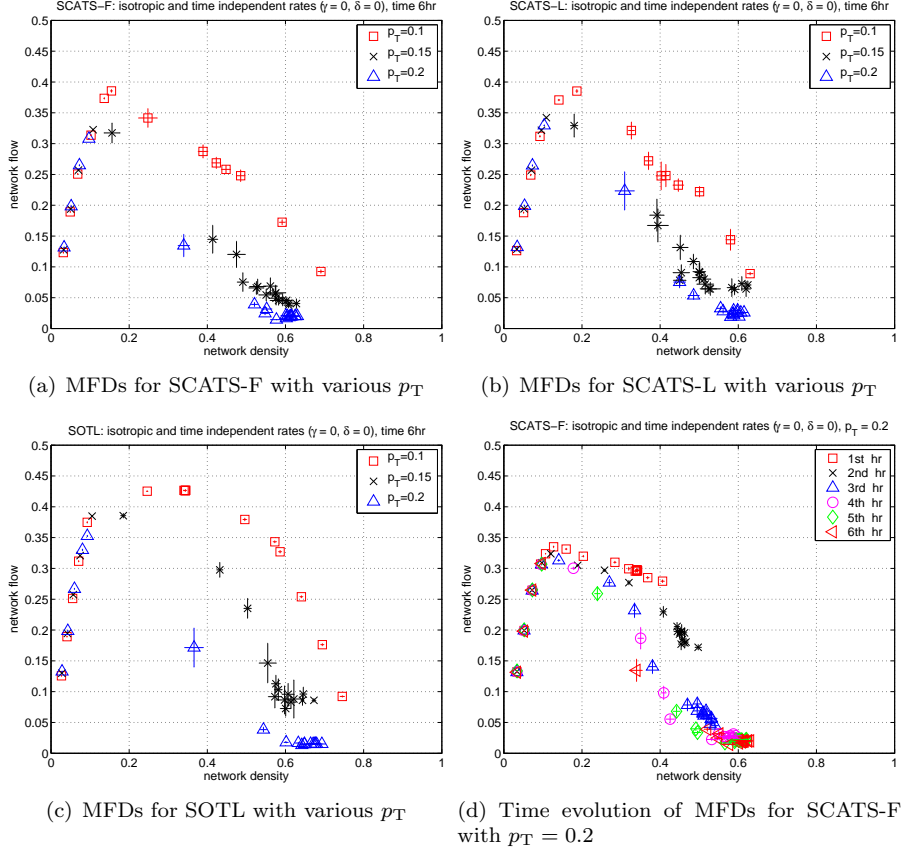

\centering
\subfigure[MFDs for \scatsf\ with various $p_{\text T}$]{\label{subfig:difT-scatsf-mfd}\fdcuscatsfturns}
\subfigure[MFDs for \scatsl\ with various $p_{\text T}$]{\label{subfig:difT-scatsl-mfd}\fdcuscatslturns}
\subfigure[MFDs for \sotl\ with various $p_{\text T}$]{\label{subfig:difT-sotl-mfd}\fdcusotlturns}
\subfigure[Time evolution of MFDs for \scatsf\  with $p_{\text T} = 0.2$]{\label{subfig:difT-scatsfhours-mfd}\fdscatsfhoursturns}
\caption{Figs.~\subref{subfig:difT-scatsf-mfd}, \subref{subfig:difT-scatsl-mfd}, and~\subref{subfig:difT-sotl-mfd} show MFDs of \scatsf, \scatsl, and \sotl,
with turning probability $p_{\text T}=0.1,0.15,0.2$, for a network with isotropic and time-independent boundary conditions.
Fig.~\subref{subfig:difT-scatsfhours-mfd} shows MFDs for \scatsf\ at hours 1,2,$\dots$,6 with $p_{\text T}=0.2$.
Error bars corresponding to one standard deviation are shown, but are often smaller than the symbol size of the data point.
}
\label{fig:fdTs}
\end{figure}

We now briefly turn our attention to the impact on MFDs of varying the turning probabilities, again using isotropic boundary conditions (Boundary Condition~(\ref{subenum:isotropic})).
Figs.~\ref{subfig:difT-scatsf-mfd}, \ref{subfig:difT-scatsl-mfd} and~\ref{subfig:difT-sotl-mfd} 
compare the MFDs produced for networks with turning probabilities $p_{\text T}=0.1,0.15,0.2$, for \scatsf, \scatsl, and \sotl, respectively.
While the low density branches are invariant with $p_{\text T}$, it is clear that the high density branches become more rapidly decaying as $p_{\text T}$ increases. 
For both $p_{\text T}=0.15$ and $p_{\text T}=0.2$, there is a critical density $\rho_{\text j}$ such that the flow remains constant for all $\overline{\rho} > \rho_{\text j}$.
This value of $\rho_{\text j}$ decreases as $p_{\text T}$ increases. Although it is not observable for $p_{\text T}=0.1$, it is presumably present in principle but at a density higher than any we simulated. 
We note that even for $\overline{\rho} > \rho_{\text j}$, the flow is non-zero, so the network is not locked into a rigid gridlock.
The value of this plateaued high-density flow can be seen to decrease as $p_{\text T}$ increases.

On a practical level, we note that for $p_{\text T}=0.15$ and $p_{\text T}=0.2$ the capacities for \scatsl\ and \scatsf\ are again very close, 
and again both are lower than the corresponding capacity observed for \sotl.

Fig.~\ref{subfig:difT-scatsfhours-mfd} shows the instantaneous MFDs at hours 1,2,\ldots,6 for a network governed by \scatsf, with $p_{\text T}=0.2$. 
The transient behavior is qualitatively the same as that shown in Fig.~\ref{subfig:uniform-scatsf-mfd} for $p_{\text T}=0.1$. 
The low density branch is stationary already by hour 1, while the high density branch decreases to its stationary limit, which is reached by hour 4 or 5. 
The extent to which the curves at hours 1 and 2 differ from their stationary limits is clearly much larger for $p_{\text T}=0.2$ than observed for $p_{\text T}=0.1$ however.
Similar behavior was observed for \scatsl\ and \sotl.

\subsubsection{Link lengths}\label{sssec:shortlinks}
As discussed in Section~\ref{ssec:links}, in general, the networks we simulate have bulk links of length 750m, with right-turning lanes of length 120m.
In this section we study the effect of varying the length of both the main lanes and the turning lanes.

We begin by considering the effect of doubling the length of the turning lanes. 
In Fig.~\ref{subfig:longTurnsSotl-mfd} we compare the MFDs produced by a system governed by \sotl\ with turning probabilities $p_{\text T}=0.15,0.20$, 
with turning lanes of length 120m (denoted short TL) and 240m (denoted long TL). All other parameters are set as in Section~\ref{sssec:turning}. 
The data for the short TL system is in fact identical to that plotted in Fig.~\ref{subfig:difT-sotl-mfd}; we repeat it here for ease of comparison with the long TL case.
For both $p_{\text T}=0.15$ and $p_{\text T}=0.2$, we see that the systems with short and long turning lanes have identical behavior at very-low and very-high densities. 
This is to be expected; at very low densities both turning lanes will be relatively unoccupied, whereas at very high densities both turning lanes will suffer from spillback.
At moderately high densities however, the MFD of the long TL system lies strictly above that of the short TL system.
We conclude that for sufficiently large turning probabilities, increasing the length of turning lanes can produce a non-trivial increase in the network's MFD at moderately high densities.

An analogous comparison of the long and short TL systems was also made with $p_{\text T}=0.1$, however there was no observable difference in this case.
This is intuitively reasonable; for fixed lane lengths, there should be a value of the turning probabilities below which the effect of congestion on the turning lanes is negligible compared to
congestion on the main lanes. In such a case, increasing the length of the turning lane further does not modify the network's MFD.
Our results show that this is already the case for $p_{\text T}=0.1$ in the short TL network.

\begin{figure}
\centering
\subfigure[MFDs using different turning lane lengths]{\label{subfig:longTurnsSotl-mfd}\fdlongturnssotl}
\subfigure[MFDs using different bulk link lengths]{\label{subfig:shortlinks-comparison-mfds}\fdshortlinks}
\caption{Fig.~\subref{subfig:longTurnsSotl-mfd} shows MFDs of \sotl\ at hour 6 with turning probabilities $p_{\text T}=0.15,0.2$, for two different lengths of the turning lanes; 120m and 240m.
The bulk links in each case are 750m.
Fig.~\subref{subfig:shortlinks-comparison-mfds} shows MFDs of \scatsf, \scatsl\ and \sotl\ at hour 6 with $p_{\text T}=0.1$, for a network with short (210m) bulk links.
In both cases, time-independent, isotropic boundary conditions are applied, with zero sources and sinks.
Error bars corresponding to one standard deviation are shown, but are often smaller than the symbol size of the data point.
}
\label{fig:fdLinkLengths}
\end{figure}

We next considered the effect of decreasing the length of the bulk links, to a length relevant to urban {\em downtown} networks.
In Fig.~\ref{subfig:shortlinks-comparison-mfds} we show the results of repeating the simulations presented in Fig.~\ref{fig:fdhoursT10} using 210m bulk links.
To facilitate comparison with Fig.~\ref{subfig:uniform-comparison-mfd}, we left the lengths of the turning lanes fixed.
A comparison of Figs.~\ref{subfig:shortlinks-comparison-mfds} and~\ref{subfig:uniform-comparison-mfd} shows a number of differences.
Firstly, for each signal system, we observe that the value of the maximum flow is lower for the network with shorter links.
For example, the maximum flow for \sotl\ drops from around $0.43$ for the network with long bulk links to around $0.39$.
Such behavior is in agreement with theoretical predictions based on variational theory presented in~\cite{DaganzoGeroliminis08}.
In addition, for all three signal systems, the slope of the low-density branch of the MFD is larger in the system with long links.
Conversely, the slope on the high-density branch is larger for the system with short links.
For the system with short links, a steep drop in the MFD is clearly observed at density $\overline{\rho}\approx 0.5$.
Beyond that density, the network becomes highly congested and the flow plateaus at a very small value, less than $0.05$.
Comparing the different signal systems, we see that again \sotl\ outperforms the \scats\ systems. 
Furthermore, the relative improvement of \sotl\ over \scats\ in the low-density branch of the MFD is significantly more pronounced for the system with short links.

Finally, we note that in reality, networks with short links would likely also have shorter turning lanes.
We therefore also repeated the simulations presented in Fig.~\ref{subfig:shortlinks-comparison-mfds} using turning lanes of length 45m.
The relationship between the resulting MFDs and Fig.~\ref{subfig:shortlinks-comparison-mfds} are found to be similar to the scenario illustrated in Fig.~\ref{subfig:longTurnsSotl-mfd}.
In this case however, we observe that the MFD of the system with the shorter turning lanes lies strictly below that of the system with longer turning lanes, even at $p_{\text T}=0.1$.
This illustrates the converse of our observation above: for a given value of $p_{\text T}$,
the high-density branch of the MFD can be decreased by shortening the length of the turning lanes sufficiently.

\subsection{Anisotropic boundary conditions}\label{ssec:biased}
In this section we present our results for simulations using Boundary Condition~(\ref{subenum:anisotropic}).
In this case, we again have two free parameters, $\alpha$ and $\beta$, 
where $\alpha$ is the input probability on each inlink on the west boundary, and $\beta$ the output probability on each outlink on the east boundary.
All other boundary inlinks have input probability $\alpha/2$ and all other boundary outlinks have output probability $2\beta$.
No internal sources or sinks are present ($\gamma=\delta=0$) and the turning probability is $p_{\text T}=0.1$.
These boundary conditions imply that the demand in the west-to-east direction is twice that of other directions.
In the presence of such anisotropy, applying linking with \scatsl\ is a very natural thing to do, and our simulations of \scatsl\ are linked in the west-to-east direction.

\begin{figure}
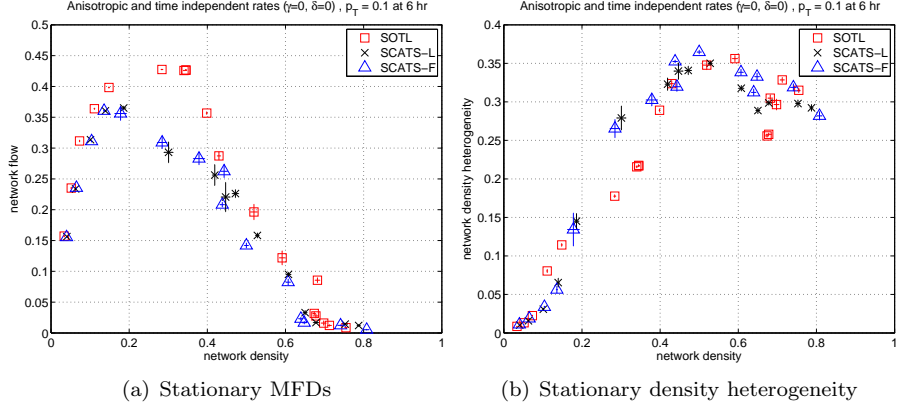

\centering
\subfigure[Stationary MFDs]{\label{subfig:biased-comparison-mfd}\fdbiasedturnten}
\subfigure[Stationary density heterogeneity]{\label{subfig:biased-comparison-heteroRho}\fdbiasedRhoturnten}
\caption{Comparison of \scatsl, \scatsf, and \sotl\ at stationarity, on a network with anisotropic $\alpha$ and $\beta$ in the west-to-east direction. 
Linking in \scatsl\ is applied along the high-demand direction.
Fig.~\subref{subfig:biased-comparison-mfd} shows the MFDs while Fig.~\subref{subfig:biased-comparison-heteroRho} shows density heterogeneity.
Error bars corresponding to one standard deviation are shown, but are often smaller than the symbol size of the data point.
}
\label{fig:fdschemesT10-aniso}
\end{figure}

In Fig.~\ref{subfig:biased-comparison-mfd} we plot a comparison of the stationary MFDs of the networks using \scatsf, \scatsl, and \sotl, corresponding to hour 6 of our simulations,
by which time each system had reached approximate stationarity. 
The transient behavior prior to stationarity is qualitatively the same as described for the case of isotropic boundary conditions discussed in Section~\ref{ssec:uniform}.
The first observation is simply that for each signal system, well-defined stationary MFDs do exist in the presence of anisotropic boundary conditions.
However, the shapes of the stationary curves are clearly quite different to those produced by isotropic boundary conditions as shown in Fig.~\ref{fig:fdhoursT10}.
While in the isotropic case the MFDs are smooth curves which qualitatively resemble a typical single-link FD, the MFDs shown in Fig.~\ref{subfig:biased-comparison-mfd} have a more intricate structure.
In particular, for all three signal systems, the stationary MFD displays two steep drops in flow. The first of these occurs for $\overline{\rho}\approx0.4$, and is most pronounced for \scats.
For \sotl, the location of this first drop is close to the density $\rho_{\text c}$ which produces maximum flow, while for \scats\ it occurs well beyond $\rho_{\text c}$.

As observed in the isotropic case for $p_{\text T}=0.2$, for densities above a critical value, $\rho_{\text j}$, an oversaturated regime with constant flow is obtained.
Again, even for $\overline{\rho} > \rho_{\text j}$ the flow remains non-zero, although it is very low. 
For all three signal systems, a second steep drop can be observed for $\overline{\rho}\approx\rho_{\text j}$. This drop is very sharp, even for \sotl.
Some insight into this behavior can be obtained by comparing with the corresponding density heterogeneities in Fig.~\ref{subfig:biased-comparison-heteroRho}. 
For all three signal systems, the location of the drop in the MFD near $\rho_{\text j}$ coincides exactly with a cusp in the density heterogeneity. 
In addition, for \scatsl\ and \scatsf, the sharp drop in the MFD near $\overline{\rho}\approx0.4$ coincides exactly with a sharp rise in the density heterogeneity.
In summary, we see that anisotropies in demand significantly affect the shape, but not the existence of the MFD curves.

We now turn to a comparison of the different signal systems. 
We begin by comparing \scatsl\ and \scatsf.
For networks with strongly anisotropic demand, one would intuitively expect signal linking to play a beneficial role.
Indeed, if one focuses only on travel times along the peak (linked) direction, the \scatsl\ system performs considerably better than \scatsf\ in the low density regime.
However, we see from Fig.~\ref{subfig:biased-comparison-mfd} that the MFDs produced by \scatsl\ and \scatsf\ are very similar.
Although \scatsl\ provides improved performance for the links in the peak direction, this beneficial treatment of the linked subsystem occurs at the expense of the non-linked routes,
and no clear net benefit is observable when using the network MFD as the performance metric.

Finally, let us compare the performance of \sotl\ with that of the \scats\ systems. At densities lower than 0.2 and higher than 0.4 all three systems perform similarly, although we note that at low densities
\sotl\ has both higher network flows and lower travel times in the linked direction than either \scatsl\ or \scatsf.
For moderate densities, however, we observe that \sotl\ has significantly higher network flow, and lower density heterogeneity, than either \scats\ system.
This suggests that a methodology of adaptive density homogenization can provide a more effective means of network control than signal linking, even in networks with inherently anisotropic demand.

\section{Simulations: Time-dependent boundary conditions}\label{sec:dependent}
\subsection{Simulations}\label{ssec:dependent}
In the previous section, we applied time-independent boundary conditions, and observed that the system took up to 6 hours to reach stationarity. 
In real-world scenarios, however, traffic demand typically varies with time, and in practice a network may never actually reach stationarity. 
Understanding transient behavior of the network dynamics is therefore of significant practical importance.

In this section we present our results for simulations using Boundary Condition~\ref{enum:time-dependent} of Section~\ref{ssec:boundary}.
We select appropriate values of $\alpha$ and $\beta$ so that we can simulate the network traffic variation during a typical weekday.
Specifically, we consider a 20 hour period, and enforce two peaks in the demand; one corresponding to the morning, the other to the afternoon.
At each instant, the same value of $\alpha$ (resp. $\beta$) is applied to each boundary inlink (resp. outlink), so the boundary conditions are isotropic. 
We consider two scenarios for the internal sources and sinks: $\gamma=\delta=0$, in which case the demand is entirely driven by the boundary; and $\gamma=\alpha$, $\delta=\beta$,
in which case the bulk input (output) occurs at the same rate as the boundary input (output). We refer to these two cases as boundary loading and uniform loading, respectively.
Unlike the time-independent case discussed in the previous section, in the time-dependent case we find that varying the relative strength of the internal sources/sinks compared with the boundary demand can 
produce rather different qualitative behavior. 

\begin{figure}
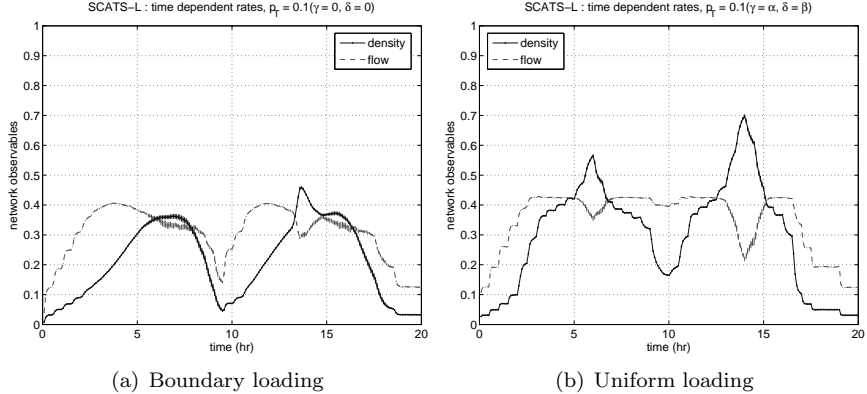

	\centering
        \subfigure[Boundary loading]{\label{scatsl_dependent_timeseries_boundary}\densityflowtdscatslturntenboundary}
        \subfigure[Uniform loading]{\label{scatsl_dependent_timeseries_uniform}\densityflowtdscatslturntenuniform}
	\caption{Time series of network-aggregated flow and density for the \scatsl\ traffic signal system under time-dependent isotropic boundary conditions.
          Error bars corresponding to one standard deviation are shown.
  }
  \label{fig:scatsl_dependent_timeseries}
\end{figure}

For each of the three signal systems, and for both boundary and uniform loading, 
the peak densities were chosen so as to drive the MFD into the high-density regime during loading, and then allow it to relax back to the stationary low-density curve during recovery.
The time series of the resulting network density and flow for \scatsl\ are shown in Fig.~\ref{fig:scatsl_dependent_timeseries};
the corresponding profiles for \scatsf\ and \sotl\ were very similar, although \sotl\ produced higher and sharper peaks in the density.

\begin{figure}
  \centering
  \subfigure[\scatsf: MFD]{\label{mfd_scatsf_dependent_bulkIO}\fdtdscatsfturntenbulkIO}
  \subfigure[\scatsf: Density heterogeneity]{\label{Heterodensity_scatsf_dependent_bulkIO}\dependentHeteroScatsfDensitybulkIO}
  \subfigure[\scatsl: MFD]{\label{mfd_scatsl_dependent_bulkIO}\fdtdscatslturntenbulkIO}
  \subfigure[\scatsl: Density heterogeneity]{\label{Heterodensity_scatsl_dependent_bulkIO}\dependentHeteroScatslDensitybulkIO}
  \subfigure[\sotla: MFD]{\label{mfd_sotl_dependent_bulkIO}\fdtdsotlturntenbulkIO}
  \subfigure[\sotla: Density heterogeneity]{\label{Heterodensity_sotl_dependent_bulkIO}\dependentHeteroSotlDensitybulkIO}
  \caption{Performance of network under time-dependent isotropic boundary conditions, when using \scatsf, \scatsl, and \sotl\ traffic signal systems with uniform loading.
    Left column: Instantaneous MFDs.
    Right column: Density heterogeneities.
    Error bars corresponding to one standard deviation are shown.
  }
  \label{fig:dependent_mfds_bulkIO}
\end{figure}

\begin{figure}
  \centering
  \subfigure[\scatsf: MFD]{\label{mfd_scatsf_dependent}\fdtdscatsfturnten}
  \subfigure[\scatsf: Density heterogeneity]{\label{Heterodensity_scatsf_dependent}\dependentHeteroScatsfDensity}
  \subfigure[\scatsl: MFD]{\label{mfd_scatsl_dependent}\fdtdscatslturnten}
  \subfigure[\scatsl: Density heterogeneity]{\label{Heterodensity_scatsl_dependent}\dependentHeteroScatslDensity}
  \subfigure[\sotla: MFD]{\label{mfd_sotl_dependent}\fdtdsotlturnten}
  \subfigure[\sotla: Density heterogeneity]{\label{Heterodensity_sotl_dependent}\dependentHeteroSotlDensity}
  \caption{Performance of network under time-dependent isotropic boundary conditions, when using \scatsf, \scatsl, and \sotl\ traffic signal systems with boundary loading.
    Left column: Instantaneous MFDs. 
    Right column: Density heterogeneities.
    Error bars corresponding to one standard deviation are shown.
  }
  \label{fig:dependent_mfds}
\end{figure}

Figs.~\ref{mfd_scatsf_dependent_bulkIO}, \ref{mfd_scatsl_dependent_bulkIO}, and~\ref{mfd_sotl_dependent_bulkIO} show the relationship between the density and flow for each of the three signal control systems studied, in
the case of uniform loading, while Figs.~\ref{mfd_scatsf_dependent}, \ref{mfd_scatsl_dependent} and~\ref{mfd_sotl_dependent}, show the analogous plots in the case of boundary loading.
Each plot is in fact a parametric plot in the density-flow plane, parameterized by time.
In each case, the density-flow curve obtained during the first 4 hours of the simulation coincides exactly with the corresponding stationary MFD.
During this period, the network is initially empty, and as the density increases the flow increases until maximum flow is obtained.
Throughout this period, the network remains uncongested, and no transient effects are observed.
However, as the morning peak in demand is approached, which occurs at around hour 6 of the simulations, we see that the density-flow curves develop non-trivial time dependences, and hysteresis effects emerge.

\subsection{Hysteresis}\label{ssec:hysteresis}
Let us analyze the observed hysteresis patterns in more detail. We begin with the case of uniform loading. Fig.~\ref{mfd_sotl_dependent_bulkIO} shows the time evolution of the MFD for \sotl. 
The system reaches capacity at approximately hour 4, after which time flow decreases as density is further increased. 
This continues until around hour 6, when the first peak in demand is reached. After this point, density drops and flow begins to increase again until it again reaches capacity, at around hour 8. 
However the curve followed during this ``recovery'' process (hours 6 to 8) lies below the curve followed in the original ``loading'' process (hours 4 to 6).
The MFD curve from hour 4 to hour 8 therefore defines a clockwise hysteresis loop.
Similar behavior occurs at later times also. Indeed, we observe a further two clockwise hysteresis loops: from capacity, to low density, back to capacity (hours 8 to 12); 
from capacity, to high density, back to capacity (hours 12 to 16). 
Finally, the system recovers from capacity to the very low density regime (hours 16 to 20); this final recovery curve is initially below the initial loading curve (hours 0 to 4), 
but coincides with it at sufficiently low densities.
Qualitatively similar behavior is also observed for \scatsl\ and \scatsf, although the size of the loops tend to be somewhat larger.

Clockwise hysteresis loops were recently observed in an empirical study of MFDs in Toulouse, presented in~\cite{BuissonLadier09}, 
where it was argued that hysteresis is caused by spatial heterogeneity in network density.
In Figs.~\ref{Heterodensity_scatsf_dependent_bulkIO}, \ref{Heterodensity_scatsl_dependent_bulkIO} and \ref{Heterodensity_sotl_dependent_bulkIO} we show plots of the density heterogeneity, under uniform loading,
for each of the three signal systems studied.
In each case, the heterogeneity displays hysteretic behavior very similar to that observed in the corresponding MFD, but with opposite orientation; the MFDs in Fig.~\ref{fig:dependent_mfds_bulkIO}
all display clockwise hysteresis loops, while their corresponding heterogeneity plots display anticlockwise loops.
This behavior is to be expected when loading is uniform;
the level of density heterogeneity produced when a network is loading is governed predominantly by the locations of the sources, whereas the level of heterogeneity produced when the network recovers will depend not only
on the locations of the sinks, but also on drivers' behavior as they disperse through the network towards those sinks.
Therefore, whenever the spatial distribution of the sources is sufficiently uniform, one would expect that the heterogeneity of loading should be smaller than that of recovery.
A recent study of the {\em two-bin} model presented in \cite{GayahDaganzo11}, under the assumption of perfectly symmetric loading, provides a simple analytical explanation of this behavior.

If the constraint of perfectly uniform loading is relaxed however, different behaviour can arise.
If loading is sufficiently non-uniform then density may in fact become more heterogeneously distributed in loading than in recovery, which would produce anticlockwise hysteresis loops in the MFD, rather than clockwise loops. 
To study this possibility, we have therefore simulated networks in which the boundary sources/sinks are stronger than the internal sources/sinks.
Situations where such behavior might arise include commuter corridors as well as arterial networks in the presence of perimeter control, or {\em gating}. 
The most extreme example of such loading is when the strength of the internal sources and sinks is set identically to zero, corresponding to the profiles in Fig.\ref{fig:dependent_mfds}.

\begin{figure}
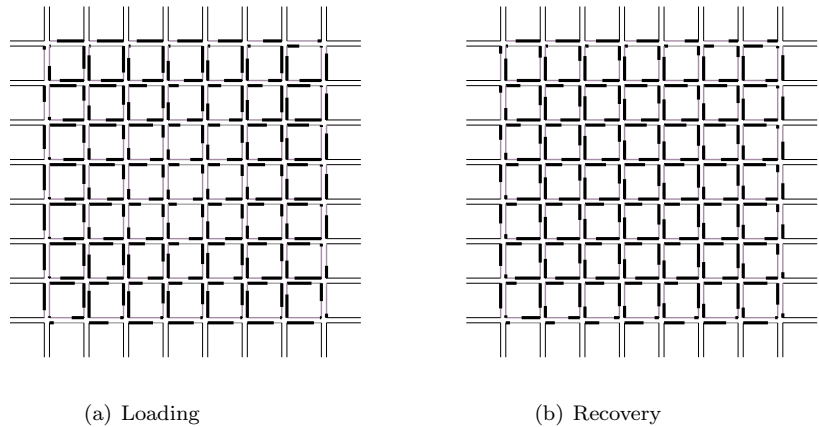

 \centering
 \subfigure[Loading]{\label{subfig:density-distribution-Loading}\densityDistributionLoading}
 \subfigure[Recovery]{\label{subfig:density-distribution-Recovery}\densityDistributionRecovery}
 \caption{Link densities in a network governed by \sotl\ with boundary-loading.
   On each link, the length of the superimposed solid line is proportional to the magnitude of that link's mean density during a specific 5-minute time interval.
   The densities on boundary links are not shown, since these are not considered part of the network, and are not included in our network-aggregated observables.
   Fig.~\subref{subfig:density-distribution-Loading} is measured during loading, 
   at time 13:25, and has $\overline{\rho}=0.504\pm0.003$, $\overline{J}=0.340\pm0.001$ and $\overline{h_{\rho}}=0.257\pm0.002$.
   Fig.~\subref{subfig:density-distribution-Recovery} is measured during recovery,
   at time 15:40, and has $\overline{\rho}=0.500\pm 0.005$, $\overline{J}=0.401\pm0.002$ and $\overline{h_{\rho}}=0.213\pm0.002$.
 }
 \label{fig:density-distributions}
 \end{figure}

Consider Fig.~\ref{mfd_sotl_dependent} for the \sotl\ signal system.
During the first 10 hours of the simulation corresponding to the lead-up to, and recovery from, the morning peak hour, we observe two distinct hysteresis loops. 
The first of these loops, which occurs at moderately high density, is traced out by the flow-density curve in an anticlockwise direction, as time evolves, while the second loop, occuring at low density, 
is traced out in a clockwise direction.
As the density then increases again due to the afternoon peak, a second, larger, anticlockwise hysteresis loop is traced out at high densities.
The low-density behavior is qualitatively similar to that observed for uniform loading, and the clockwise loops in the MFD again coincide with anticlockwise loops in the density heterogeneity,
Fig.~\ref{Heterodensity_sotl_dependent}; c.f. Figs.~\ref{mfd_scatsf_dependent_bulkIO} and~\ref{Heterodensity_scatsf_dependent_bulkIO} for example.
The behavior at moderately high density is the opposite of that observed for uniform loading, but it can again be understood from Fig.~\ref{Heterodensity_sotl_dependent} as follows.
An initially empty network is loaded just above its capacity, in a non-uniform manner which leaves some links very congested, with others still uncongested.
After the morning peak (around hour 6), as the inflow drops and outflow increases, vehicles disperse through the network, and the density becomes more evenly distributed.
Fig.~\ref{fig:density-distributions} illustrates this effect by comparing the spatial density distibrutions during loading and recovery, at a fixed value of the network density, during the afternoon peak.
We clearly observe that the centre of the network is less congested than the boundary region during loading, whereas the distribution during recovery is reasonably uniform.
These heterogeneity differences cause the anticlockwise loops seen in the MFD at high density.
In Section~\ref{sec:twobin} we show that this behavior can be predicted using a heterogeneous version of the two-bin model discussed in~\cite{GayahDaganzo11}.

The two scenarios, boundary loading and uniform loading, can be considered opposite extremes; 
in practice, one would likely expect that the strength of the internal sources and sinks would be non-zero but smaller than the boundary rates. 
We therefore also simulated a scenario with $\gamma=\alpha/3$ and $\delta=\beta/3$.
The results are intermediate between the uniform loading and boundary loading cases discussed above.
In particular, for the \sotl\ system, we observe both anticlockwise and clockwise hysteresis at high density:
anticlockwise hysteresis during hours 4 to 8 (as observed for boundary loading); and clockwise hysteresis during hours 12 to 16 (as observed for uniform loading).

Finally, let us compare the behavior of the \sotl\ simulations with the corresponding simulations of \scats. 
For both boundary loading and uniform loading, we again find that \sotl\ has lower values of density heterogeneity and higher capacities than both the \scats\ systems.
For the specific case of boundary loading, we note that while
Figs.~\ref{mfd_scatsf_dependent} and \ref{mfd_scatsl_dependent} display clockwise loops in the MFDs of \scatsl\ and \scatsf, as observed for \sotl, the anticlockwise loops appear to be absent.
By zooming in on Fig.~\ref{Heterodensity_scatsl_dependent}, one finds that there is, in fact, a small clockwise hysteresis loop in the density heterogeneity plot for \scatsl,
at the start of the final recovery process, and similarly zooming in on Fig.~\ref{mfd_scatsl_dependent} one observes a corresponding anticlockwise loop in the MFD.
However the loops are comparable to the size of the error bars in the simulations and so their existence cannot be firmly established. 
The presence of strong anticlockwise loops for \sotl\ but not for \scats\ can be understood as a consequence of \sotl's generally stronger ability to homogenize the network density.

\subsection{Two-Bin model}\label{sec:twobin}
The {\em two-bin} model was introduced by~\cite{DaganzoGayahGonzales11}. It represents the interaction between two subnetworks, or {\em bins}, in a larger road network. 
A state of the system consists of the pair $(\rho_1,\rho_2)$, and the expression for the flow $J(\cdot)$ in each bin is assumed to be given by the same triangular MFD, 
with capacity $(\rho_{\rm c},J_{\rm c})$ and jam density $\rho_{\rm j}$. The dynamical evolution of the system is defined by the following system of ordinary differential equations
\begin{subequations}\label{equ:dynamic}
  \begin{equation}
    \begin{split}
      \frac{\diff \rho_1}{\diff t} &= \frac{a_1 - b_1J(\rho_1) + p_2 J(\rho_2) - p_1 J(\rho_1) }{L_1},\\
      \frac{\diff \rho_2}{\diff t} &= \frac{a_2 - b_2J(\rho_2) + p_1 J(\rho_1) - p_2 J(\rho_2) }{L_2},
    \end{split}
  \end{equation}
  for $\rho_1,\rho_2<\rho_{\rm j}$, and
  \begin{equation}
    \frac{ d\rho_1}{d t} = \frac{d \rho_2}{d t} =0\mbox{ if } \rho_1 \text{ or } \rho_2=\rho_{\rm j}.
  \end{equation}
\end{subequations}
The model as stated in~\eqref{equ:dynamic} has 8 free parameters: $a_i$, the rate of inflow into bin $i$;
$b_i$, the proportion of traffic flowing out of the network from bin $i$; $p_i$, the proportion of traffic {\em turning} out of bin $i$ into the other bin; $L_i$, the total network length of bin $i$.

The perfectly symmetric case, in which $a_1=a_2$, $b_1=b_2$, $p_1=p_2$, and $L_1=L_2$, was studied in detail by~\cite{GayahDaganzo11}. 
It was found that if hysteresis was observed in the aggregated MFD, it was typically oriented clockwise. 
The results for the uniform loading scenario studied in Sections~\ref{ssec:dependent} and~\ref{ssec:hysteresis} are in perfect agreement with these results.

In this section, we consider instead a highly asymmetric case, in which $a_2=b_2=0$, and $p_1$ and $p_2$ are not necessarily equal.
This can be viewed as a two-bin model of the boundary loading scenario studied in Sections~\ref{ssec:dependent} and~\ref{ssec:hysteresis}.
In this interpretation, bin 1 corresponds to the links adjacent to the boundary, while bin 2 corresponds to the remainder of the bulk links, see Fig.~\ref{fig:lattice8a}.
This implies that $L_1$ is slightly smaller than $L_2$.
The aggregated density and flow are then
\begin{equation}
\rho=\frac{L_1\rho_1+L_2\rho_2}{L_1+L_2}, \qquad J = \frac{L_1J(\rho_1)+L_2J(\rho_2)}{L_1+L_2}.
\label{two bin aggreged variables}
\end{equation}

Figs.~\ref{subfig:twobinLoadingA} and~\ref{subfig:twobinUnloading} show phase plots for the system~\eqref{equ:dynamic} during loading and recovery, respectively, for a typical choice of the model parameters. 
The loading process has $a_1=0.1$ and $b_1=0$, while the recovery process has $a_1=0$ and $b_1=0.1$, so that no vehicles exit the network during loading, and no vehicles enter the network during recovery.
In both cases $p_1=0.08$ and $p_2=0.02$. 
For the perfectly symmetric case studied by~\cite{GayahDaganzo11}, trajectories that begin on the line $\rho_1=\rho_2$, remain on that line.
We note that for the asymmetric system studied here, however, this is not the case.

\begin{figure}
  \centering
  \subfigure[Loading: $a_1=0.1$, $b_1=0$, $p_1=0.08$.]{\label{subfig:twobinLoadingA}\twobinLoadingA}
  \subfigure[Recovery: $a_1=0$, $b_1=0.1$, $p_1=0.08$.]{\label{subfig:twobinUnloading}\twobinUnloading}
  \subfigure[MFD corresponding to \subref{subfig:twobinLoadingA} and \subref{subfig:twobinUnloading}.]{\label{subfig:twobinMFD}\twobinMFD}
  \subfigure[Loading: $a_1=0.1$, $b_1=0$, $p_1=0.12$.]{\label{subfig:twobinLoadingB}\twobinLoadingB}
  \caption{Solutions to \eqref{equ:dynamic} with $a_2=b_2=0$. The loading paths have $b_1=0$ while the recovery paths have $a_1=0$. 
    Fig.~\subref{subfig:twobinMFD} shows the trajectory in the $\rho-J$ plane corresponding to a typical loading/recovery process, given by the black paths 
    in~\subref{subfig:twobinLoadingA} and~\subref{subfig:twobinUnloading}.
    The star represents the maximum value of density obtained during this loading/recovery process, and the diamond (triangle) represents the maximum flow attained during loading (recovery).
    Fig.~\subref{subfig:twobinLoadingB} shows the effect on Fig.~\subref{subfig:twobinLoadingA} of increasing $p_1$.}
  \label{fig:loadingrecovery}
\end{figure}

To illustrate the effect on the MFD of combining such loading and recovery processes, let us now consider the black trajectories shown in Figs.~\ref{subfig:twobinLoadingA} and~\ref{subfig:twobinUnloading}.
In Fig.~\ref{subfig:twobinLoadingA}, the black trajectory starts with $\rho_1=0$ and $\rho_2=0$, which corresponds to the
initial state of the boundary loading scenario simulated in Section~\ref{ssec:dependent}.
Suppose we stop the loading process at a finite time, denoted by the star in Fig.~\ref{subfig:twobinLoadingA}, and then begin a recovery process.
The resulting recovery trajectory is shown in black in Fig.~\ref{subfig:twobinUnloading}.
The diamond (triangle) in Fig.~\ref{subfig:twobinLoadingA} (Fig.~\ref{subfig:twobinUnloading}) shows the location at which maximum flow was attained on the black loading (recovery) trajectory.
In Fig.~\ref{subfig:twobinMFD} we show the trajectory in the $\rho-J$ plane resulting from combining these loading and recovery processes, superimposed on the underlying MFD of the model.
There is a clearly visible anticlockwise hysteresis loop as the system recovers from high density, precisely as observed for the boundary loading scenario simulated in Section~\ref{ssec:dependent}.

Since the recovery path in Fig.~\ref{subfig:twobinUnloading} initially moves towards more balanced states, i.e. moves toward the diagonal,
the higher flow during the initial stages of recovery can be seen as a consequence of lower heterogeneity, as observed in Section~\ref{ssec:hysteresis}.
Note that the location of the maximum flow during recovery occurs precisely at zero heterogeneity, $\rho_1=\rho_2$.
As the recovery trajectory continues further, Fig.~\ref{subfig:twobinUnloading} shows that the heterogeneity increases again, 
and the corresponding trajectory in the MFD then drops below the original loading curve and a clockwise hysteresis loop results. 
This combination of anticlockwise hysteresis loops at high density and clockwise loops at low density agrees exactly with the behavior observed in Fig.~\ref{mfd_sotl_dependent}.
This behavior is quite generic, and similar MFD hysteresis is observed for any similar pairs of loading and recovery trajectories, 
whenever loading begins with $\rho_2\ll\rho_{\rm c}$ and ends with $\rho_1\gg\rho_{\rm c}$.

As a final observation, Fig.~\ref{subfig:twobinLoadingB} shows an alternative loading process, 
for which all parameters are the same as Fig.~\ref{subfig:twobinLoadingA} except that $p_1=0.12$ is slightly higher. 
While the trajectories are qualitatively similar, 
it is apparent that even a small variation in the rate that vehicles from the boundary can enter the network can produce quite different loading trajectories, starting from a given initial state. 
One significant factor that would contribute to the effective value of $p_1$ in an actual arterial network is the type of traffic signal system used. 
From the perspective of the two-bin model, it is therefore unsurprising that different types of signal systems can have rather different levels of hysteresis in their MFDs.

\section{Discussion}\label{sec:concl}
We have studied macroscopic fundamental diagrams (MFD) on a square-lattice traffic network for a variety of traffic scenarios. 
In particular we have studied networks as a function of both anisotropy (directional bias) and uniformity (presence of bulk sinks and sources) in demand. 
We furthermore studied the impact of various turning probabilities, and we have studied these networks with three different traffic signal systems, \scatsf, \scatsl\ and \sotl.

Our main findings include the following:

\begin{itemize}
\item For time-independent demand, MFDs do exist even when demand is not uniform, but their shapes depend on the nature of the non-uniformity:
\begin{itemize}
  \item Systems with more uniformly distributed demand achieve similar capacities, but capacities occur at higher densities;
  \item Systems subject to anisotropic exogenous demand display a steep drop in the flow just beyond the maximum of the MFD;
  \item As turning rates increase, capacity and jamming occur at lower densities, capacity decreases, and the congested branch of the MFD decreases;
\end{itemize}
\item For time-dependent demand, MFDs show clear hysteresis which is strongly correlated with the spatial heterogeneity of the density.
The qualitative behaviour of this hysteresis is strongly dependent on the level of uniformity of loading and unloading;
\item The choice of traffic signal system plays a crucial role in determining a network's performance.
  The idealized control scheme \sotl, which is designed to uniformize the network density distribution, always results in a higher MFD compared to the commonly used \scats\ system.
  \sotl\ increases the network capacity and produces higher flows in the congested regime.
\end{itemize}

We finally remark that given the strong dependence of the choice of traffic signal system on the shape of the MFD, one can in principle use MFDs as a metric for comparing the performance of different 
traffic signal systems.

\section*{Acknowledgments}
We gratefully acknowledge the financial support of the Roads Corporation of Victoria (VicRoads), and we thank VicRoads staff, in particular Adrian George, Andrew Wall and Hoan Ngo,
for numerous valuable discussions. We also thank Carlos Daganzo, Vikash Gayah, Yibing Wang and John Gaffney for useful discussions, and we thank two anonymous referees for their invaluable comments.
This work was supported under the Australian Research Council's Linkage Projects funding scheme (project number LP120100258), and
T.G. is the recipient of an Australian Research Council Future Fellowship (project number FT100100494).
This research was undertaken with the assistance of resources provided at the NCI National Facility through the National Computational Merit Allocation Scheme supported by the Australian Government.
We also gratefully acknowledge access to the computational facilities provided by the Monash Sun Grid.

\appendix

\section{Details of Traffic Signal Control Systems}\label{app:system}

\subsection{\scats}\label{asec:scats}

The strategy for adapting the cycle length $\mc$ based on the volume ratio $R$, defined in~\eqref{volume ratio definition}, of a master node is presented as below.
\begin{samepage}
\begin{myalg}\label{alg:scats}\scats\ cycle length decision\\
\smallskip
\begin{tabular}{l}
\hline
{\bf Case 1: \quad if }  $\mc=$ \tt{MIN}  $\& \,R> 0.4$, {\bf then} $\mc = $ \tt{STOPPER} \\
{\bf Case 2: \quad if }  $\mc=$ \tt{STOPPER}  $\& \,R < 0.2 $, {\bf then} $\mc = $ \tt{MIN} \\
{\bf Case 3: \quad if }  $R> 0.95$, {\bf then} $\mc=\min\{\mc+$ \tt{STEP}, \tt{MAX}$\}$ \\
{\bf Case 4: \quad if } $R <0.85$, {\bf then} $\mc=\max\{\mc-$ \tt{STEP}, \tt{STOPPER}$\}$ \\
{\bf Otherwise: } \quad$C$ remains unchanged.\\
\hline
\end{tabular}
\end{myalg}

\noindent The parameters used in Algorithm~\ref{alg:scats} are set as follows:

\indent\tt{MIN}: minimum cycle length $44$ seconds;

\indent\tt{STOPPER}: stopper cycle length $64$ seconds;

\indent\tt{MAX}: maximum cycle length $130$ seconds;

\indent\tt{STEP}: fixed amount of increment/decrement $6$ seconds.\\
\end{samepage}

\begin{figure}
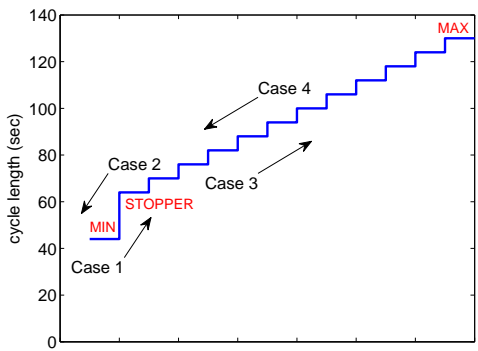

\centering
\scatsclselection
\vspace{-0.3cm}
\caption{Cycle length selection by \scats-like systems.}
\label{fig:scatsclselection}
\end{figure}

Fig.~\ref{fig:scatsclselection} illustrates the cycle length decision process implemented by Algorithm~\ref{alg:scats}.
The main strategy underlying the above algorithm is to attempt to keep the volume ratio within the range $[0.85,0.95]$.
If the volume ratio was high during the previous cycle (over $0.95$) the cycle length is increased by a fixed amount.
Otherwise if green time was wasted on the previous cycle, signaled by $R<0.85$, the cycle length is decreased.
The \tt{STOPPER} is included to allow a steep increase in the cycle length due to a sudden increase in traffic volume, when the cycle length is at its minimum.

The volume ratio of a master node, $m$, is given by
\be
R(m)=R(\ml^*,\mp^*),
\ee
where $\ml^*$ is the unique inlink flowing in the linked direction, and $\mp^*$ is the linked phase.
The cycle length of each slave node is equal to that of its master.

For a non-subsystem node, $n$, the adaptive cycle length strategy remains valid, except that the volume ratio is defined by maximizing $R$ over all inlinks and phases,
\be
R(n)=\max_{\mp}\max_{\ml} R(\ml,\mp).
\ee

\medskip
Given the cycle length $\mc$, the split time of phase $\mp$ for a master or non-subsystem node is given by
\be
\ms=\frac{d(\mp)}{\sum_{\mp} d(\mp)}[\mc-\mbox{number of phases}\times\ms_{\min}-\mbox{total amber time}] + \ms_{\min},
\ee
with the demand function $d(\mp)=\max_{\ml} \mv(\ml,\mp)$ where $\ml$ is an inlink of phase $\mp$. 
We impose a fixed delay $T_{\rm wait}$ to the nodes each time there is a phase change and the next phase does not share any path with the current one. 
During that time, only right-turning vehicles that have been giving way to others may traverse the intersection. This delay mimics the amber time for phase change.
We set $T_{\rm wait}$ to 2 seconds in our simulations, however the precise value does not impact greatly on the simulation results provided that the split times are not too small. 
For \scats, the total amber time in a cycle is 4 second.
The minimum split time in our simulations was set to $\ms_{\min}=5$ seconds.

For slave nodes, we demand that the split time of the linked phase must be the same as that of its master node. 
The remaining portion of the cycle is then shared between the other phases according to their maximum inlink volumes.

Initially, at the beginning of each simulation, the cycle length of each node is set to the minimum value and the split time plan is chosen based on the turning probabilities. 
We note that since split times are adaptive, the initial condition for the splits is unimportant.

\subsection{\sotla}\label{asec:sotl}
\sotl\ is an acyclic signal system, in the sense that no fixed ordering of the phases is imposed. 
The following algorithm provides a precise description of how \sotl\ operates at each time step and node.
The observable $\tau(n)$ acts as a clock for node $n$, recording how long the current phase has been activate for.

\begin{samepage}
{\begin{myalg}\label{alg:sotl}\sotla\ \\
\smallskip
\begin{tabular}{l}
\hline
Increment $\tau(n)$\\
{\bf for } each phase $\mp\neq \mp_{\mbox{active}}$ {\bf do}\\
\hspace{0.5cm} Increment $\tau(\mp)$\\
{\bf end for}\\
{\bf if } $\tau(n)>S_{\min}$ {\bf then}\\
\hspace{0.5cm} Let $\Pi'=\{\mp\in\Pi=\{\mp_1,\mp_2,\dots\}: \kappa(\mp)>\theta\}$\\
\hspace{0.5cm} {\bf if } $\Pi'\neq \emptyset$ {\bf then}\\
\hspace{1cm} Let $\Pi''=\{\mp\in\Pi':\kappa(\mp)=\max_{\mp'\in\Pi'} \kappa(\mp')\}$\\
\hspace{1cm} Let $\Pi'''=\{\mp\in\Pi'':\tau(\mp)=\max_{\mp''\in\Pi''} \tau(\mp'')\}$\\
\hspace{1cm} Uniformly at random choose $\mp\in\Pi'''$ and set $\mp_{\mbox{active}}=\mp$\\
\hspace{1cm} Set $\tau(\mp_{\mbox{active}})=0$\\
\hspace{1cm} Set $\tau(n)=0$\\
\hspace{0.5cm} {\bf end if}\\
\hspace{0.cm} {\bf end if}\\
\hline
\end{tabular}\medskip
\end{myalg}}
\end{samepage}

When the idle time of node $n$ is greater than the minimal split time, $S_{\min}$, \sotl\ chooses the phases for which the threshold functions $\kappa$ are greater than the threshold $\theta$, 
and among those it selects the phases with the maximal $\kappa$, then among those it selects the phases with the longest idle time. 
If there is more than one element in this latter set, then the next active phase will be chosen at random from it, however in practice there is typically only one such phase to choose from. 
The fixed delay $T_{\rm wait}$ is also applied to \sotl.
In our simulations, the minimal split time was set to $S_{\min}=5$ seconds, as was done for \scats.

\end{document}